\documentclass[9pt,bestpractices]{livecoms}
\usepackage[normalem]{ulem}
\usepackage{lipsum} 
\usepackage[version=4]{mhchem}
\usepackage{siunitx}
\DeclareSIUnit\Molar{M}
\usepackage[italic]{mathastext}
\graphicspath{{figures/}}
\usepackage{tikz,lipsum,lmodern}
\usepackage[most]{tcolorbox}
\usepackage[utf8]{inputenc}
\usepackage[T1]{fontenc}
\usepackage{amsmath}
\usepackage{hyperref}
\hypersetup{
    colorlinks=true,
    linkcolor=blue,
    filecolor=magenta,      
    urlcolor=blue,
    citecolor=cyan,
    }
\newcommand{\R}{\mathbf{R}}
\newcommand{\Rprime}{\mathbf{R'}}
\newcommand{\re}{\mathbf{r}}
\newcommand{\nacv}[2]{\mathbf{d}_{#1#2}(\R)}

\newcommand{\g}[2]{\mathbf{g}_{#1#2}(\R)}
\newcommand{\gua}[2]{\overline{\mathbf{g}}_{#1#2}(\R)}
\newcommand{\h}[2]{\mathbf{h}_{#1#2}(\R)}
\newcommand{\hua}[2]{\overline{\mathbf{h}}_{#1#2}(\R)}
\newcommand{\s}[2]{\mathbf{s}_{#1#2}(\R)}

\newcommand{\xua}[2]{\overline{\mathbf{x}}_{#1#2}(\R)}

\newcommand{\yua}[2]{\overline{\mathbf{y}}_{#1#2}(\R)}

\newcommand{\versionnumber}{1.0}  
\newcommand{\githubrepository}{\url{https://github.com/ispg-group/best-practices-namd}}  

\newcommand{\bra}[1]{\langle #1|}
\newcommand{\ket}[1]{|#1\rangle}
\newcommand{\braket}[2]{\langle #1|#2\rangle}
\newcommand{\bs}{\mathbf}

\usepackage{ulem}
\title{Best practices for nonadiabatic molecular dynamics simulations [Article v\versionnumber]}

\author[1*]{Antonio Prlj}
\corr{antonio.prlj@irb.hr}{AP}
\author[2]{Jack T. Taylor}
\author[3,4]{Ji\v{r}\'{i} Jano\v{s}}%
\author[4]{Elise Lognon}%
\author[4]{Daniel Hollas}%
\author[3]{Petr Slav\'{i}\v{c}ek}
\author[5]{Federica Agostini}
\author[4*]{Basile F. E. Curchod}
\affil[1]{Division of Physical Chemistry, Ru{\dj}er Bo\v{s}kovi\'{c} Institute, Bijeni\v{c}ka cesta 54, 10000 Zagreb, Croatia}
\affil[2]{Department of Physics, Rutgers University, Newark 07102 New Jersey, United States}
\affil[3]{Department of Physical Chemistry, University of Chemistry and Technology, Prague, Technick\'{a} 5, 16628 Prague, Czech Republic}
\affil[4]{Centre for Computational Chemistry, School of Chemistry, University of Bristol, Bristol BS8 1TS, United Kingdom}
\affil[5]{Universit\'e Paris-Saclay, CNRS, Institut de Chimie Physique UMR8000, 91405, Orsay, France}

\corr{basile.curchod@bristol.ac.uk}{BFEC}

\orcid{Antonio Prlj}{0000-0002-5589-9776}
\orcid{Jack T. Taylor}{0009-0008-2139-1916}
\orcid{Ji\v{r}\'{i} Jano\v{s}}{0000-0001-5903-8538}
\orcid{Elise Lognon}{0000-0003-0895-1188}
\orcid{Daniel Hollas}{0000-0003-4075-6438}
\orcid{Petr Slav\'{i}\v{c}ek}{0000-0002-5358-5538}
\orcid{Federica Agostini}{0000-0003-2951-4964}
\orcid{Basile F. E. Curchod}{0000-0002-1705-473X}

\blurb{This LiveCoMS document is maintained online on GitHub at \githubrepository; to provide feedback, suggestions, or help improve it, please visit the GitHub repository and participate via the issue tracker.}

\pubDOI{10.XXXX/YYYYYYY}
\pubvolume{<volume>}
\pubissue{<issue>}
\pubyear{<year>}
\articlenum{<number>}
\datereceived{Day Month Year}
\dateaccepted{Day Month Year}

\begin{document}

\begin{frontmatter}
\maketitle

\begin{abstract}
Nonadiabatic molecular dynamics simulations aim to describe the coupled electron-nuclear dynamics of molecules in excited electronic states, beyond the celebrated Born-Oppenheimer approximation. These simulations have been applied to understand a plethora of photochemical and photophysical processes and to support the interpretation of ultrafast spectroscopy experiments at advanced light sources. As a result, the number of nonadiabatic dynamics simulations has been growing significantly over the past decade. Yet, the field remains in its infancy, and a potential user may find it difficult to approach this type of simulation, given their complexity and the number of elements that should be considered for a (hopefully) successful nonadiabatic dynamics simulation. Nonadiabatic molecular dynamics relies on several key steps: finding a level of electronic-structure theory to describe the molecule in its Franck-Condon region and beyond, describing the photoexcitation process, selecting a method to perform the nonadiabatic dynamics, and analyzing the final results before calculating observables for a more direct comparison with experiment. This Best Practices guide aims to provide a general guide for the user of nonadiabatic molecular dynamics by (i) discussing the fundamentals of nonadiabatic molecular dynamics and the various trajectory-based methods developed for molecular systems, (ii) introducing the different electronic-structure methods and concepts – adiabatic/diabatic representation, conical intersections – that can be used with nonadiabatic molecular dynamics (or for benchmarking), (iii) providing details on the various steps required to perform a nonadiabatic dynamics simulation and their practical use, as well as guided examples and a discussion on the calculation of observables, (iv) proposing a FAQ with the typical questions a user may have when performing nonadiabatic dynamics, and (v) sketching a checklist for the key practical steps when performing a (trajectory-based) nonadiabatic molecular dynamics. Each section is self-contained, but we endeavor to provide additional key references for each concept discussed, making this Guide a starting point for the interested reader to dig further into the field of nonadiabatic dynamics.  
\end{abstract}

\end{frontmatter}

\tableofcontents

\section{Introduction}

Over the centuries, chemists have mastered the development of simple phenomenological rules to predict, or at least anticipate, the outcome of chemical reactions. The advent of quantum mechanics, and more specifically the Born-Oppenheimer approximation, provided a sort of framework to rationalize some of the more empirical rules of chemistry. The Born--Oppenheimer approximation provided the concept of potential energy surfaces (PESs) for the nuclear degrees of freedom of molecules, focusing in particular on the ground electronic state, where electrons are arranged in their most stable configuration around the nuclei. The development of quantum chemistry (in combination with the application of statistical mechanics to chemical problems) further sedimented a global understanding of the fate and reactivity of molecules in their ground-electronic state, with some notable exceptions and challenges.\cite{carpenter2005nonstatistical,tantillo2021beyond} 

Determining the critical points of the ground-state PES is a central task in studying the mechanisms of chemical reactions. Using an approximation to the electronic Schr\"{o}dinger equation -- an electronic-structure method -- allows us to obtain electronic energies for different geometries of a molecule. Equilibrium constants or reaction rate coefficients can be approximated based on the energy differences at different critical points on the PES, combined (often) with a harmonic approximation to describe the nuclear degrees of freedom of a molecular system. The connection between these characteristic features of the PES and the chemical reactivity of a molecule is due to the equilibrium nature of a molecule in the ground state, but also the validity of the Born-Oppenheimer approximation. The sampling of configuration or phase space using, for instance, molecular dynamics simulations is used only when the dimensionality of the problem (and therefore the complexity of its configuration space) is such that other strategies (localization of critical points of the ground-state PES, i.e., minima and transition states) are unfeasible to determine thermodynamics quantities.

\begin{figure*}[h!]
    \centering
    \includegraphics[width=1.0\linewidth]{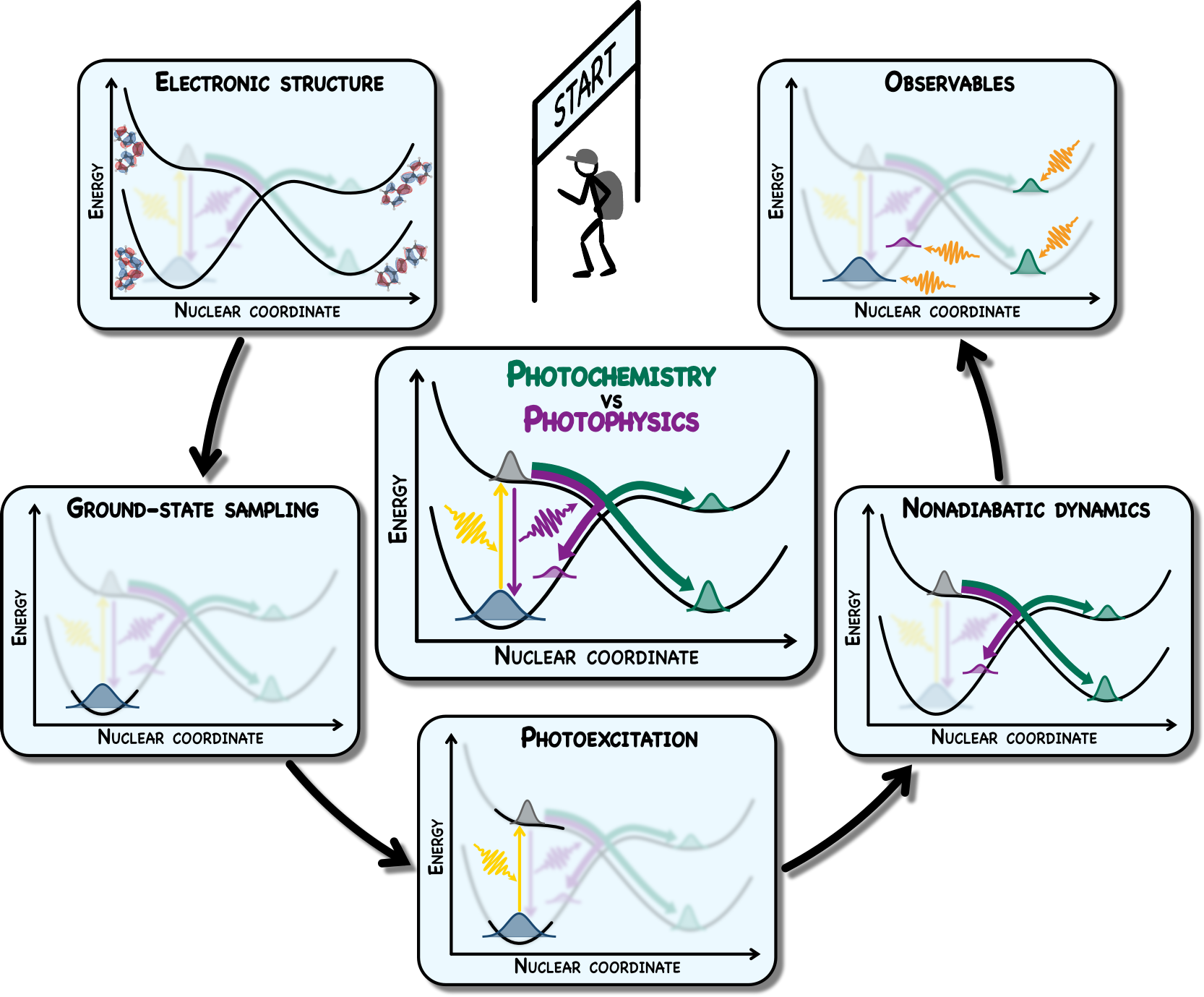}
    \caption{Schematic representation of a photochemical (green) and photophysical (purple) process (center), surrounded by the different key steps necessary to perform a nonadiabatic molecular dynamics simulations and discussed in this Guide. Concepts of electronic-structure theory for excited electronic states (adiabatic/diabatic representation, conical intersections) and various electronic-structure methods are discussed in Section~\ref{elecstructproblem}, and practical considerations on the selection of an electronic-structure methods can be found in Section~\ref{sec:banchmarkelstr}. Section~\ref{sec:initconds} offers a discussion on the generation of initial conditions for nonadiabatic molecular dynamics, namely ground-state sampling and the description of photoexcitation. The fundamental equations of nonadiabatic dynamics are discussed in Section~\ref{setting-the-scene}, the approximations leading to methods for nonadiabatic molecular dynamics are introduced in Section~\ref{nonadiabaticdynproblem}, and their practical use is described in Section~\ref{sec:performingNAMD}. Section~\ref{sec:analysisandobservables} offers a brief survey on the calculation of experimental observables from nonadiabatic molecular dynamics simulations. The Guide also includes a FAQ (Section~\ref{sec:faq} and a Checklist for nonadiabatic molecular dynamics (Section~\ref{sec:checklist}) with its rationale (Section~\ref{sec:rationale}).}
    \label{fig:guidescheme}
\end{figure*}

Add a single photon of visible or UV light to this mix, triggering the electronic excitation of a molecule, and the predictive power of the chemists' toolbox decreases significantly. Photophysical reactions -- when a molecule absorbs a photon of light but its nuclei, overall, do not change their position in the process significantly (blue path in the central panel of Figure~\ref{fig:guidescheme}) -- and photochemical reactions -- when a molecule absorbs a photon of light and its nuclei undergo a significant change in their position, leading to the formation of photoproducts (red path in the central panel of Figure~\ref{fig:guidescheme}) -- are two cases of light-triggered processes often resulting in unexpected outcomes. The past decade has seen a dramatic increase in the interest of chemists in light-induced processes such as the capture or storage of sunlight,\cite{ponseca2017ultrafast} the emission of light at low electricity cost,\cite{hong2021brief} the use of light as a green reactant in chemical synthesis,\cite{protti2009contribution} the use of light in medical applications like imaging and sensing,\cite{gautier2022fluorescence,shi2019recent} or the interpretation of time-resolved experiments conducted at advanced light sources like free-electron lasers.\cite{chergui2009electron}

As alluded to in the previous paragraph, triggering the formation of excited electronic states breaks some of the key assumptions of ground-state chemical reactivity. First of all, the molecule finds itself in an out-of-equilibrium state upon light absorption, and it may visit various regions of nuclear configuration space it would not visit while in its ground electronic state. We should stress that the precise strategy employed to excite the molecule will lead to different molecular (excited) states and a different response of the molecule: a continuous-wave (CW) laser may trigger the formation of a vibrational stationary state in an excited electronic state, while a much shorter laser pulse could trigger the formation of a nuclear wavepacket (a superposition of multiple vibrational states) in the excited electronic states\cite{suchan2018importance} -- or even the formation of an electronic wavepacket when an attosecond pulse is employed. The previous statement makes it clear that light-induced processes often unravel the quantum nature of the nuclei, an aspect typically not considered for ground-state reactivity. But that is not all: excited electronic states can come close in energy, meaning that nuclear motion can trigger a change of electronic states: the coupling between electronic and nuclear motion, often referred to as 'nonadiabatic coupling', is nothing but a breakdown of the Born-Oppenheimer approximation. 

The previous paragraph may sound as if the study of excited electronic states requires only an extension to the usual ground-state protocol and the calculation of critical points for more than one PES. But accounting for the fact that light-induced reactions are inherently out-of-equilibrium, at least in their early times, the main concepts of minimum energy paths, successful in predicting chemical reactions in the ground state, will run out of steam when the actual nuclear dynamics plays a critical role in the nonradiative relaxation process, breaking the Born-Oppenheimer approximation.\cite{levine2007isomerization,liu2016dyncorrel} This observation constitutes one of the reasons why the predictive power of Lewis structures cannot be easily extended to photochemical processes. Hence, understanding light-induced processes often requires a description of their underlying nuclear dynamics or, more precisely, their nonadiabatic molecular dynamics -- the dynamics allowing electron and nuclear motion to couple beyond the Born-Oppenheimer approximation, leading to transfer of the molecule between electronic states without the emission of light.

Despite the growing number of applications of (trajectory-based) nonadiabatic molecular dynamics in chemistry, it is surprising that no single go-to reference or book exists to date. By go-to reference, we mean here a reference for any researchers interested in computational photochemistry (newcomer in the field or expert) who is looking for (i) a brief survey of the various available techniques in (excited-state) electronic-structure methods and nonadiabatic dynamics, (ii) a summary of the main concepts and vocabulary encountered in nonadiabatic molecules dynamics, (iii) a step-by-step guide to conduct a nonadiabatic molecular dynamics simulations, (iv) practical advice and best practices on how to best conduct such complex simulations, and (v) a FAQ of common questions in the field. This Guide is meant to fill this gap and will discuss all the steps necessary for a nonadiabatic molecular dynamics simulation as depicted in Figure~\ref{fig:guidescheme}, from a fundamental and practical perspective. The caption of Figure~\ref{fig:guidescheme} provides the different sections where a given topic is covered. The living nature of this Guide hopefully means that it will contains the most up-to-date recommendations on how to best perform nonadiabatic molecular dynamics simulations. We encourage the Community to contribute to this effort and stimulate the evolution of this Guide by submitting suggestions and updates via our GitHub repository at \githubrepository.

We note at this point that this Guide mostly focuses on photochemical and photophysical processes taking place in the gas phase, simulated using trajectory-based nonadiabatic approaches and various electronic-structure methods. In the FAQ (Section~\ref{sec:faq}), we provide some background discussion and key references on various topics not covered in detail in the original version of this work. Future iterations of this Guide will hopefully contain more information on the inclusion of an environment and the use of quantum-dynamics methods.

\section{Prerequisites}

This Guide assumes background knowledge of quantum chemistry, (time-dependent) quantum mechanics, and photochemistry. Several monographs and reviews cover these aspects, and we refer the reader to the following examples for quantum chemistry,\cite{mcquarrie2008quantum,atkins2011molecular,szabo1989modern,tddftcarsten} time-dependent quantum mechanics,\cite{schatz2002quantum,tannor_book} and (theoretical) photochemistry.\cite{turro2009principles,michl,persico2018photochemistry} 

\section{Underlying concepts and methods for computational photochemistry}
\subsection{Setting the formal scene for nonadiabatic molecular dynamics}
\label{setting-the-scene}

Before defining the main ingredients necessary to perform nonadiabatic molecular dynamics, a detour is required to define where such ingredients are coming from and what the theoretical background of nonadiabatic molecular dynamics is.

Studying the dynamics of a photoexcited molecule implies that we follow the evolution of the \textit{state} of said molecule over time. Hence, nonadiabatic molecular dynamics is concerned with solving the time-dependent molecular Schr\"{o}dinger equation for the system of interest,
\begin{equation}
i\hbar  \frac{\partial}{\partial t} \Psi(\mathbf{r},\mathbf{R},t)=\hat{H}(\mathbf{r},\mathbf{R})\Psi(\mathbf{r},\mathbf{R},t) \, ,
\label{eq:tdse}
\end{equation}
where $\Psi(\mathbf{r},\mathbf{R},t)$ is the molecular time-dependent wavefunction describing the state of our molecular system, with $\mathbf{R}$ and $\mathbf{r}$ being collective variables for the coordinates of the $N_\text{n}$ nuclei and $N_{\text{el}}$ electrons forming the molecule. $\hat{H}(\mathbf{r},\mathbf{R})$ is the full molecular Hamiltonian and includes the nuclear kinetic energy $\hat{T}_{\text{n}}(\mathbf{R})$ and the so-called electronic Hamiltonian $\hat{H}_{\text{el}}(\mathbf{r},\mathbf{R})$
\begin{align}
\hat{H}(\mathbf{r},\mathbf{R})&=\hat{T}_{\text{n}}(\mathbf{R})+ \hat{H}_{\text{el}}(\mathbf{r},\mathbf{R})  \notag \\
&= \sum_\nu^{N_{\text{n}}}\frac{-\hbar^2}{2M_\nu}\nabla_{\mathbf{R}_\nu}^2+\hat{T}_{\text{e}}(\mathbf{r})+\hat{V}_{\text{ee}}(\mathbf{r})+\hat{V}_{\text{nn}}(\mathbf{R})+\hat{V}_{\text{en}}(\mathbf{r},\mathbf{R})  \, .
\label{eq:fullHamiltonian}
\end{align}
The electronic Hamiltonian consists of the kinetic energy operator for the electrons $\hat{T}_{\text{e}}(\mathbf{r})$, the electron-electron $\hat{V}_{\text{ee}}(\mathbf{r})$, nucleus-nucleus $\hat{V}_{\text{nn}}(\mathbf{R})$, and electron-nucleus $\hat{V}_{\text{en}}(\mathbf{r},\mathbf{R})$ interaction. Each nucleus is labeled by $\nu$ with corresponding mass $M_\nu$. We note that Eq.~\eqref{eq:tdse} is non-relativistic, and any relativistic effects required for the molecule of interest must be added as a patch to the formalism described here and below. $\hat{H}(\mathbf{r},\mathbf{R})$ can also become time-dependent if the interaction between a molecule and a (classical) time-dependent electromagnetic field (like a laser pulse) is included explicitly.\cite{tannor_book} 
 
By focusing solely on the electronic Hamiltonian $\hat{H}_{\text{el}}(\mathbf{r},\mathbf{R})$, one can solve an electronic structure problem for a fixed nuclear configuration, the so-called electronic time-independent Schr\"{o}dinger equation:
\begin{equation}
\hat{H}_{\text{el}}(\mathbf{r},\mathbf{R}) \Phi_J(\mathbf{r};\mathbf{R}) = E^{\text{el}}_{J}(\mathbf{R})\Phi_J(\mathbf{r};\mathbf{R}) \, .
\label{eq:tise}
\end{equation}
$\Phi_J(\mathbf{r};\mathbf{R})$ is the electronic wavefunction for electronic state $J$, and the use of a semi-colon ';' indicates that this electronic wavefunction is defined in Eq.~\eqref{eq:tise} for a given nuclear configuration.  $E^{\text{el}}_{J}(\mathbf{R})$ is the corresponding adiabatic electronic energy for electronic state $J$ at the nuclear configuration $\mathbf{R}$. Solving Eq.~\ref{eq:tise}, or more precisely, finding approximations to this equation, is the work of quantum chemistry and the development of electronic-structure methods -- this will be further detailed in Section~\ref{elecstructproblem}.

The electronic time-independent Schr\"{o}dinger equation, Eq.~\eqref{eq:tise}, is solved for a given nuclear configuration $\mathbf{R}$. We can vary this nuclear configuration along all possible nuclear coordinates of our molecular system to obtain 'electronic energies for each electronic state as a function of the nuclear configuration', the definition of PESs. Following this procedure of solving Eq.~\ref{eq:tise} for any possible nuclear configurations also provides electronic wavefunctions defined for any point of the nuclear configuration space. This realization is critical to offer a \textit{representation} for the molecular wavefunction: if we know the electronic wavefunction for each electronic state for any nuclear configuration, we can use these functions as a basis to express the time-dependent nuclear wavefunction $\Psi(\mathbf{r},\mathbf{R},t)$:

\begin{equation}
\Psi(\mathbf{r},\mathbf{R},t)=\sum_J\Phi_J(\mathbf{r};\mathbf{R})\chi_J(\mathbf{R},t) \, .
\label{eq:BH}
\end{equation}
Eq.~\eqref{eq:BH} is called the \textit{Born-Huang representation} and expresses the time-dependent molecular wavefunction in products of a time-independent electronic wavefunction $\Phi_J(\mathbf{r};\mathbf{R})$ and a corresponding time-dependent nuclear wavefunction (coefficient) $\chi_J(\mathbf{R},t)$, summed over all electronic states. 

The Born-Huang representation gives a formally exact representation of the time-dependent molecular wavefunction\footnote{For an overview of the mathematical treatment of the Born-Huang representation and a discussion of its formal limitations, see Ref.~\citenum{huggett2024quantumtheorymoleculesrigour}.} and is at the heart of our way to picture photochemical processes: time-dependent nuclear wavefunctions evolved on (time-independent) PESs given by the electronic wavefunctions. To make this observation more apparent, we can substitute the time-dependent molecular wavefunction in Eq.~\eqref{eq:tdse} by the Born-Huang representation and perform the following steps of algebra: (1) apply the molecular Hamiltonian onto the Born-Huang representation (note that the electronic wavefunctions are eigenfunctions of the electronic Hamiltonian and that the nuclear kinetic energy operator acts not only on the nuclear wavefunctions but also on the electronic ones), (2) apply the partial derivative with respect to time on the Born-Huang representation, (3) left multiply both sides of the equation by $\Phi_I^\ast(\mathbf{r};\mathbf{R})$, (4) integrate over the electronic coordinates (remembering that the electronic wavefunctions form an orthornormal basis). As a result, we obtain a set of coupled equations of motion (for $I=0, 1, 2, \dots$) for each nuclear amplitude:

\begin{align}
i\hbar\frac{\partial}{\partial t} \chi_0(\mathbf{R},t) & = \left(\sum_\nu^{N_{\text{n}}}\frac{-\hbar^2}{2M_\nu} \nabla_{\mathbf{R}_\nu}^2 + E^{\text{el}}_{0}(\mathbf{R})\right)\chi_0(\mathbf{R},t) + \sum_J \hat{C}_{0J}(\mathbf{R}) \chi_J(\mathbf{R},t) \notag \\
i\hbar\frac{\partial}{\partial t} \chi_1(\mathbf{R},t) & = \left(\sum_\nu^{N_{\text{n}}}\frac{-\hbar^2}{2M_\nu} \nabla_{\mathbf{R}_\nu}^2 + E^{\text{el}}_{1}(\mathbf{R})\right)\chi_1(\mathbf{R},t) + \sum_J \hat{C}_{1J}(\mathbf{R}) \chi_J(\mathbf{R},t) \notag \\
i\hbar\frac{\partial}{\partial t} \chi_2(\mathbf{R},t) & = \left(\sum_\nu^{N_{\text{n}}}\frac{-\hbar^2}{2M_\nu} \nabla_{\mathbf{R}_\nu}^2 + E^{\text{el}}_{2}(\mathbf{R})\right)\chi_2(\mathbf{R},t) + \sum_J \hat{C}_{2J}(\mathbf{R}) \chi_J(\mathbf{R},t) \notag \\
i\hbar\frac{\partial}{\partial t} \chi_3(\mathbf{R},t) & = \left(\sum_\nu^{N_{\text{n}}}\frac{-\hbar^2}{2M_\nu} \nabla_{\mathbf{R}_\nu}^2 + E^{\text{el}}_{3}(\mathbf{R})\right)\chi_3(\mathbf{R},t) + \sum_J \hat{C}_{3J}(\mathbf{R}) \chi_J(\mathbf{R},t) \notag \\
& \vdots
\end{align}
which can be rewritten in a matrix-vector form,
\begin{align} 
 i\hbar & \frac{\partial}{\partial t} 
     \begin{pmatrix}
        \chi_0(\mathbf{R},t)  \\
        \chi_1(\mathbf{R},t) \\
        \vdots
    \end{pmatrix} =  \notag \\
   & \begin{pmatrix}
        -\sum_\nu^{N_{\text{n}}}\frac{\hbar^2}{2M_\nu} \nabla_{\mathbf{R}_\nu}^2 + C_{00}(\mathbf{R})  & \hat{C}_{01}(\mathbf{R}) & \cdots \\
        \hat{C}_{10}(\mathbf{R}) & -\sum_\nu^{N_{\text{n}}}\frac{\hbar^2}{2M_\nu} \nabla_{\mathbf{R}_\nu}^2 + C_{11}(\mathbf{R}) & \cdots  \\
        \vdots & \vdots & \ddots
    \end{pmatrix}  \notag \\
    & \times \begin{pmatrix}
        \chi_0(\mathbf{R},t)  \\
        \chi_1(\mathbf{R},t) \\
        \vdots
    \end{pmatrix} + 
    \begin{pmatrix}
        E^{\text{el}}_{0}(\mathbf{R})   & 0 & \cdots \\
        0 & E^{\text{el}}_{1}(\mathbf{R})  & \cdots  \\
        \vdots & \vdots & \ddots
    \end{pmatrix} 
    \begin{pmatrix}
        \chi_0(\mathbf{R},t)  \\
        \chi_1(\mathbf{R},t) \\
        \vdots
    \end{pmatrix}
    \, ,
\label{eq:matrixtdse}
\end{align}
or, in compact notation,
\begin{equation} 
 i\hbar \frac{\partial}{\partial t} \boldsymbol{\chi} = \textbf{T}_\text{n} \boldsymbol{\chi} + \textbf{H}_\text{el}  \boldsymbol{\chi}
    \, .
\label{eq:matrixtdsecomp}
\end{equation}
The evolution of the nuclear wavefunction assigned to a given electronic state is dictated by the nuclear kinetic energy operator and the PES for this electronic state (the evolution term that depends only on one electronic state) \textit{and} a term responsible for the coupling between the nuclear wavefunction in the electronic state of interest with the nuclear amplitudes for all the other electronic states. In the literature, this set of coupled equations is often written in a shorthand notation:
\begin{equation}
i\hbar\frac{\partial}{\partial t} \chi_I(\mathbf{R},t)  = \left(\sum_\nu^{N_{\text{n}}}\frac{-\hbar^2}{2M_\nu} \nabla_{\mathbf{R}_\nu}^2 + E^{\text{el}}_{I}(\mathbf{R})\right)\chi_I(\mathbf{R},t) + \sum_J \hat{C}_{IJ}(\mathbf{R}) \chi_J(\mathbf{R},t) \, 
\label{eq:tdseBH}
\end{equation}
and the terms $\hat{C}_{IJ}(\mathbf{R})$ are responsible for all the so-called nonadiabatic couplings, that is, the couplings between electronic and nuclear motion responsible for the transfer of nuclear amplitude between electronic states. This term is defined as
\begin{align}
 \hat{C}_{IJ}(\mathbf{R}) = &  \sum_\nu^{N_{\text{n}}}-\frac{\hbar^2}{2M_\nu}\bra{\Phi_I(\mathbf{R})}\nabla_{\mathbf{R_\nu}}^2\ket{\Phi_J(\mathbf{R})}_{\mathbf{r}} \notag \\ 
 & -\sum_\nu^{N_{\text{n}}}\frac{\hbar^2}{M_\nu} \bra{\Phi_I(\mathbf{R})}\nabla_{\mathbf{R_\nu}}\ket{\Phi_J(\mathbf{R})}_{\bs r}\cdot\nabla_{\mathbf{R}_\nu} \, 
\label{eq:tdseBH2}
\end{align}
with $D_{IJ}(\mathbf{R})=\bra{\Phi_I(\mathbf{R})}\nabla_{\mathbf{R}}^2\ket{\Phi_J(\mathbf{R})}_{\mathbf{r}}$ being the second-order nonadiabatic couplings and $\mathbf{d}_{IJ}(\mathbf{R})=\bra{\Phi_I(\mathbf{R})}\nabla_{\mathbf{R}}\ket{\Phi_J(\mathbf{R})}_{\bs r}$ being the first-order nonadiabatic coupling vectors (NACVs). We note that the diagonal terms $\mathbf{d}_{II}(\mathbf{R})=0$ for real electronic wavefunctions, while $D_{II}(\mathbf{R})$ are not necessarily zero. The nonadiabatic coupling terms $\mathbf{d}_{IJ}(\mathbf{R})$ and $D_{IJ}(\mathbf{R})$ are responsible for non-Born-Oppenheimer effects, that is, effects resulting from the coupling between electronic and nuclear motion. We will discuss these terms and their behavior in further details below.

The nonadiabatic coupling terms defined above couple electronic states with nuclear motion and are the terms neglected within the Born-Oppenheimer approximation. If we were to neglect all the $\mathbf{d}_{IJ}(\mathbf{R})$ and  $D_{IJ}(\mathbf{R})$ terms with $I \neq J$, we would obtain the following equation of motion for the nuclear wavefunction in a given electronic state $I$  
\begin{equation}
i\hbar\frac{\partial}{\partial t} \chi_I(\mathbf{R},t)  = \left(\sum_\nu^{N_{\text{n}}}\frac{-\hbar^2}{2M_\nu} \nabla_{\mathbf{R}_\nu}^2 + E^{\text{el}}_{I}(\mathbf{R}) + C_{II}(\mathbf{R}) \right) \chi_I(\mathbf{R},t) \, 
\label{eq:tdseBO}
\end{equation}
which means that the nuclear amplitude evolves only within a given electronic state $I$ and cannot be transferred to another electronic state -- the electrons adapt instantaneously to any nuclear motion to remain in an adiabatic electronic state $I$ and the molecule evolves adiabatically in this state; this is the Born-Oppenheimer approximation. Eq.~\eqref{eq:tdseBO} is strictly equivalent to the equation of motion one would obtain by substituting the time-dependent molecular wavefunction in Eq.~\eqref{eq:tdse} by the Ansatz $\Psi(\mathbf{r},\mathbf{R},t) \approx \Phi_I(\mathbf{r};\mathbf{R})\chi_I(\mathbf{R},t)$, the well-known Born-Oppenheimer approximation. Using the Born-Oppenheimer Ansatz leads to the appearance of  $C_{II}(\mathbf{R})$ in the equation of motion for the nuclear wavefunction (see Eq.~\eqref{eq:tdseBO}) -- this term is coined the diagonal Born-Oppenheimer correction and one can include its effect on the (Born-Oppenheimer) PESs, $E^{\text{el}}_{I}(\mathbf{R})$, to produce so-called Born-Huang surfaces, i.e., $E^{\text{BH}}_{I}(\mathbf{R}) = E^{\text{el}}_{I}(\mathbf{R}) + C_{II}(\mathbf{R})$.\cite{Levine_JCP2016,hush2017cusp,ibele2021diabolical} If one neglects the diagonal Born-Oppenheimer correction -- which is the most commonly employed framework in practice -- we obtain the following equation, 
\begin{equation}
i\hbar\frac{\partial}{\partial t} \chi_I(\mathbf{R},t)  = \left(\sum_\nu^{N_{\text{n}}}\frac{-\hbar^2}{2M_\nu} \nabla_{\mathbf{R}_\nu}^2 + E^{\text{el}}_{I}(\mathbf{R}) ) \right) \chi_I(\mathbf{R},t) \, 
\label{eq:tdseBOA}
\end{equation}
which is often called the \textit{adiabatic Born-Oppenheimer approximation} in the literature.\cite{worth2004beyond} We note that we cannot suggest an approximation to the molecule wavefunction that would lead to Eq.~\eqref{eq:tdseBOA} upon insertion in the molecular time-dependent Schr\"{o}dinger equation.

We finally note that, in practice, the summation in Eq.~\eqref{eq:tdseBH} is truncated to a subset of electronic states that are sufficient to describe the process of interest. The approximation of retaining only a subset of $N_\text{states}$ coupled electronic states is often referred to as the \textit{group Born-Oppenheimer approximation}\cite{conicalintersection2004} and leads to the following coupled equations of motion:
\begin{equation}
i\hbar\frac{\partial}{\partial t} \chi_I(\mathbf{R},t)  = \left(\sum_\nu^{N_{\text{n}}}\frac{-\hbar^2}{2M_\nu} \nabla_{\mathbf{R}_\nu}^2 + E^{\text{el}}_{I}(\mathbf{R})\right)\chi_I(\mathbf{R},t) + \sum_J^{N_{\text{states}}} \hat{C}_{IJ}(\mathbf{R}) \chi_J(\mathbf{R},t) \, .
\label{eq:tdsegBH}
\end{equation}

Where are we at this stage? We defined the \textit{formally exact} framework of the Born-Huang representation, where nuclear amplitudes are propagated in time thanks to the set of coupled equations of motion given by Eq.~\eqref{eq:tdsegBH}. This propagation, however, requires electronic-structure quantities: the electronic energies $E^{\text{el}}_{I}(\mathbf{R})$ for the electronic states considered as well as the nonadiabatic coupling terms to describe nonadiabatic effects. These quantities are obtained by solving, for any necessary nuclear configurations, the time-independent electronic Schr\"{o}dinger equation, Eq.~\eqref{eq:tise}. Hence, nonadiabatic molecular dynamics, that is, the practical solution of an approximate form of Eq.~\eqref{eq:tdsegBH} for a molecule in its full dimensionality, will require getting these electronic-structure quantities (and others, like nuclear gradients) -- the electronic-structure problem discussed in Section~\ref{elecstructproblem} -- and performing the nuclear (nonadiabatic) dynamics based on approximations to Eq.~\eqref{eq:tdsegBH} -- the nuclear dynamics problem summarized in Section~\ref{nonadiabaticdynproblem}.

\subsection{Electronic structure for excited electronic states}
\label{elecstructproblem}

To determine the electronic structure of a molecule, one needs to calculate a manifold composed of a ground and (multiple) electronically excited states. Electronic states of molecules, characterized by their electronic energies and electronic wavefunctions, are obtained by solving an eigenvalue problem of the stationary electronic Schr{\"o}dinger equation with the nuclei fixed -- Eq.~\eqref{eq:tise}. By solving the electronic Schr\"{o}dinger equation for various nuclear geometries, one can scan the PES of each electronic state. Electronic states mapped in this way are called adiabatic electronic states, and they will be our primary focus throughout the text. In the following, we start by providing a discussion on a series of concepts of key importance when characterizing the (in-principle exact) electronic states of a molecule, namely the adiabatic and diabatic representation for the electronic states (Section~\ref{sec_adiabatic_diabatic}) and conical intersections (CXs) (Section~\ref{ch2_cx}). 

We then discuss practical approximations to the electronic Schr{\"o}dinger equation (Eq.~\eqref{eq:tise}), at the heart of the different electronic-structure methods for excited electronic states, focusing on their pros and cons for nonadiabatic molecular dynamics. Two broad families of electronic structure methods are used to calculate electronic quantities of importance for nonadiabatic dynamics (e.g., electronic energies, nuclear gradient, nonadiabatic coupling terms): the approaches based on an approximation of the electronic wavefunction -- the wavefunction-based methods -- and the techniques focusing on the electronic density -- the density-based methods, within which we can find formal, rigorous approaches as well as semiempirical strategies. Section~\ref{sec:wft} discusses the wavefunction-based methods and Section~\ref{sec:dftbased} focuses on the density-based approaches. We also note that Section~\ref{sec:banchmarkelstr} will address the challenge of selecting an appropriate electronic-structure method for a molecule of interest.

\subsubsection{Adiabatic and diabatic representations}
\label{sec_adiabatic_diabatic}

\begin{figure*}[h!]
    \centering
    \includegraphics[width=0.8\linewidth]{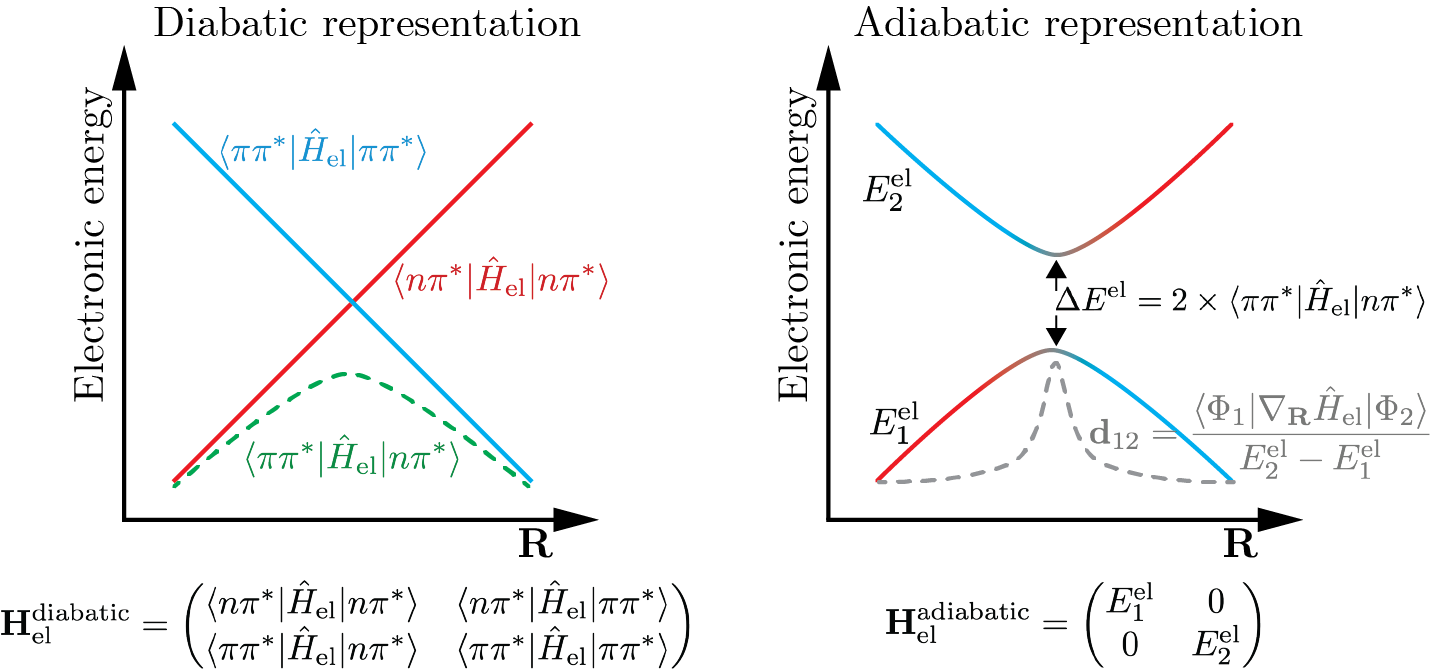}
    \caption{Schematic representation of potential energy curves in the diabatic (left panel) and adiabatic representation (right panel). Left panel: the electronic energy (expectation value of the electronic Hamiltonian) for two diabatic electronic states, $n\pi^\ast$ (red) and $\pi\pi^\ast$ (blue), is represented along a given nuclear coordinate, together with the diabatic coupling as a dashed green curve. The electronic Hamiltonian matrix in this basis of two diabatic electronic states is given under the panel. Right panel: the adiabatic electronic energy $E_1^{\text{el}}$ and $E_2^{\text{el}}$ (eigenvalues of the electronic Hamiltonian) for two adiabatic electronic states, $\Phi_1$ and $\Phi_2$, are represented along a given nuclear coordinate, together with the NACV as a dashed grey curve. The color code used for the adiabatic electronic energies reflects the underlying character of the electronic state at a specific nuclear configuration (red for $n\pi^\ast$ and blue for $\pi\pi^\ast$, in between when the character is mixed near the avoided crossing). The diagonal electronic Hamiltonian matrix in this basis of two adiabatic electronic states is given under the panel. }
    \label{adiadia}
\end{figure*}

At the beginning of this Section, we introduced the notion of \textit{adiabatic} electronic states. Adiabatic electronic states (or wavefunctions) are the eigenfunctions of the electronic Hamiltonian $\hat{H}_{\text{el}}$, with corresponding eigenvalues called adiabatic electronic energies. In other words, the adiabatic electronic wavefunctions make the electronic Hamiltonian diagonal but not the nuclear kinetic energy matrix, as described by Eq.~\ref{eq:matrixtdse} and repeated here for the case where only two adiabatic electronic states $\Phi_1(\bs r; \bs R)$ and $\Phi_2(\bs r; \bs R)$ are considered:
\begin{align} 
 i\hbar & \frac{\partial}{\partial t} 
     \begin{pmatrix}
        \chi_1(\mathbf{R},t)  \\
        \chi_2(\mathbf{R},t) \\
    \end{pmatrix} =  \notag \\
   & \begin{pmatrix}
        -\sum_\nu^{N_{\text{n}}}\frac{\hbar^2}{2M_\nu} \nabla_{\mathbf{R}_\nu}^2 + C_{11}(\mathbf{R})  & \hat{C}_{12}(\mathbf{R})  \\
        \hat{C}_{21}(\mathbf{R}) & -\sum_\nu^{N_{\text{n}}}\frac{\hbar^2}{2M_\nu} \nabla_{\mathbf{R}_\nu}^2 + C_{22}(\mathbf{R}) & \\
    \end{pmatrix}  \notag \\
    & \times \begin{pmatrix}
        \chi_1^{\text{(a)}}(\mathbf{R},t)  \\
        \chi_2^{\text{(a)}}(\mathbf{R},t) \\
    \end{pmatrix} + 
    \begin{pmatrix}
        E^{\text{el}}_{1}(\mathbf{R})   & 0  \\
        0 & E^{\text{el}}_{2}(\mathbf{R})    \\
    \end{pmatrix} 
    \begin{pmatrix}
        \chi_1(\mathbf{R},t)  \\
        \chi_2(\mathbf{R},t) \\
    \end{pmatrix}
    \, .
\label{eq:matrixtdse2states}
\end{align}
The first matrix on the right-hand side of Eq.~\eqref{eq:matrixtdse2states} is the nuclear kinetic energy matrix (non-diagonal) and the second matrix is the electronic Hamiltonian matrix (diagonal).

Another electronic basis is possible, though, often referred to as the \textit{diabatic} electronic states. The diabatic electronic states are directly connected to the character of an electronic state, particularly liked by chemists when assigning electronic transitions. The diabatic electronic wavefunctions do not diagonalize the electronic Hamiltonian, but diagonalize the nuclear kinetic energy matrix\footnote{Even though the nuclear kinetic energy is not a hermitian operator in the electronic space, an equation can be solved to minimize, or set to zero, the off-diagonal elements of the nuclear kinetic energy matrix representation in the electronic diabatic basis.} -- from an equation perspective, the two-electronic state description offered in the adiabatic basis by Eq.~\eqref{eq:matrixtdse2states} would look as follows in the diabatic basis:

\begin{align} 
 i\hbar & \frac{\partial}{\partial t} 
     \begin{pmatrix}
        \chi_1^{\text{(dia)}}(\mathbf{R},t)  \\
        \chi_2^{\text{(dia)}}(\mathbf{R},t) \\
    \end{pmatrix} =  \notag \\
   & \begin{pmatrix}
        -\sum_\nu^{N_{\text{n}}}\frac{\hbar^2}{2M_\nu} \nabla_{\mathbf{R}_\nu}^2   & 0 \\
        0  & -\sum_\nu^{N_{\text{n}}}\frac{\hbar^2}{2M_\nu} \nabla_{\mathbf{R}_\nu}^2  & \\
    \end{pmatrix} 
    \begin{pmatrix}
        \chi_1^{\text{(dia)}}(\mathbf{R},t)  \\
        \chi_2^{\text{(dia)}}(\mathbf{R},t) \\
    \end{pmatrix}
    \notag \\
    &  + 
    \begin{pmatrix}
        \langle \Phi_1^{\text{(dia)}}(\bs R')  | \hat{H}_{\text{el}} | \Phi_1^{\text{(dia)}}(\bs R')  \rangle_{\bs r}    & \langle \Phi_1^{\text{(dia)}}(\bs R')  | \hat{H}_{\text{el}} | \Phi_2^{\text{(dia)}}(\bs R')  \rangle_{\bs r}   \\
        \langle \Phi_2^{\text{(dia)}}(\bs R')  | \hat{H}_{\text{el}} | \Phi_1^{\text{(dia)}}(\bs R')  \rangle_{\bs r}  & \langle \Phi_2^{\text{(dia)}}(\bs R')  | \hat{H}_{\text{el}} | \Phi_2^{\text{(dia)}}(\bs R')  \rangle_{\bs r}    \\
    \end{pmatrix} \notag \\
    & \times \begin{pmatrix}
        \chi_1^{\text{(dia)}}(\mathbf{R},t)  \\
        \chi_2^{\text{(dia)}}(\mathbf{R},t) \\
    \end{pmatrix}
   ,
\label{eq:matrixtdse2statesdiab}
\end{align}
\noindent where $\Phi_J^{\text{(dia)}}(\bs R')$ is the electronic wavefunction for the diabatic state $J$, with $\bs R'$ signifying that such a diabatic state is defined only at a single specified nuclear geometry. The diabatic electronic energies are in principle different from the adiabatic ones. We note that both the diabatic and the adiabatic electronic states constitute an adequate basis to describe the molecular wavefunction and could be used interchangeably. As we will see below, there are concepts that solely belong to one representation or the other. 

The beginning of this section on adiabatic and diabatic states is quite technical, and we can actually obtain a simpler picture of the key differences between adiabatic and diabatic electronic states by looking at a simple example with two electronic states. Let us consider a molecule with a carbonyl (\ce{C=O}) group. This molecule possesses two low-lying electronic states, which can be described by their character: $n\pi^\ast$ (the character of the electronic state is described by an electron leaving a $n$ orbital to populate a $\pi^\ast$ orbital) and $\pi\pi^\ast$ (the character of the electronic state is described by an electron leaving a $\pi$ orbital to populate a $\pi^\ast$ orbital). As we describe here the electronic states by their character, we use a diabatic representation, and the eigenstate for the $n\pi^\ast$ state is $|n\pi^\ast\rangle$ and that of the $\pi\pi^\ast$ is $|\pi\pi^\ast\rangle$. The electronic Hamiltonian matrix resulting from these two electronic states will have diagonal elements, that is, \textit{diabatic electronic energies}, given by $\langle n\pi^\ast | \hat{H}_{\text{el}} | n\pi^\ast\rangle$ for the energy of the  $n\pi^\ast$ state and $\langle \pi\pi^\ast | \hat{H}_{\text{el}} | \pi\pi^\ast\rangle$ for the energy of the  $\pi\pi^\ast$ state. The interaction between these two diabatic states is given by the \textit{diabatic coupling} term $\langle n\pi^\ast | \hat{H}_{\text{el}} | \pi\pi^\ast\rangle$ that indicates how the electronic wavefunction of $n\pi^\ast$ is coupled to the electronic wavefunction of $\pi\pi^\ast$ via the electronic Hamiltonian. The overall curves for the diabatic electronic energies (red and blue for each electronic character) and the diabatic coupling (dashed green) are represented schematically on the left panel of Figure~\ref{adiadia}. Notice that the diabatic electronic energies \textit{can cross} and that a given electronic state always conserves its electronic character for any variation of the nuclear configuration $\mathbf{R}$. The electronic Hamiltonian in this diabatic basis is not diagonal (see the expression at the bottom of the left panel in Fig.~\ref{adiadia}). 

\begin{figure*}[h]
    \centering
    \includegraphics[width=0.8\linewidth]{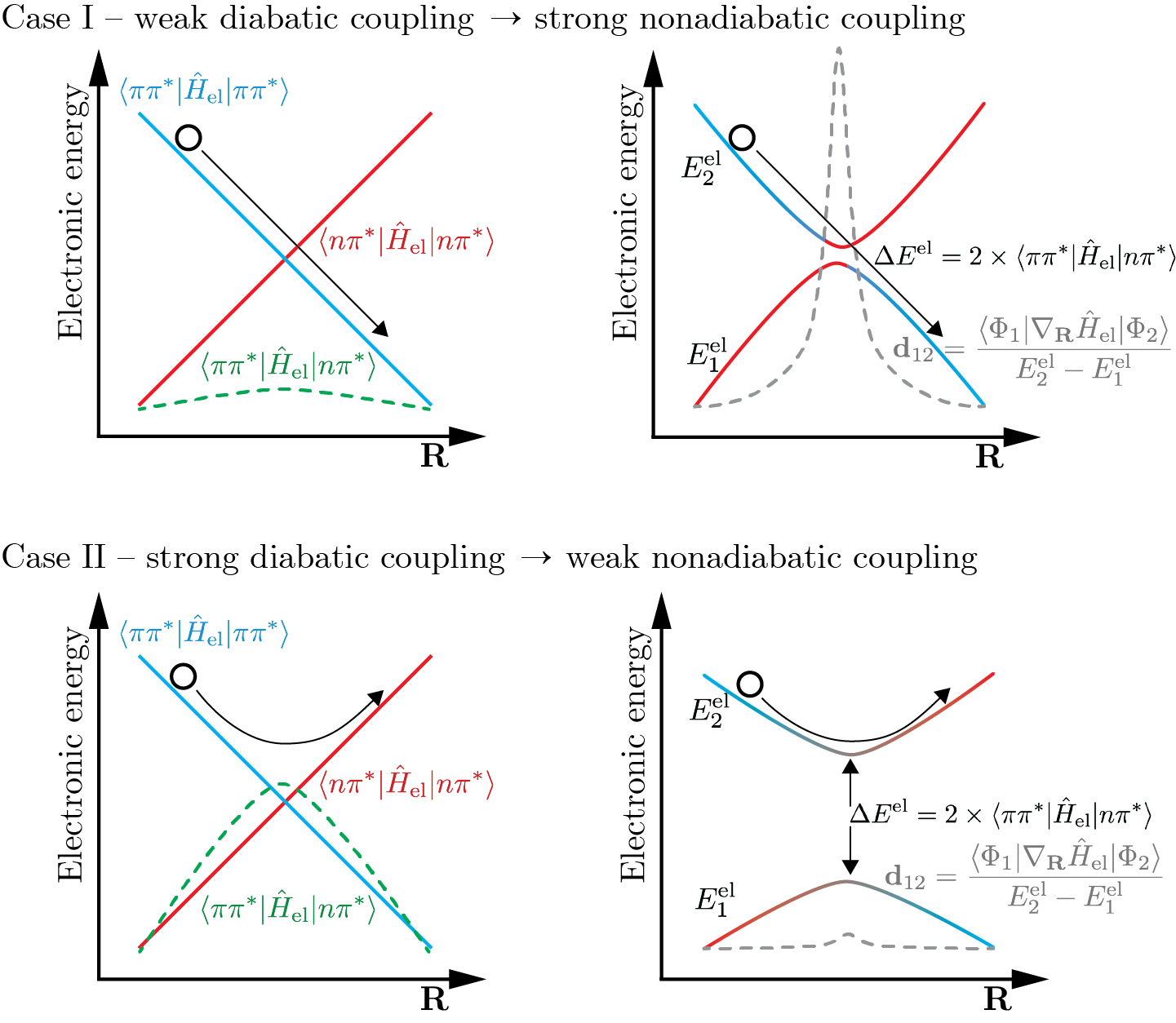}
    \caption{Two different cases for nonadiabatic processes in the diabatic and adiabatic representation. Case I exhibits a system with a weak diabatic coupling, resulting in a strong nonadiabatic coupling in the adiabatic representation. Case II highlights a system with a strong diabatic coupling, meaning a weak nonadiabatic coupling in the adiabatic representation. For both cases, the nuclear dynamics of the system is symbolized by a circle following the arrow.}
    \label{adiadiacases}
\end{figure*}

How does this diabatic picture connect to the adiabatic picture? Formally, we could diagonalize the electronic Hamiltonian expressed in the diabatic representation to obtain adiabatic electronic states and energies. The adiabatic electronic energies, $E_1^{\textit{el}}$ and $E_2^{\textit{el}}$, are labeled no more by an electronic character but by their energy ordering --  $E_1^{\textit{el}}$ is the lowest electronic energy, $E_2^{\textit{el}}$ the second lowest -- for any nuclear configuration $\mathbf{R}$. The adiabatic electronic energies are represented in the right panel of Figure~\ref{adiadia}. In regions of the nuclear configuration space $\mathbf{R}$ where the adiabatic electronic energies are well separated, one often can assign an electronic character to a given adiabatic electronic states -- this is represented in Fig.~\ref{adiadia} by a curve with a clear blue or red color, symbolizing a dominant $\pi\pi^\ast$ or $n\pi^\ast$ character, respectively. However, when one focuses on a region of the nuclear configuration space where the adiabatic electronic states get close in energy, each adiabatic electronic state will have a mixed character, i.e., a mix of $\pi\pi^\ast$ and $n\pi^\ast$ character (symbolized in Fig.~\ref{adiadia} by a gradient of red/blue). More importantly, the ordering of the adiabatic electronic states implies that adiabatic electronic energies \textit{cannot} cross! $E_1^{\textit{el}}$ will always be the first lowest adiabatic electronic energy and $E_2^{\textit{el}}$ the second -- at the cost that their electronic character may change along $\mathbf{R}$. At best, adiabatic electronic energies can become degenerate, leading to the formation of a CX for molecules composed by more than two atoms -- we will come back to CXs shortly (Section~\ref{ch2_cx}). The right panel of Fig.~\ref{adiadia} shows that the splitting between adiabatic electronic energy at the point in configuration space where diabatic electronic energy would cross ($\langle \pi\pi^\ast | \hat{H}_{\text{el}} | \pi\pi^\ast\rangle = \langle n\pi^\ast | \hat{H}_{\text{el}} | n\pi^\ast\rangle$) is exactly equal to twice the diabatic coupling $\langle n\pi^\ast | \hat{H}_{\text{el}} | \pi\pi^\ast\rangle$. We will soon discuss the implications of this observation with two limiting cases. But first, how do we measure the coupling between adiabatic electronic states? We need to focus on the off-diagonal element of the nuclear kinetic energy operator, which features terms called NACVs defined for the present case as $\mathbf{d}_{12} = \langle \Phi_1| \nabla_{\mathbf{R}}| \Phi_2 \rangle_{\bs r} = \frac{\langle \Phi_1| \nabla_{\mathbf{R}}\hat{H}_{\text{el}}| \Phi_2 \rangle_{\bs r} }{E_2^{\text{el}}-E_1^{\text{el}}}$, with $|\Phi_1\rangle$ and $|\Phi_2\rangle$ the adiabatic electronic wavefunctions (eigenfunctions of the electronic Hamiltonian). In words, the NACVs indicate how electronic state 1 can be coupled to electronic state 2 via a nuclear displacement ($\nabla_{\mathbf{R}}$), and the second equality shows that these vectors are inversely proportional to adiabatic electronic energy gap between the two coupled electronic states: a large energy gap between $E_1^{\text{el}}$ and $E_2^{\text{el}}$ means small NACVs, while a small gap makes these vectors large.

To better understand the connection between the diabatic and adiabatic representation, let us look at two limiting cases, depicted in Figure~\ref{adiadiacases}. Case I describes a molecular system where the diabatic coupling $\langle n\pi^\ast | \hat{H}_{\text{el}} | \pi\pi^\ast\rangle$ is weak (small dashed curve in the left panel of Fig.~\ref{adiadiacases}). In other terms, the diabatic states $\pi\pi^\ast$ and $n\pi^\ast$ do not interact strongly. We know that the splitting of the adiabatic electronic energies (at the point of configuration space where $\langle \pi\pi^\ast | \hat{H}_{\text{el}} | \pi\pi^\ast\rangle$ is equal to $\langle n\pi^\ast | \hat{H}_{\text{el}} | n\pi^\ast\rangle$ in the diabatic representation) is twice the diabatic coupling, meaning here a small value. Hence, the adiabatic electronic energies will only be weakly split, and the NACVs would be large as a result. Despite the apparent discrepancy between the diabatic and adiabatic picture, they both tell the same story! In the diabatic representation, a weak diabatic coupling means that a system initiated in the $\pi\pi^\ast$ state will remain in this electronic state (blue curve in Figure~\ref{adiadiacases}) in its trajectory along $\mathbf{R}$, as the coupling with the other diabatic state is weak. In the adiabatic representation, a molecule would start in $E_2^{\text{el}}$ away from the intersection region (the adiabatic state exhibits there a dominant $\pi\pi^\ast$ character -- blue in the schematic of Fig.~\ref{adiadiacases}) and reach the point in $\mathbf{R}$ where the nonadiabatic couplings are strong, meaning that it will change (adiabatic) electronic state to land in $E_1^{\text{el}}$ and move away from the intersection, where $E_1^{\text{el}}$ now exhibit a strong $\pi\pi^\ast$ character. (We note that the switch of electronic character for the adiabatic states is more localized in nuclear configuration space due to the small diabatic coupling presented in this case.)  In other words, the molecule preserves its electronic character as in the diabatic representation, but to achieve this goal in the adiabatic representation the molecule will transfer from one adiabatic electronic state to another: a weak diabatic coupling implies a strong nonadiabatic coupling.

Case II is the opposite scenario with a molecular system displaying a very strong diabatic coupling (lower panels of Figure~\ref{adiadiacases}). A large diabatic coupling implies that the two electronic characters may strongly mix and that the dynamics of a molecule initiated (like for Case I) on the diabatic electronic energy $\langle \pi\pi^\ast | \hat{H}_{\text{el}} | \pi\pi^\ast\rangle$ (blue curve) is likely to transfer to the diabatic electronic energy $\langle n\pi^\ast | \hat{H}_{\text{el}} | n\pi^\ast\rangle$ during its path along $\mathbf{R}$. This diabatic transition means an overall change of the electronic character of the molecule along its trajectory. In the adiabatic picture, a large diabatic coupling means a large adiabatic splitting between adiabatic electronic energies and, as a result, a weak nonadiabatic coupling. A molecule initiated in $E_2^{\text{el}}$ (with a strong $\pi\pi^\ast$ character in this region) will evolve along the $E_2^{\text{el}}$ potential \textit{adiabatically}, meaning that the character of the electronic state will slowly change from $\pi\pi^\ast$ to $n\pi^\ast$ (and no jump will occur to $E_1^{\text{el}}$ as the NACVs are small). Hence, the diabatic and adiabatic picture shows the very same process: a molecule changes its electronic character along its nuclear dynamics. An important comment is required at this stage: case II actually represents a case where the Born-Oppenheimer approximation is valid. The Born-Oppenheimer approximation neglects the NACVs, meaning that a molecule in a given adiabatic electronic state will always remain in this state, independently of the nuclear dynamics. In case II, the molecule remains on $E_2^{\text{el}}$ as the (adiabatic) energy gap between the two (adiabatic) electronic states is large and the NACVs are negligible. 

We finish this Section with a few considerations regarding the adiabatic and diabatic representations. First of all, only expectation values of operators are observables, and both the adiabatic and the diabatic representations are not \textit{per se} directly related to observables. In this sense, adiabatic or diabatic electronic populations -- an important quantity often plotted from nonadiabatic molecular dynamics simulations -- are not experimental observables. Similarly, CXs designate regions of the nuclear configuration space where two (or more) adiabatic electronic energies are degenerate (as will be discussed below in detail). CXs can be seen as ultra-efficient funnels to transfer an electronically-excited molecule from one adiabatic electronic state to another, as NACVs will be singular at this very point of degeneracy. However, CXs are purely a product of the adiabatic representation, and as such are not \textit{per se} observables, as any observable should be independent from the electronic representation. Finally, it is important to stress that diabatic states can only be strictly defined for diatomic molecular systems, and only quasi-diabatic states can be produced for larger molecules.\cite{mead1982conditions} The reader interested in diabatization strategies for electronic states can consult Refs.~\cite{conicalintersection2004,Wittenbrink2016diab}, for example.

\subsubsection{Conical intersections}
\label{ch2_cx}

Mathematical considerations of CXs first entered the chemical literature as early as 1929, with von Neumann and Wigner\cite{vonneumann1929eigenvalues} presenting the conditions for the existence of electronic degeneracies between two adiabatic states of the same spatial symmetry, as well as, a few years later, Teller\cite{teller_crossing_1937} highlighting the potential for CXs to facilitate efficient internal conversion on ultrafast timescales.
Nevertheless, the history of CXs within the 20th century, for the most part, was one of scepticism.
In contrast to the views of Teller and others,\cite{zimmerman_molecular_1966,michl_physical_1974} the most common understanding of photochemical reactions was based on the concept of a small energy gap at an avoided crossing (AC) minimum\cite{van_der_lugt_symmetry_1969} along a linear path between reactants and products.\cite{schapiro_using_2011}
CXs were instead considered by many to be rare, mathematical curiosities, unlikely to occur at chemically relevant energies, unless their existence was required by symmetry.\cite{levine2007isomerization}
Even then, since most molecules do not exhibit point group symmetry, not to mention high enough symmetry to impose electronic degeneracies, the prevalence of CXs was thought to be severely limited.\cite{yarkony_nonadiabatic_2012}

It was not until the early 1990s (some 60 years after their initial conception) that CXs began to receive the recognition they do today.\cite{domcke_front_2004}
Thanks to increased computer power, implementation of gradients in multiconfigurational electronic-structure methods, and improved algorithms for geometry optimisation, CXs could be located without the aid of symmetry constraints.\cite{lipkowitz_conical_2007}
The respective efforts of the Ruedenberg\cite{atchity_potential_1991, xantheas_potential_1991} and Yarkony\cite{riad_manaa_noncrossing_1990} groups (for small molecules), as well as the international collaboration of Bernardi, Olivucci and Robb\cite{bernardi_mechanism_1990, bernardi_predicting_1990} (for larger molecules) showed not only that 'non-symmetry-required' CXs exist (which was known before), but are in fact omnipresent in numerous organic molecules.\cite{domcke_role_2012} 
More recently, Truhlar and Mead\cite{truhlar_relative_2003} further demonstrated that, in contrast to previous opinions, CXs are indeed much more likely to exist than ACs. 
They proved that if a local minimum in the energy gap (i.e., a supposed AC) between two electronic PESs is encountered upon traversing a path through nuclear configuration space, rather than being a true AC, it is much more likely to instead be associated with the neighbourhood (or shoulder) of a CX. 
It is now widely agreed that, far from being arcane theoretical concepts,\cite{zhu_non-adiabaticity_2016,boeije_one-mode_2023} CXs constitute the bedrock of our mechanistic understanding of a wide array of photochemical and photophysical processes.\cite{gonzalez2012progress,domcke_role_2012,gozem2017theory,10.1039/9781782626954-00016,blancafort_photochemistry_2014,levine2019conical}

\paragraph{Foundations: the non-crossing rule, the branching space and the seam space}
\label{ch2_non-cross_branch_seam}

In order to determine the conditions for the existence of a CX, it is instructive to consider again the matrix of the two-state diabatic electronic Hamiltonian previously introduced in Eq.~\eqref{eq:matrixtdse2statesdiab},\cite{matsika_electronic_2021} 

\begin{equation}
\label{dia_elec_ham}  
    \mathbf{H}_{\text{el}}^{\text{(dia)}}(\R) =
    \begin{pmatrix}
        V_{11}(\R) & V_{12}(\R) \\
        V_{12}(\R) & V_{22}(\R)
    \end{pmatrix} \, ,
\end{equation}
where here we define $V_{IJ}(\R) = \langle \Phi_I^{\text{(dia)}}(\Rprime) | \hat{H}_{\text{el}}(\R) | \Phi_J^{\text{(dia)}}(\Rprime) \rangle_{\re}$ for brevity and $V_{21}(\R) = V^*_{12}(\R) = V_{12}(\R)$ as $\mathbf{H}_{\text{el}}^{\text{(dia)}}(\R)$ is a Hermitian matrix defined with real matrix elements.
Diagonalization of $\mathbf{H}_{\text{el}}^{\text{(dia)}}(\R)$ can be achieved via the unitary transformation matrix,

\begin{equation}
\label{uni_trans_mat}
    \mathbf{U}(\R) = 
    \begin{pmatrix}
        \cos[\theta(\R)] & \sin[\theta(\R)] \\
        -\sin[\theta(\R)] & \cos[\theta(\R)]
    \end{pmatrix} \, .
\end{equation}
The eigenfunctions of $\mathbf{H}_{\text{el}}^{\text{(dia)}}(\R)$ are thus given as the adiabatic electronic states expanded in terms of the two diabatic electronic states, 

\begin{equation}
\label{adia_elec_eigenvec}
\begin{split}
    \Phi_1(\re;\R) &= \cos[\theta(\R)] \Phi_1^{\text{(dia)}}(\re; \Rprime) + \sin[\theta(\R)] \Phi_2^{\text{(dia)}}(\re; \Rprime)\\
    \Phi_2(\re;\R) &= -\sin[\theta(\R)] \Phi_1^{\text{(dia)}}(\re; \Rprime) + \cos[\theta(\R)] \Phi_2^{\text{(dia)}}(\re; \Rprime) \, ,
\end{split}
\end{equation}
with the rotation angle, $\theta(\R)$, defining the diabatic-to-adiabatic transformation,

\begin{equation}
\label{mix_ang}
    \theta(\R) = \frac{1}{2} \arctan \left(\frac{2V_{12}(\R)}{V_{11}(\R) - V_{22}(\R)}\right) \, .
\end{equation}

Equivalently, one yields the adiabatic electronic energies (written in terms of the diabatic electronic energies) as the eigenvalues of $\mathbf{H}_{\text{el}}^{\text{(dia)}}(\R)$,\cite{atchity_potential_1991, domcke_conical_2004}

\begin{equation}
\label{adia_elec_eigenval}
    E_{1,2}^{\text{el}}(\R) = \overline{V}(\R) \pm \sqrt{\left[\Delta V(\R)\right]^2 + \left[V_{12}(\R)\right]^2} \,
\end{equation}
where
\begin{equation}
\label{terms}
    \overline{V}(\R) = \frac{V_{11}(\R) + V_{22}(\R)}{2} \;\;\;\;\; \text{and} \;\;\;\;\; \Delta V(\R) = \frac{V_{11}(\R) - V_{22}(\R)}{2} \, .
\end{equation}

\begin{figure*}[h!]
    \centering
    \includegraphics[width=\textwidth]{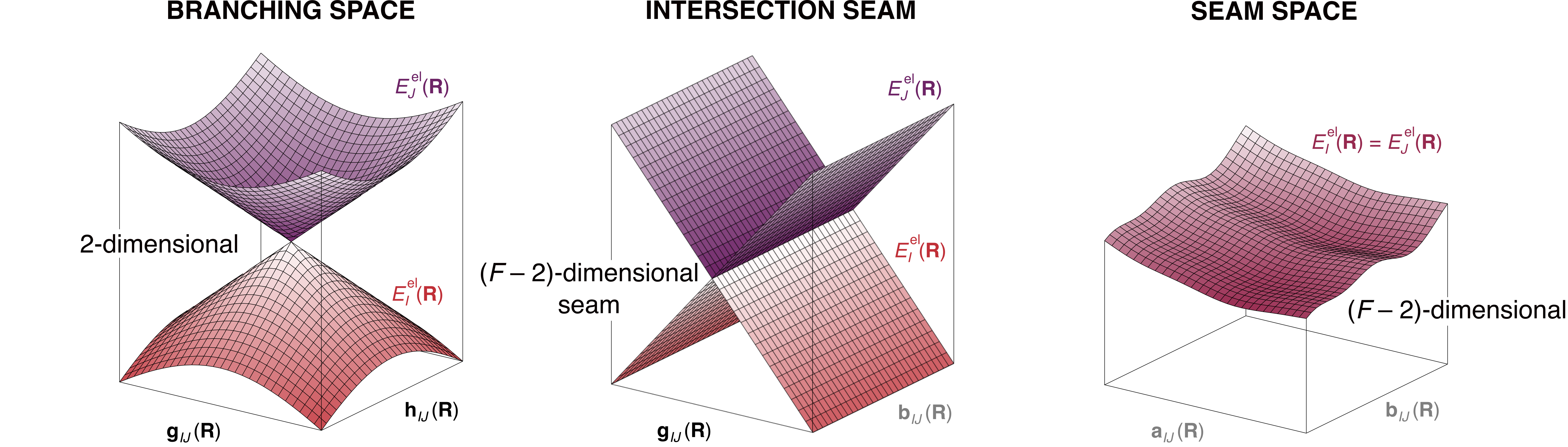}
    \caption{Schematic representation of the adiabatic PESs in the branching space (left) and seam space (right) of a two-state CX, where $\g{I}{J}$ and $\h{I}{J}$ are the branching space vectors (expressed in terms of adiabatic states) and $\mathbf{a}_{IJ}(\R)$ and $\mathbf{b}_{IJ}(\R)$ are two arbitrarily chosen seam space coordinates. Plotting the PESs along one branching space vector (here $\g{I}{J}$) and one seam space coordinate reveals the intersection seam (middle) -- see text for discussion.}
    \label{fig:branch_vs_seam}
\end{figure*}

Inspecting Eqs.~\eqref{adia_elec_eigenval} and \eqref{terms}, it is clear that for the eigenvalues of $\mathbf{H}_{\text{el}}^{\text{(dia)}}(\R)$ (i.e., two adiabatic electronic states) to be degenerate, the following two conditions must be satisfied,

\begin{equation}
\label{degen_cond}
\begin{split}
    V_{11}(\R_{\text{CX}}) \: =& \:\: V_{22}(\R_{\text{CX}})\\
    V_{12}(\R_{\text{CX}}) \: =& \:\: 0 \, ,
\end{split}
\end{equation}
where $\R_\text{CX}$ is the geometry at the CX. For a nonlinear molecule with $F = 3N_\text{n} - 6$ nuclear degrees of freedom, these two conditions are fulfilled in an ($F - 2$)-dimensional subspace (i.e., one independent nuclear degree of freedom must be lost to satisfy each of the two conditions).\cite{malhado2014non}
Therefore, the full $F$-dimensional space associated with a given molecule can be divided into two subspaces: (i) this ($F - 2$)-dimensional seam (or intersection) space\cite{atchity_potential_1991} (Fig.~\Ref{fig:branch_vs_seam}, right), where the adiabatic electronic states are degenerate and (ii) a two-dimensional branching\cite{atchity_potential_1991} (or $g$-$h$)\cite{yarkony_conical_2001} space (Fig.~\Ref{fig:branch_vs_seam}, left) orthogonal to it, where the degeneracy is lifted.
Therefore, rather than occurring at an isolated geometry (as may appear to be the case in the branching space), CXs exist as an infinite number of points (with necessarily different electronic energies, associated with different nuclear geometries) connected\cite{B407959K, coe2008extent, morreale2024topology, barbatti2004photochemistry} along the so-called intersection seam (Fig.~\ref{fig:branch_vs_seam}, middle).
The idea of an 'intersection \textit{seam}' compared to a 'seam (or intersection) \textit{space}' is exactly equivalent and thus the two are interchangeable; the former simply refers to an ($F - 2$)-dimensional hyperline, which may be more appropriate to consider given the context.
Von Neumann and Wigner first presented these conditions for electronic degeneracy in the context of diatomic molecules.
Since only one nuclear degree of freedom exists in an isolated diatomic (i.e., the bond distance between the two nuclei), two adiabatic electronic states with the same spatial symmetry can never become degenerate\footnote{
The same does not apply to electronic states of different spatial symmetry. 
In this case, the second condition in Eq.~\eqref{degen_cond} is trivially satisfied (i.e., $V_{12}(\R_{\text{CX}})$ is identically zero), so only one nuclear degree of freedom must be lost to fulfil the remaining condition, something generally possible in diatomic molecules. 
It was this observation that caused many to erroneously assume that CXs could only form between electronic PESs because of point group symmetry and that intersections between electronic PESs of the same spatial symmetry were primarily avoided.\cite{domcke_conical_2004}
} 
-- this is the origin of their 'non-crossing rule'.\cite{vonneumann1929eigenvalues}
Polyatomic molecules, on the other hand, possess many degrees of freedom (or at the very least, the required two), so fulfilment of both conditions in Eq.~\eqref{degen_cond} is indeed highly possible, but not necessarily guaranteed.\cite{yarkony_conical_1998}

Complete characterization of a given CX requires the determination (and subsequent visualization) of its branching space. 
To do so for a two-state CX, it is useful to first redefine the matrix of the diabatic electronic Hamiltonian [Eq.~\eqref{dia_elec_ham}] in terms of the quantities given in Eq.~\eqref{terms},\cite{domcke_conical_2004, ferre_description_2015}

\begin{equation}
\label{dia_elec_ham_redefine}
    \mathbf{H}_{\text{el}}^{\text{(dia)}}(\R) = \overline{V}(\R)\mathbb{I}_2 + 
    \begin{pmatrix}
        \Delta V(\R) & V_{12}(\R) \\
        V_{12}(\R) & -\Delta V(\R)
    \end{pmatrix} \, ,
\end{equation}
where $\mathbb{I}_2$ is a ($2 \times 2$) identity matrix.
Consider two electronic states that are energetically degenerate at nuclear geometry $\R_\text{CX}$.
At a neighbouring geometry, $\R = \R_\text{CX} + \delta\R$, it is possible to expand $\mathbf{H}_{\text{el}}^{\text{(dia)}}(\R)$ in a Taylor expansion around the intersection point $\R_\text{CX}$. This is given as

\begin{align}
\label{dia_elec_ham_taylor}
\begin{split}
    \mathbf{H}_{\text{el}}^{\text{(dia)}}(\R) & \\ 
    =& \mathbf{H}_{\text{el}}^{\text{(dia)}}(\R_{\text{CX}}) + \nabla_{\R} \mathbf{H}_{\text{el}}^{\text{(dia)}}(\R_{\text{CX}}) \cdot \delta\R + \dots \\
    =& V(\R_{\text{CX}})\mathbb{I}_2 + \Bigg[\nabla_{\R} \overline{V}(\R_{\text{CX}})\mathbb{I}_2 \\ 
    & + 
    \begin{pmatrix}
        \nabla_{\R} \Delta V(\R_{\text{CX}}) & \nabla_{\R} V_{12}(\R_{\text{CX}}) \\
        \nabla_{\R} V_{12}(\R_{\text{CX}}) & -\nabla_{\R} \Delta V(\R_{\text{CX}})
    \end{pmatrix}
    \Bigg] \cdot \delta \R + \dots \\
    =& \left(V(\R_{\text{CX}})  + \nabla_{\R} \overline{V}(\R_{\text{CX}}) \cdot \delta\R \right) \mathbb{I}_2 \\ 
    &+ 
    \begin{pmatrix}
        \nabla_{\R} \Delta V(\R_{\text{CX}}) & \nabla_{\R} V_{12}(\R_{\text{CX}}) \\
        \nabla_{\R} V_{12}(\R_{\text{CX}}) & -\nabla_{\R} \Delta V(\R_{\text{CX}})
    \end{pmatrix}
   \cdot \delta \R + \dots \, ,
\end{split}
\end{align}
where $V_{11}(\R) = V_{22}(\R) = V(\R)$ at $\R = \R_{\text{CX}}$.\cite{sicilia2007quadratic}
Given the first term in the last line of Eq.~\eqref{dia_elec_ham_taylor} corresponds to a diagonal matrix (and thus has no effect on the coupling between the two electronics states), it is apparent that, to first order in $\R$, the conditions for the degeneracy to remain upon moving from $\R_{\text{CX}}$ to $\R$ become 

\begin{equation}
\label{new_degen_cond}
\begin{split}
    \nabla_{\R} \Delta V(\R_{\text{CX}}) \cdot \delta \R \: =& \:\: 0\\
    \nabla_{\R} V_{12}(\R_{\text{CX}}) \cdot \delta \R \: =& \:\: 0 \, .
\end{split}
\end{equation}

To retain the degeneracy, the nuclear displacement vector, $\delta \R$, is restricted to the subspace orthogonal to that spanned by the vectors $\nabla_{\R} \Delta V(\R_{\text{CX}})$ and $\nabla_{\R} V_{12}(\R_{\text{CX}})$.
The conditions in Eq.~\eqref{degen_cond} have therefore been extended and the branching space vectors, $\nabla_{\R} \Delta V(\R_{\text{CX}})$ and $\nabla_{\R} V_{12}(\R_{\text{CX}})$, defined.
The branching space vectors (or any linear combination between them)\cite{boeije_one-mode_2023} constitute the two possible\footnote{
Strictly speaking, the branching space vectors collectively constitute the branching plane, which spans an infinite number of linearly-dependent nuclear coordinates that can lift the degeneracy. 
} nuclear coordinates along which the molecule can distort in order to exit the strong nonadiabatic region at the CX; the degeneracy is lifted linearly to first order upon an infinitesimal displacement in either direction.
Movement along any of the remaining $F - 2$ nuclear coordinates retains the degeneracy and simply translates the molecule along the intersection seam (or within the seam space), which can be characterized in terms of minima and transition states.\cite{tuna2015assessment,pieri2021namdreactor}

Since the definition of diabatic electronic states is arbitrary,\cite{ferre_description_2015, Persico2018, nikiforov_assessment_2014} it is customary to express the branching space vectors in terms of adiabatic electronic states instead. 
As such, using Yarkony’s notation,\cite{yarkony_conical_2001} the two branching space vectors can now be defined as
\begin{equation}
\label{g_vect}
    \g{I}{J} = \frac{1}{2} \nabla_{\R} \left[E_I^{\text{el}}(\R) - E_J^{\text{el}}(\R)\right] \, ,
\end{equation}
which is referred to as the gradient difference vector, and

\begin{equation}
\label{h_vect}
    \h{I}{J} = \left< \Phi_I(\R) | \nabla_{\R}\hat{H}_{\text{el}} | \Phi_J(\R) \right>_{\re} \, ,
\end{equation}
which is referred to as the derivative coupling vector.
Another useful quantity to define for the characterization of the branching space is the seam coordinate,\cite{zhang_nonadiabatic_2021} 

\begin{equation}
\label{s_vect}
    \s{I}{J} = \frac{1}{2} \nabla_{\R} \left[E_I^{\text{el}}(\R) + E_J^{\text{el}}(\R)\right] \, ,
\end{equation}
where the projections of $\s{I}{J}$ onto the branching plane are subsequently given by

\begin{equation}
\label{sx_and_sy}
    s_x = \s{I}{J} \cdot \xua{I}{J} \;\;\;\;\; \text{and} \;\;\;\;\; s_y = \s{I}{J} \cdot \yua{I}{J} \, ,
\end{equation}
with
\begin{equation}
\label{int_adapt_coords}
    \xua{I}{J} = \frac{\gua{I}{J}}{\overline{g}} \;\;\;\;\; \text{and} \;\;\;\;\; \yua{I}{J} = \frac{\hua{I}{J}}{\overline{h}} \, 
\end{equation}
being the orthonormalized versions of the $\g{I}{J}$ and $\h{I}{J}$ vectors introduced in Eqs.~\eqref{g_vect} and \eqref{h_vect}, respectively; they are the so-called intersection-adapted coordinates.
Here, $\overline{g} = \|\gua{I}{J}\|$ and $\overline{h} = \|\hua{I}{J}\|$ are the norms of the respective orthogonalized branching space vectors.
The electronic Hamiltonian matrix of Eq.~\eqref{dia_elec_ham} can now be recast within the branching space as 

\begin{equation}
\label{adia_elec_ham_bs}
    \mathbf{H}_{\text{el, bs}}(x, y) = \left(E^\times + s_x x + s_y y\right)\mathbb{I}_2 + 
    \begin{pmatrix}
        \overline{g}x & \overline{h}y \\
        \overline{h}y & -\overline{g}x
    \end{pmatrix} \, ,
\end{equation}
where $x$ and $y$ are displacements along the $\gua{I}{J}$ and $\hua{I}{J}$ directions.\cite{matsika_nonadiabatic_2011}
The eigenvalues of Eq.~\eqref{adia_elec_ham_bs} are thus given by

\begin{equation}
\label{adia_elec_eigenval_bs}
    E_{1,2}^{\text{el, bs}}(x,y) = \left(E^\times + s_x x + s_y y\right) \pm \sqrt{\left(\overline{g}x\right)^2 + \left(\overline{h}y\right)^2} \, ,
\end{equation}

where $V(\R_{\text{CX}}) = E^\times$ is the energy at the point of degeneracy.\cite{sicilia2007quadratic, fdez_galvan_analytical_2016}
This results in the two adiabatic electronic energies, $E_{1,2}^{\text{el, bs}}(x,y)$, plotted around the CX along the two branching space directions exhibiting the characteristic double-cone (or diabolical)\cite{yarkony_diabolical_1996, yarkony_conical_1998} shape.\footnote{
If we take a random one-dimensional slice through the adiabatic PESs in the branching space of a CX, it is unlikely that it will pass directly through the point of intersection, $\R_\text{CX}$. 
We are much more likely to observe a supposed AC, as was argued and proven by Truhlar and Mead (mentioned above).\cite{truhlar_relative_2003}
ACs are formally different to CXs, in that the energy gap is lifted at second (or higher) order in all directions around the former, whereas the degeneracy is lifted at first order only in two directions around the latter.\cite{levine2007isomerization}
}

Having laid out over the last few pages (i) the origin of the “non-crossing rule”, (ii) the definitions of the complementary branching and seam spaces, as well as (iii) the double-cone shape given by adiabatic electronic energies within the branching space, it is now necessary to characterize CXs further. 
There are a number of ways this can be achieved, namely by symmetry, by topography, and by topology. 
Each will be discussed in turn below.

\paragraph{Characterizing conical intersections by symmetry}
\label{ch2_sym}

One way to classify two-state CXs is by the part played by point group symmetry in satisfying the degeneracy conditions given in Eq.~\eqref{degen_cond}.
There are three main groups: symmetry-required, accidental symmetry-allowed, and accidental same-symmetry CXs.
Symmetry-required CXs arise when the molecule exhibits a particular point group symmetry where the intersecting electronic states belong to the same doubly-degenerate irreducible representation.
For nuclear geometries for which this holds, both conditions in Eq.~\eqref{degen_cond} are trivially satisfied, guaranteeing the degeneracy by symmetry.\cite{yarkony_nonadiabatic_2012}
The Jahn-Teller effect\cite{jahn1937stability} is a symmetry-lowering process intrinsically related to this class of CX.\cite{Persico2018}
Accidental symmetry-allowed CXs occur when one of the intersecting electronic states possesses a distinct spatial symmetry to the other (i.e., they belong to different irreducible representations).
As a result, the second condition in Eq.~\eqref{degen_cond} is automatically fulfilled for all nuclear geometries where this is the case;\cite{zhu_non-adiabaticity_2016} the off-diagonal matrix elements of the totally-symmetric $\mathbf{H}_{\text{el}}^{\text{(dia)}}(\R)$ are zero by symmetry.\cite{kjonstad2017crossing}
Accidental same-symmetry CXs involve electronic states that exhibit the same spatial symmetry (i.e., belong to the same (non-degenerate) irreducible representation).
Unlike the two cases already discussed, neither condition in Eq.~\eqref{degen_cond} is therefore fulfilled by group theoretical arguments.\cite{kjonstad2017crossing} 
Given that most molecules do not possess sufficiently high symmetry to form the above two classes of CXs, accidental same-symmetry CXs are by far the most common in photochemistry.\cite{matsika_electronic_2021}

\paragraph{Characterizing conical intersections by topography}
\label{ch2_topog}
So far in Section~\ref{ch2_cx}, the discussion of CXs has centred around what we now refer to as their topology, that is, the dimensionality of the CX branching and seam spaces.
However, an equally important aspect to consider is the topography of a given CX,\cite{matsika_electronic_2021, farfan_systematic_2020} which relates instead to the shape of the PESs in the vicinity of the intersection point within the branching space (of typically a two-state CX).
A number of ways exist to characterise the local topography of a CX, each of which make use of the four parameters ($\overline{g}, \overline{h}, s_x, s_y$) given in Eq.~\eqref{sx_and_sy} and \eqref{int_adapt_coords}.

Firstly, the parameters $\overline{g}$ and $\overline{h}$ characterise the slope of a CX in the directions of the two branching space vectors, respectively.\cite{yarkony_nuclear_2001}
The average of these is related to the pitch,

\begin{equation}
\label{pitch}
    \delta_{\text{gh}} = \sqrt{\frac{1}{2}\left({\overline{g}}^2 + {\overline{h}}^2\right)} \, ,
\end{equation}

\noindent{which} defines the overall steepness of the PESs within the branching space, whereas their relative difference relates to the asymmetry,

\begin{equation}
\label{asymmetry}
    \Delta_{\text{gh}} = \frac{{\overline{g}}^2 - {\overline{h}}^2}{{\overline{g}}^2 + {\overline{h}}^2} \, ,
\end{equation}

\noindent{which} defines the extent to which the shape of the CX differs from a reference double cone of perfect radial symmetry.\cite{fdez_galvan_role_2022}
A CX is classified as symmetric (asymmetric) if it gives a zero (non-zero) value of $\Delta_{\text{gh}}$. Visually, symmetric CXs possess PESs with the same slope in both the $\gua{I}{J}$ and $\hua{I}{J}$ vector directions (i.e., $\overline{g} = \overline{h}$), whereas the slopes differ in asymmetric CXs (i.e., $\overline{g} \neq \overline{h}$).
It is seen in certain cases, like in trivial unavoided crossings,\cite{MEEK2015117, shchepanovska2021nonadiabatic} that if the CX is strongly asymmetric, it may appear as (but is not) a one-dimensional, or linear, intersection in the branching space, with a ($F - 1$)-dimensional seam space.\cite{boeije_one-mode_2023}
Such intersections are common in systems involving long-range charge or energy transfer.

\begin{figure}[t!]
    \centering
    \includegraphics[width=0.5\textwidth]{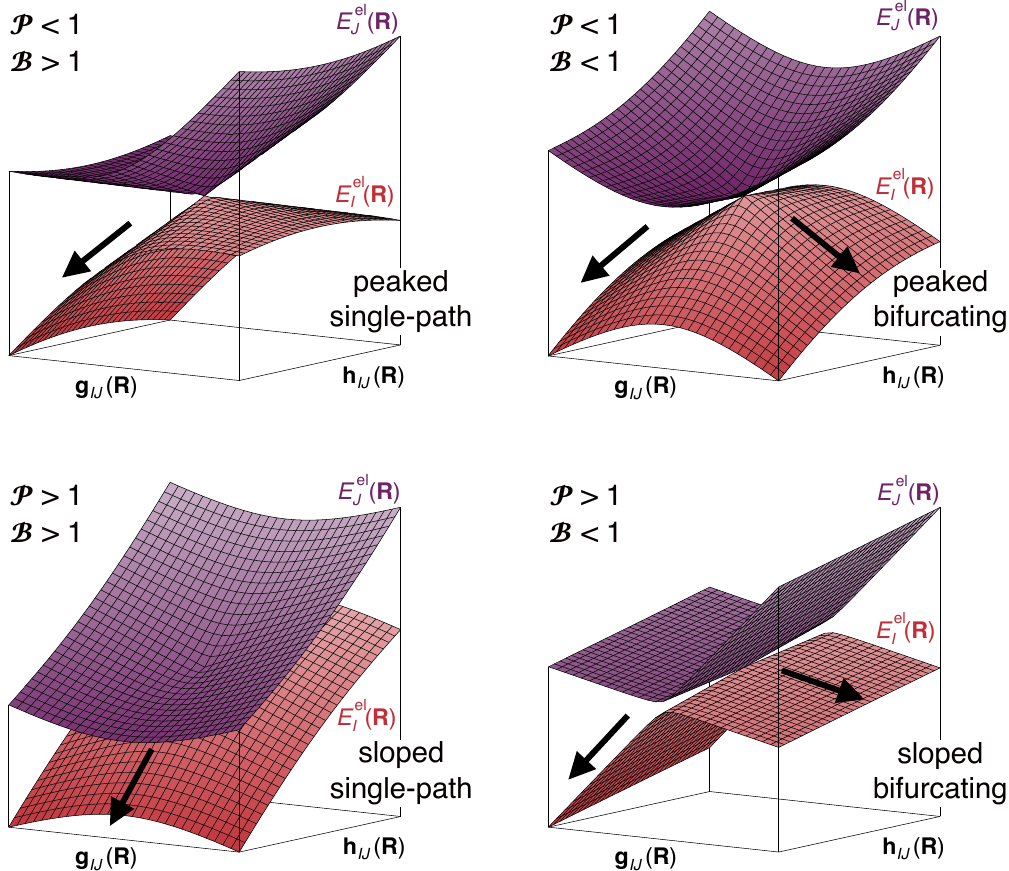}
    \caption{Schematic representation of the adiabatic PESs in the branching space of four two-state CXs characterised by different local topographies: (top left) peaked and single-path ($\delta_{\text{gh}}$ = 0.0949, $\Delta_{\text{gh}}$ = 0.5320, $s_x$ = 0.9550, $s_y$ = 0.0000); (top right) peaked and bifurcating ($\delta_{\text{gh}}$ = 0.1249, $\Delta_{\text{gh}}$ = 0.3402, $s_x$ = 0.0000, $s_y$ = 0.7133); (bottom left) sloped and single-path ($\delta_{\text{gh}}$ = 0.1326, $\Delta_{\text{gh}}$ = 0.2680, $s_x$ = 0.0000, $s_y$ = 2.1588); and (bottom right) sloped and bifurcating ($\delta_{\text{gh}}$ = 0.1399, $\Delta_{\text{gh}}$ = 0.8753, $s_x$ = 0.0000, $s_y$ = 0.4553). The numerical CX branching space topography parameters used to generate the plots in this figure were taken from Ref.~\cite{fdez_galvan_analytical_2016}. We note that, here, in Fig.~\ref{fig:topog_param_schematic}, and in Fig.~\ref{fig:branch_vs_seam}, we use the (raw) $\g{I}{J}$ and $\h{I}{J}$ vector notation instead of the $x$ and $y$ (orthonormalized) scalar quantities for schematic simplicity.}
    \label{fig:topog_param_schematic}
\end{figure}

Secondly, the parameters $s_x$ and $s_y$ characterise the tilt of a CX and lead to further classifications of CX topography, that of peaked and sloped (Fig.~\ref{fig:topog_param_schematic}).\cite{atchity_potential_1991}
For a CX to be defined as peaked, $s_x$ and $s_y$ must be (close or equal to) zero; for a sloped CX, either one or both of $s_x$ and $s_y$ must be non-zero.
Depending on which is larger, the CX will be tilted either more in the $\gua{I}{J}$ or $\hua{I}{J}$ vector directions, respectively.
Visually, a CX is peaked if the intersection point in the branching space is the lowest (highest) energy point on the upper (lower) PES. 
Although, in this case, the point of degeneracy relates to a minimum on the upper PES, this should not be confused with a stationary point, as the nuclear gradient at such a geometry is not zero, but in fact discontinuous.\cite{boeije_one-mode_2023}
A sloped CX, in contrast, can be recognised visually when the principle axis of the CX is tilted to such an extent that now, in some directions along the branching plane, the energy of the upper (lower) PES becomes lower (higher) in energy than the intersection point.\cite{doi:https://doi.org/10.1002/9781119417774.ch1}

Rewriting Eq.~\eqref{adia_elec_eigenval_bs} explicitly in terms of the pitch, asymmetry and tilt parameters [Eqs~\eqref{pitch}, \eqref{asymmetry} and \eqref{sx_and_sy}, respectively] yields 
\begin{equation}
\label{adia_elec_eigenval_bs_topog_params}
\begin{split}
E_{1,2}^{\text{el, bs}}(x,y)
    &= E^\times + \delta_\text{gh}\Bigl(s_x x + s_y y \\
    &\qquad \pm \sqrt{(x^2 + y^2) + \Delta_\text{gh}(x^2 - y^2)}\Bigr)\, ,
\end{split}
\end{equation}
which provides a direct means in plotting the branching space of CXs exhibiting different topographies (Fig.~\ref{fig:topog_param_schematic}).
It may be argued, however, that in their current form the aforementioned CX branching space topography parameters do not bear an immediate connection to the possible reactive outcomes of nonadiabatic dynamics near a CX.\cite{wang2025minimum}
Fdez. Galv\'an \textit{et al.}\cite{fdez_galvan_analytical_2016, fdez_galvan_role_2022} recognised this by defining the following composite parameters,
\begin{equation}
\label{p_topog}
    \mathcal{P} = \frac{\sigma^2}{{\delta_{\text{gh}}}^2\left(1 - {\Delta_{\text{gh}}}^2\right)} \left(1 - \Delta_{\text{gh}} \cos(2\theta_\text{s})\right)
\end{equation}
and
\begin{equation}
\label{b_topog}
\begin{split}
\mathcal{B} &= \sqrt[3]{\frac{\sigma^2}{4\left(\delta_{\text{gh}}\Delta_{\text{gh}}\right)^2}}
\left(\sqrt[3]{\left(1 + \Delta_{\text{gh}}\right)\cos^2\left(\theta_\text{s}\right)} \right. \\
&\qquad\left. + \sqrt[3]{\left(1 - \Delta_{\text{gh}}\right)\sin^2\left(\theta_\text{s}\right)}\right) \, ,
\end{split}
\end{equation}
where 
\begin{equation}
\label{rel_tilt}
    \sigma = \sqrt{{s_x}^2 + {s_y}^2} 
\end{equation}
is a collective tilt parameter and
\begin{equation}
\label{tilt_heading}
    \theta_\text{s} = \arctan\left(\frac{s_y}{s_x}\right)
\end{equation}
is the tilt heading, i.e., the polar angle within the $xy$-plane (defined with respect to the $x$-axis) for which the average energy of the two PESs is maximum.
Depending on the numerical value of these composite parameters, a CX can be characterised as\cite{cuellar2023characterizing}

\begin{equation}
    \mathcal{P} = \begin{cases}
    \;\; < 1 \;\;\;\; \text{peaked} \\
    \;\; > 1 \;\;\;\; \text{sloped}
    \end{cases} 
    \:\:\: \text{and} \:\:\:\:\:\:\:\:\:\:
    \mathcal{B} = \begin{cases}
    \;\; < 1 \;\;\;\; \text{bifurcating} \\
    \;\; > 1 \;\;\;\; \text{single-path}
    \end{cases} \, .
\end{equation}

Here, the terms peaked and sloped are the exact same as those already discussed above.\cite{atchity_potential_1991}
However, additionally a CX can be classified as bifurcating if there are two minima on the lower PES, affording two preferred relaxation paths from the CX; single-path refers to when there is just one (Fig.~\ref{fig:topog_param_schematic}).
By defining these composite parameters, it is possible to get a rough, but immediate feeling for the possible affect a given CX may have on nonadiabatic dynamics within its vicinity, simply based on its topography.

\paragraph{Characterizing conical intersections by topology}
\label{ch2_topol}

Finally, the most fundamental characteristic of a CX, i.e., its topology, can also be used to classify it, with different CXs (or, more generally, different intersections) possessing different topologies.
The most common type of CX encountered in photochemical investigations, as discussed already, is that between two electronic states of the same multiplicity (considered within a non-relativistic framework).
Such CXs afford the archetypical ($F - 2$)-dimensional seam space and orthogonal two-dimensional branching space, in which the electronic PESs form the distinctive double-cone.
For completeness, we now consider two alternative cases: (i) CXs between three (or more) electronic states and (ii) CXs between two electronic states considered within a relativistic framework.

To start, we address case (i). 
Let us consider the three-state analogue of Eq.~\eqref{dia_elec_ham},\cite{matsika_nonadiabatic_2011}

\begin{equation}
\label{dia_elec_ham_3_state}
     \mathbf{H}_{\text{el}}^{\text{(dia)}}(\R) =
    \begin{pmatrix}
        V_{11}(\R) & V_{12}(\R) & V_{13}(\R)\\
        V_{21}(\R) & V_{22}(\R) & V_{23}(\R)\\
        V_{31}(\R) & V_{32}(\R) & V_{33}(\R)
    \end{pmatrix} \, .
\end{equation}

Using the same procedure as for the two-state case, the eigenvalues of $\mathbf{H}_{\text{el}}^{\text{(dia)}}(\R)$ (i.e., the three adiabatic electronic energies) can only become degenerate if the following five conditions are met,

\begin{equation}
\label{degen_cond_3_state}
\begin{split}
    V_{11}(\R_{\text{CX}}) \: = \: V_{22}(\R_{\text{CX}}) \: =& \:\: V_{33}(\R_{\text{CX}})\\
    V_{12}(\R_{\text{CX}}) \: = \: V_{13}(\R_{\text{CX}}) \: = \: V_{23}(\R_{\text{CX}}) \: =& \:\: 0 \, .
\end{split}
\end{equation}

In other words, degeneracy requires (a) all diagonal matrix elements to be equal and (b) all off-diagonal matrix elements to be zero.
Three-state CXs, therefore, have a five-dimensional branching space and an ($F - 5$)-dimensional seam space.
These arguments can be generalized to consider the branching space dimensionality, denoted $\eta$, of a CX involving an arbitrary number of $M$ electronic states of the same spin-multiplicity.
In order for $M$-fold degeneracy, $M - 1$ diagonal conditions and $M(M - 1)/2$ off-diagonal conditions must be fulfilled [considering an ($M \times M$) matrix].\cite{lipkowitz_conical_2007}
Thus, the total number of conditions that need to be obeyed, that is, the branching space dimensionality of the $M$-state CX is

\begin{equation}
\label{num_degen_conds}
    \eta = \frac{\left( M - 1 \right)\left( M + 2 \right)}{2} \, ,
\end{equation}
where the dimensionality of the corresponding seam space can then also be given as $F - \eta$.

For a discussion relating to case (ii), it is prudent to distinguish between two sub-cases, that of intersections between electronic states of different spin-multiplicity (i.e., S$_n$/T$_n$) and that of intersections exhibited by systems of an odd number of electrons.
In both cases, relativistic effects should not be neglected and, as such, the Hamiltonian that needs to be considered is $\hat{H}_{\text{el}}^\text{full}(\re,\R) = \hat{H}_{\text{el}}(\re,\R) + \hat{H}_{\text{SO}}(\re,\R)$, where $\hat{H}_{\text{el}}(\re,\R)$ is the non-relativistic electronic Hamiltonian in Eq.~\eqref{eq:fullHamiltonian} and $\hat{H}_{\text{SO}}(\re,\R)$ is the operator accounting for spin-orbit coupling.
It is well understood that intersections between electronic states of different spin-multiplicity, when viewed in a non-relativistic framework, exhibit linear ($F - 1$)-dimensional intersections, where both states cross one another.\cite{matsika2002conical, matsunaga2004modelling}
In other words, CXs do \textit{not} occur between states of different spin-multiplicity within a spin-diabatic representation of the electronic wavefunction (the eigenstates of $\hat{H}_{\text{el}}(\re,\R)$).
The inclusion of spin-orbit coupling, however, allows electronic states of differing spin-multiplicity to couple and mix, affording a new set of electronic states (now labelled by total angular momentum instead of spin-multiplicity, as the latter is no longer a 'good' quantum number\cite{Curchodatmophotochem2024}), where, for example, the components of a triplet state are no longer degenerate, and in which no states cross.
It may be possible, in such a case, for a previously designated singlet state to form a conical ($F - 2$)-dimensional intersection with one of the previously designated components of a triplet state.
This situation involves a spin-adiabatic representation of the electronic wavefunction (the eigenstates of $\hat{H}_{\text{el}}^\text{full}(\re,\R)$).

As stated several times before, in a non-relativistic framework, the electronic Hamiltonian is real-valued and, as such, the branching space of a CX between two electronic states of the same spin multiplicity is two-dimensional.
The same holds true within a relativistic framework so long as the molecule has an even number of electrons.
For a two-state CX in a system of an odd number of electrons, inclusion of the spin-orbit interaction leads to five conditions needing to be satisfied in order to allow degeneracy.\cite{matsika2001effects1, matsika2001effects2, matsika2002spin, han2003properties}
Three conditions arise solely from the electronic Hamiltonian now being complex-valued as a result of including $\hat{H}_{\text{SO}}(\re,\R)$, which can be complex.\cite{mead1979noncrossing}
Two further conditions arise due to combination of $\hat{H}_{\text{el}}^\text{full}(\re,\R)$ being complex-valued and as a consequence of time-reversal symmetry, which dictates that all eigenvalues of $\hat{H}_{\text{el}}^\text{full}(\re,\R)$ be doubly-degenerate irrespective of the nuclear geometry, a subtlety known as Kramer’s degeneracy.\cite{kramers1930theorie}
This means that such CXs, in the general case of no spatial symmetry, have a five-dimensional branching space.
If, however, the molecule exhibits $C_s$ or higher spatial symmetry, the two extra conditions related to Kramer’s degeneracy are trivially satisfied.
As such, the dimensionality of the branching space reduces to three (i.e., the value corresponding to an otherwise non-degenerate complex-valued Hermitian matrix).\cite{matsika2001effects1}

\paragraph{Conical intersections and representations of the molecular wavefunction}
As stated in Section~\ref{sec_adiabatic_diabatic}, CXs belong to the adiabatic representation of the electrons. More generally, the multi-electronic picture inherent to photochemistry emerges from the Born-Huang representation for the molecular wavefunction. Another representation of the molecular wavefunction is possible, such as the exact factorization, which describes the total molecular wavefunction as a product of a single time-dependent nuclear wavefunction and a single time-dependent electronic wavefunction (that depends parametrically on the nuclear positions).\cite{abedi_exact_2010,abedi_correlated_2012} The multiple, time-independent, and coupled PESs in the Born-Huang picture are replaced by a single, time-dependent PES and a time-dependent vector potential in the exact factorization.\cite{abedisteps} As such, CXs do not exist in the exact factorization\cite{curchod_dynamics_2017,agostini_when_2018} and the exact factorization does not depend on a representation for the electronic states (even if a connection between the exact factorization and the diabatic/adiabatic basis exists\cite{schurger2025efbh}). We stress here that CXs only appear within the Born-Huang picture of the total molecular wavefunction \textit{and} when the adiabatic representation is used, i.e., CXs do not appear between diabatic electronic states.

The previous paragraph should also serve to highlight that our representation of chemistry and photochemistry, where a molecule and its (photo)reactivity is depicted as nuclear wavefunction(s) evolving on time-independent (electronic) potential energy surface(s), emerges from a certain representation of the molecular wavefunction: the Born-Huang representation.

\subsubsection{Approximation to the electronic Sch\"{o}dinger equation: wavefunction-based methods}
\label{sec:wft}

Strictly speaking, an electronic wavefunction should be explicitly calculated for each electronic state of interest with a given wavefunction-based method. However, various linear-response and equation-of-motion techniques have been proposed where knowledge of an electronic wavefunction is needed only for the reference state (typically the ground state), and information about the other (excited) electronic states is obtained from a response formalism. Here, we provide a high-level overview of the most common approaches available in widely-used software packages, focusing on methods that are routinely used in nonadiabatic dynamics simulations.\cite{crespo2018recent,gonzalez2012progress} We do not provide the mathematical details of these methods, which can be found in the suggested references (Section~\ref{sec:elecstructbook}). Practical considerations on the selection of an electronic-structure method can be found in Section~\ref{sec:banchmarkelstr}. 

We use the following vocabulary throughout this guide: a \textit{single-reference} electronic-structure method characterizes an approach built upon a single (Hartree-Fock) Slater determinant and aiming to include dynamic correlation (we note that we will avoid to use the term 'single-reference' for methods emanating from Kohn-Sham density-functional theory), a \textit{multiconfigurational} electronic-structure approach describes electronic state(s) with a combination of Slater determinants (each having a significant weight) to capture static correlation for the electronic state(s) of interest, and a \textit{multireference} electronic-structure technique attempts to capture dynamic correlation using a a multiconfigurational wavefunction as a reference. 

\paragraph{Configuration-interaction methods}
The simplest method for calculating electronically excited states is configuration interaction singles (CIS), which is effectively an excited-state extension of Hartree-Fock (HF). In CIS, single-excitation configurations are generated directly from a ground-state HF reference wavefunction.\cite{DH05} The electronic wavefunction for each excited electronic state is described as a linear combination of such singly-excited configurations. Due to its low accuracy rooted in the lack of correlation effects (correlation arises from doubly-excited configurations and beyond), the CIS method is hardly ever used on its own in nonadiabatic simulations. It is occasionally used for large molecules where CIS is combined with semiempirical Hamiltonians.\cite{nelson2014nonadiabatic} 
Attempts to extend CIS to multiple excitation levels (CISD, CISDT, etc) to capture the missing dynamic correlation effects have generally not been very successful,\cite{gonzalez2012progress} being also computationally cumbersome. The computational burden is even a bigger issue for multireference CI methods (MRCI), which, instead of a HF reference, employ a set of configurations (a multiconfigurational wavefunction) as a reference for the CI expansion. Multireference treatment allows the inclusion of static correlation effects, which, along with the treatment of dynamic correlation (achieved through higher CI excitation levels), makes the MRCI approach potentially very accurate.\cite{lischka2018multireference} However, steep scaling and slow convergence with respect to the excitation level limit the use of MRCI electronic structure methods to very small systems. Illustrative examples of MRCI-based excited-state dynamics simulations are provided in works of Barbatti, Lischka, and coworkers (see, for example, Refs.~\cite{vazdar2009excited,szymczak2008mechanism}).

\paragraph{Coupled-cluster and algebraic-diagrammatic-construction methods}
Within the single-reference domain, a notable competitor of the CI approach is the family of coupled cluster (CC) methods (though still seldom used, multireference CC methods are also being developed\cite{evangelista2018perspective}). The CC hierarchy is a powerful tool for systematically increasing the level of dynamic correlation by incorporating single, double, triple, and higher excitations through the so-called cluster operator. Unlike CI methods, CC methods are generally size-extensive upon truncation of excitation level, ensuring that the energy scales correctly with system size. The CC sequence of CCSD, CCSDT, CCSDTQ, etc., can be implemented in both linear-response and equation-of-motion fashions.\cite{watts2008introduction,krylov_equation--motion_2008} A more approximate counterpart with lower computational scaling is in the sequence of CC2, CC3, CC4, etc.\cite{sneskov2012excited} The CC3 method is presently regarded as a highly accurate benchmark method for (dominantly single) excitations in small to medium-sized organic molecules,\cite{loos2020adc} while CC4 can only be used for very small systems.\cite{loos2022mountaineering} The CC2 method has, however, been the primary workhorse in computational photochemistry/physics due to its favourable balance of accuracy and efficiency, availability of analytic gradients,\cite{kohn2003analytic} and other relevant properties. Unfortunately, while scanning the excited-state PESs, CC2 may suffer from instabilities in regions of quasi-degeneracies,\cite{plasser2014surface,parker2016unphysical} while CXs between excited states have improper dimensionality\cite{tuna2015assessment}. Solutions to these problems have recently been proposed,\cite{kjonstad2017crossing,kjonstad2017resolving} which also paved the way to employ new CC developments in nonadiabatic simulations.\cite{kjonstad2024photoinduced,hait2024prediction} Nevertheless, the issues associated with conventional CC2 (which are known to stem from its non-Hermitian foundations) can be mitigated by using a related Hermitian approach -- the algebraic diagrammatic construction (ADC) method. Despite the distinct origins of the CC and ADC formalisms, CC2, and its ADC(2) counterpart, are closely related methods\cite{hattig2005structure} sharing a similar accuracy with respect to excitation energies. ADC(2) is somewhat faster, numerically more stable,\cite{plasser2014surface} and has correct dimensionalities for excited-state CXs.\cite{tuna2015assessment,taylor2023description} However, this is not true for intersections between the ground state and the first excited state\cite{tuna2015assessment,taylor2023description}, which are essentially calculated at different levels of theory  -- ADC(2) excited states are constructed on top of the MP2 (M\o ller-Plesset perturbation theory of second-order) ground state. Despite this limitation, ADC(2) has been very popular in simulations of nonadiabatic processes of small to medium-sized organic molecules, particularly when `problematic' doubly excited states are not of key importance. Spin-component scaling (SCS) can further enhance the accuracy of ADC(2) PESs.\cite{tajti2019accuracy} Unlike for organic molecules, applying ADC(2) to systems containing transition metals may be more challenging, and the method should be used with caution.\cite{plasser2015high} ADC(2)/MP2 can also become problematic when used for molecules containing a carbonyl group, as the method can predict artificially low energy gaps between the S$_1(n\pi^\ast)$ and S$_0$ state.\cite{marsili2021caveat} Finally, one should stress that, while being useful for photophysical processes, single-reference methods have their inherent limitations when used to describe photochemical reactions, and only multireference electronic structure methods can, in principle, handle all the challenges associated with the ab initio nonadiabatic dynamics of such processes.

\paragraph{Multiconfigurational and multireference methods}
 While it is often (rightly) argued that CI and CC methods converge toward a full CI limit,\cite{bartlett2017cc} which makes the distinction between single and multireference somewhat artificial, in practice, single-reference methods require exceedingly high excitation levels to address typical `multireference problems'. Illustrative multireference situations include bond breaking, near-degeneracies in transition metals and in the vicinity of CXs -- all very common in nonadiabatic molecular dynamics and hard to capture with a response formalism based on a single-reference description of the ground electronic state. Multireference and multiconfigurational methods enable the description of challenging electronic structures and, if properly designed by a smart selection of active space orbitals (that is, a subset of molecular orbitals that are important to describe the multiconfigurational nature of the electronic states of interest), they accurately portray various types of excited states, including the notoriously challenging doubly-excited states. The freedom of active space construction presents both a challenge and an opportunity for the widespread application of these methods, keeping in mind that large active spaces are also often prohibitively expensive.

In the complete active space self-consistent field (CASSCF) method, a set of active orbitals is treated in a full CI manner, whereas both orbital coefficients and orbital shapes are being (variationally) optimized. Inactive orbitals, on the other hand, are treated in a mean-field manner similar to HF. The state-averaged CASSCF (SA-CASSCF) method is typically employed to calculate multiple electronic states and ensure their orthogonality (whilst avoiding a variational collapse of the CASSCF wavefunction towards the ground electronic state). SA-CASSCF is a powerful multiconfigurational method that offers analytic gradients and NACVs, while it also ensures the correct treatment of interstate crossings between the ground and excited electronic states. Consequently, SA-CASSCF has been a method of choice in numerous nonadiabatic studies so far.\cite{crespo2018recent,gonzalez_quantum_2021,gonzalez2012progress} However, the method does not necessarily provide highly accurate excitation energies -- it accounts only for static electron correlation but lacks most of the dynamic electron correlation. Dynamic correlation can be incorporated by additional corrections added on top of the (SA-)CASSCF wavefunctions.

MRCI and MRCC methods mentioned above often use the CASSCF as a reference wavefunction for adding dynamic correlation.\cite{lischka2018multireference} However, the most popular approach is complete active-space second-order perturbation theory (CASPT2), which introduces dynamic correlation through perturbative corrections. Multi-state (MS-CASPT2) and extended multi-state (XMS-CASPT2) CASPT2 methods are often combined with nonadiabatic simulations, although their computational cost currently limits their application to smaller systems,\cite{park2020multireference} and MS-CASPT2 can exhibit artifacts when electronic states come close in energy.\cite{shiozaki2013pyrazine} The newly developed RMS-CASPT2 (R standing for rotated)\cite{nishimoto2022analytic} is an emerging method that is the least sensitive to the number of included states among all CASPT2 variants.\cite{bezabih2023comparative} Ref.~\citenum{ibele2024aimspt2} studied the impact of various flavours of multistate CASPT2 on nonadiabatic dynamics. 

Finally, we also mention methods designed as cost-effective alternatives to SA-CASSCF, such as RASSCF (restricted active space self-consistent field),\cite{malmqvist1990restricted} FOMO-CI (floating occupation molecular orbitals configuration interaction),\cite{granucci2000molecular,granucci2001direct} FOMO-CASCI (floating occupation molecular orbitals complete active space configuration interaction),\cite{slavicek2010fomocasci,hohenstein2015fomocasci,hollas2017nonadiabatic} scaled $\alpha$-CASSCF,\cite{snyder2017alpha,frutos_tracking_2007} as well as a plethora of multireference semiempirical methods,\cite{thiel2014semiempirical} which can all be used in nonadiabatic dynamics simulations.

\subsubsection{Approximation to the electronic Sch\"{o}dinger equation: density-based methods}
\label{sec:dftbased}

An entirely different family of electronic structure methods for electronically excited states are based on density-functional theory (DFT). It is known from the Hohenberg-Kohn theorems\cite{hohenberg64} that DFT ensures that the ground-state (electronic) properties of a system are fully determined by its electron density. While the ground-state energy can be calculated by minimizing the energy functional of the density, DFT itself does not provide information about the electronically excited states of the system. The time-dependent extension of DFT (TDDFT),\cite{tddftcarsten} formalized by the Runge-Gross theorems,\cite{Runge84} opened the DFT framework to time-dependent processes and gives access to strategies to calculate excited electronic states and their properties. Both DFT and TDDFT are formally exact theories. However, practical calculations are routinely subject to a number of approximations. In practice, DFT is used with approximate exchange-correlation (xc) density functionals, which also limits the accuracy of TDDFT calculations. TDDFT itself is typically employed within the ubiquitous adiabatic approximation, which neglects the role played by the history of the time-dependent electron density on the time-dependent xc potential.\cite{casida2012progress,tddftcarsten} The adiabatic approximation should not be confused with the term adiabatic as used in the context of adiabatic/diabatic states -- it rather signifies the locality of the time-dependence in TDDFT xc potentials. This lack of memory effects allows the application of conventional DFT xc-functionals in TDDFT as well, but at the expense of limiting the accuracy of TDDFT. Casida\cite{casida95}, and Gross and coworkers\cite{petersilka_excitation_1996} showed how in principle exact excitation energies and transition dipole moments (from the ground electronic state) can be obtained from the poles and residues of linear response functions (the dynamic polarizability or the density-density response function, respectively) by expressing these quantities in a TDDFT framework.\cite{tddftcarsten,agostini2018tddft} This approach, coined linear-response TDDFT (LR-TDDFT), is by far the most commonly employed DFT-based framework in the context of photochemistry and photophysics.

The popularity of LR-TDDFT stems from its relatively low computational cost and favourable scaling with the system size, which allows the treatment of molecular systems that are difficult to deal with higher-level wavefunction-based methods. The accuracy of LR-TDDFT depends on the types of excited states under consideration. The lack of double excitations in the LR-TDDFT spectrum,\cite{elliott2011perspectives,levine06} as well as the difficulty of treating Rydberg, charge-transfer,\cite{dreuw2003long} and charge-transfer-like excitations,\cite{kuritz2011charge} are all essentially rooted in the use of the adiabatic approximation in LR-TDDFT.\cite{tddftcarsten} The choice of xc-functionals may help alleviate some of these problems. For instance, range-separated hybrid functionals are known to improve the accuracy of charge-transfer excited states,\cite{laurent2013td} which are severely underestimated with conventional GGA (generalized gradient approximation) and hybrid functionals. The use of the Tamm-Dancoff approximation (TDA) in LR-TDDFT calculations is beneficial\cite{hirata99} for improving the energy gaps between singlet and triplet states,\cite{peach2011influence} while also increasing the numerical stability of excited state calculations.\cite{oxirane_I,tapavicza08} However, there are no simple solutions when dealing with 'inherently multireference' electronic states. The approximations behind the xc-functionals, along with the adiabatic approximation atop, are overall less systematic than approximations used in standard wavefunction-based approaches (e.g., truncation of excitation levels). Hence, the results obtained with LR-TDDFT implementations should always be carefully benchmarked with respect to more rigorous excited-state methods.

LR-TDDFT has rapidly become a popular approach for combination with nonadiabatic dynamics simulations.\cite{curchod2013trajectory,barbatti2014surface} Excited-state analytic gradients are widely available in electronic structure codes, while nonadiabatic couplings are often calculated using auxiliary many-electron wavefunctions\cite{tapavicza07,tavernelli09b,Tavernelli2009,tavernelli2010nonadiabatic,curchod2013trajectory,mitric2008nonadiabatic} or other response approaches\cite{chernyak2000density,hu08,send_first-order_2010}, although more general derivations, including quadratic response for quantities between excited states, exist.\cite{li2014first,ou_first-order_2015,wang_nac-tddft_2021} LR-TDDFT correctly describes CXs between excited states, while the description of crossings between the ground and excited states has been more problematic.\cite{levine06,taylor2023description} The latter has been one of the limitations in practical simulations involving nonradiative decay from the excited to the ground state.\cite{barbatti2014surface} We will discuss this issue later.

\paragraph{Alternative density-based approaches}
In the following, we mention other density-based methods used for excited-state calculations and, occasionally, nonadiabatic dynamics. In most cases, LR-TDDFT uses a (singlet) ground-state Slater determinant composed of Kohn-Sham orbitals as a reference state for the linear-response formalism. Conversely, spin-flip TDDFT (SF-TDDFT)\cite{herbert2022spin} uses the lowest triplet state as a reference state and permits spin flips for the (de-)excited electron, allowing for the description of doubly-excited singlet states that are otherwise absent in standard LR-TDDFT and the description of CXs between the ground and excited states.\cite{winslow2020sftddftcxs} A study on the photoisomerization pathway of the DHP/CPD system also demonstrated the accuracy of SF-TDDFT to treat ionic/covalent electronic-state characters in a balanced way.\cite{lognon2023sftddft} SF-TDDFT has been employed in nonadiabatic dynamics,\cite{minezawa2019trajectory,yue2018performance} but the general problem of the SF method is its often limited accuracy and the extent of spin-contamination in excited states. The more recent MR-SF-TDDFT\cite{lee2018eliminating} (where MR stands for mixed-reference here and should not be confused with multireference) variant allows to eliminate spin-contamination with great potential for applications in nonadiabatic dynamics.\cite{lee2025expanding}  Hole-hole Tamm-Dancoff approximation (hh-TDA) employs a doubly anionic reference state, enabling a consistent description of ground and excited states of a neutral system by subsequent removal of two electrons.\cite{bannwarth_holehole_2020} As it can accurately describe CXs, hh-TDA is a promising candidate for nonadiabatic dynamics simulations (as long as the excited electronic states of interest are characterized by the same populated virtual orbital).\cite{yu_ab_2020}
DFT with various constraints on spin and charge of arbitrary molecular fragments, called CDFT (constrained DFT),\cite{kaduk2012constrained} may also be used to calculate electronic-structure quantities needed to propagate nonadiabatic trajectories\cite{krenz2020photochemical}. Related approaches include $\Delta$SCF with constraints of the occupancy of Kohn-Sham orbitals, where the ground-state density is readapted and variationally optimized to mimic the density of an excited state.\cite{kaduk2012constrained,vandaele2022deltascf} Such methods have been used since the early days of ab initio nonadiabatic simulations,\cite{doltsinis02b} but they do not necessarily offer high accuracy, while also treating a limited number of electronic states. We note that methods such as CDFT and $\Delta$SCF yield quasi-diabatic states by construction -- adiabatic states may be calculated through linear combinations via CI approach. 
Other density-based techniques, like the semiempirical DFT-MRCI\cite{grimme1999combination} or MC-PDFT (multiconfiguration pair-density functional theory),\cite{gagliardi2017multiconfiguration} suggest a multireference extension of DFT. Both approaches show good performance for calculating excited states of systems with challenging electronic structures. MC-PDFT has also been used with nonadiabatic molecular dynamics.\cite{calio2022nonadiabatic} 
Ensemble DFT-based approaches, which go beyond consideration of a pure-state Kohn-Sham reference,\cite{lieb_density_1983,gross_density-functional_1988} similarly allow access to excited states variationally, with the SI-SA-REKS (state-interaction state-averaged spin-restricted ensemble-referenced Kohn-Sham) method\cite{filatov1999spin,filatov_spin-restricted_2015} experiencing practical application in nonadiabatic dynamics simulations.\cite{filatov2019theoretical,filatov2019non,janos2023controls}
Finally, to deal with large systems, semiempirical tight-binding (TB) methods such as TD-DFTB have also been developed\cite{niehaus2001tight} and employed in excited-state dynamics simulations.\cite{mitric2009nonadiabatic,humeniuk2017dftbaby}


\subsubsection{Wishlist for electronic-structure methods to be used in nonadiabatic molecular dynamics}

We finish this Section by summarizing here the wishlist for an optimal electronic-structure method to be used in nonadiabatic molecular dynamics. 

An electronic-structure method should, ideally,
\begin{itemize}
\item describe equally well the different (excited) electronic states for the photochemical or photophysical process of interest;
\item be able to provide the different electronic-structure quantities required for the nonadiabatic molecular dynamics -- electronic energies, nuclear gradients, and various quantities between the ground and the first excited electronic state and also between excited electronic states: nonadiabatic coupling vectors, nonadiabatic couplings, spin-orbit couplings -- and possibly the calculation of electronic quantities related to experimental observables (e.g., ionization energies and intensities for photoelectron spectroscopy, transition dipole moments for transient absorption spectroscopy);
\item describe the topology and the topography of conical intersections adequately;
\item be computationally efficient to be used with on-the-fly (direct) nonadiabatic molecular dynamics;
\item not require an extremely large basis set for convergence of its electronic quantities;
\item be robust (i.e., stable) when visiting different regions of the nuclear configuration space;
\item describe all regions of the nuclear configuration space with the same accuracy.
\end{itemize}

Needless to say that none of the existing electronic-structure methods will tick all the boxes of this wishlist, and a compromise is always required for the electronic structure and depends on the molecular problem of interest. We will explore some of these compromises in Section~\ref{sec:banchmarkelstr}.

\subsubsection{Books and reviews on electronic-structure theory}
\label{sec:elecstructbook}

Additional information on electronic-structure methods for excited-state dynamics can be found in the following selected books\cite{gonzalez_quantum_2021,tddftcarsten,GONZALEZ20241,conicalintersection2004}, book chapters,\cite{barbatti2014surface} and reviews.\cite{park2020multireference,lischka2018multireference,dreuw2015algebraic,crespo2018recent,curchod2013local,matsika_electronic_2021}

\subsection{Nonadiabatic molecular dynamics techniques}
\label{nonadiabaticdynproblem}

Section \ref{elecstructproblem} introduced various concepts related to the electronic-structure challenges associated with nonadiabatic molecular dynamics, and more specifically strategies to calculate the key electronic-structure quantities that enter the equations of motion for the nuclear amplitudes (Eq.~\eqref{eq:tdsegBH}), namely the electronic energies and the nonadiabatic coupling terms. We reached a point where we can consider propagating in time the nuclear amplitudes to determine the actual nonadiabatic molecular dynamics. A wide range of techniques exists for this task and we will focus here more specifically on trajectory-based approaches, compatible with an on-the-fly determination of the electronic-structure information required and, as such, favoured to describe (approximately) the excited-state dynamics of molecular systems in their full dimensionality. The reader interested in quantum dynamics methods can refer to the following Refs.~\citenum{gatti2017applications,gatti2014molecular,tannor_book,meyer2009multidimensional}. We also note that practical details related to the numerical implementation of these methods for nonadiabatic dynamics of molecules will be discussed later, as part of Section~\ref{sec:performingNAMD}.

\subsubsection{Methods based on trajectory basis functions}
\label{sec:gaussmethod}
A first strategy to allow for simplification of the molecular time-dependent Schr\"{o}dinger equation, within a Born-Huang framework, consists of expressing the nuclear amplitudes in a basis of multidimensional, traveling Gaussian functions, denoted $\{\tilde{\chi}_k^{(J)}\}_{k=1}^{N_{\text{TBFs}}^{(J)}}$, where $\tilde{\chi}_k^{(J)} \equiv \tilde{\chi}_k^{(J)}\left(\bs R;\overline{\bs R}_k^{(J)}(t),\overline{\bs P}_k^{(J)}(t),\overline{\gamma}_k^{(J)}(t),\boldsymbol{\omega}(t)\right)$. The traveling nature of each of these trajectory basis functions (TBFs) is apparent from the definition of its center in position space, $\overline{\bs R}_k^{(J)}(t)$, and momentum space, $\overline{\bs P}_k^{(J)}(t)$, as well as its phase, $\overline{\gamma}_k^{(J)}(t)$, and width matrix $\boldsymbol{\omega}(t)$.

Using this basis to express the nuclear wavefunction as well as the electronic basis suggested by the Born-Huang representation, we obtain a new expression for the molecular wavefunction,
\begin{equation}
\Psi(\bs r,\bs R,t)=\sum_J^\infty \sum_k^{N_{\text{TBFs}}^{(J)}} C_k^{(J)}(t) \tilde{\chi}_k^{(J)}\left(\bs R;\overline{\bs R}_k^{(J)}(t),\overline{\bs P}_k^{(J)}(t),\overline{\gamma}_k^{(J)}(t),\boldsymbol{\omega}(t)\right) \Phi_J(\bs r;\bs R) \, ,
\label{wfbasis}
\end{equation}
with $N_{\text{TBFs}}^{(J)}$ being the total number of TBFs used to describe the nuclear amplitude for electronic state $J$ and $\{C_k^{(J)}\}_{k=1}^{N_{\text{TBFs}}^{(J)}}$ the complex time-dependent expansion coefficients associated with the basis functions $k$ describing the wavefunction of electronic state $J$.

Bringing this new form of the Born-Huang representation into the time-dependent Schr\"{o}dinger equation (Eq.~\eqref{eq:tdse}) (as done earlier to reach Eq.~\eqref{eq:tdseBH}), multiplying both sides of the result by $\left(\tilde{\chi}_{k'}^{(I)}(\bs R;\overline{\bs R}_k^{(J)}(t),\overline{\bs P}_k^{(J)}(t),\overline{\gamma}_k^{(J)}(t),\boldsymbol{\omega}(t)) \Phi_I(\bs r;\bs R)\right)^*$ on the left, and integrating over the electronic and the nuclear coordinates, we obtain a set of coupled equations of motion for the complex time-dependent expansion coefficients which, in a matrix form, reads:
\begin{equation}
\dot{\bs C}=-i\bs S^{-1}[(\bs H - i \dot{\bs S})\bs C] \, .
\label{eomec}
\end{equation}
$(\bs S)_{kk'}^{IJ}=\braket{\tilde{\chi}_k^{(I)}}{\tilde{\chi}_{k'}^{(J)}}_{\bs R}\delta_{IJ}$ is an overlap matrix of the nuclear basis functions, and $\bs H$ the Hamiltonian matrix composed of matrix elements $(\bs H)_{kk'}^{IJ}=\bra{\Phi_I\tilde{\chi}_k^{(I)}}\hat{H}\ket{\tilde{\chi}_{k'}^{(J)}\Phi_J}_{\bs r,\bs R}$. $\dot{\bs S}$ is a non-Hermitian time-derivative matrix with elements defined as $(\dot{\bs S})_{kk'}^{IJ}=\braket{\tilde{\chi}_k^{(I)}}{\frac{\partial}{\partial t}\tilde{\chi}_{k'}^{(J)}}_{\bs R}\delta_{IJ}$. Importantly, Eq.~\ref{eomec} \textit{couples} the TBFs.
Eq.~\eqref{eomec} is strictly equivalent to the time-dependent molecular Schr\"odinger equation, expressed now in a basis of (orthonormal) adiabatic electronic eigenfunctions and (nonorthogonal) time-dependent nuclear functions (the TBFs defined above). In other words, in the limit of a large electronic and nuclear basis,  Eq.~\eqref{eomec} is formally exact. 

A central idea of all TBF-based methods is to truncate the size of the traveling basis set and approximate their couplings to offer approximations to the time-dependent molecular Schr\"odinger equation amenable to molecular systems. Keeping in mind that the TBFs evolve in phase space, we could keep their number low if the equations of motion for propagating the TBFs ensure that they would always offer a good support to describe the time-dependent nuclear amplitudes -- the TBFs will form a moving grid. Different methods emerged from these considerations and differ in the way they evolve the TBFs in phase space and approximate the Hamiltonian matrix elements between TBFs. In the following, we will offer a brief overview of the three main frameworks used in nonadiabatic molecular dynamics. 

\paragraph{Variational multiconfigurational Gaussian}
The method called variational Multiconfigurational Gaussian (vMCG) propagates the TBFs by applying the Dirac-Frenkel variational principle to their time-dependent parameters.\cite{worth2004novel,lasorne2006direct,lasorne2007direct,worth2008solving,doi:10.1098/rsta.2020.0386,doi:https://doi.org/10.1002/9781119417774.ch13} The TBFs are not only coupled via the equations of motion for their complex coefficients, but also by their phase-space dynamics. Despite leading to possibly unstable equations of motion, this coupled propagation ensures that TBFs are covering the region of nuclear configuration space important to describe the nuclear dynamics. The coupling between TBFs is achieved via a local harmonic approximation, requiring the calculation of Hessians from the electronic-structure method used for on-the-fly (direct) dynamics -- performed in a diabatic representation. The cost of calculating these Hessians can be alleviated by using a database, storing and reusing the electronic-structure quantities when running a swarm of coupled TBFs.\cite{10.1063/5.0043720} A frozen-width approximation for the TBFs, i.e., $\boldsymbol{\omega}(t) \rightarrow \boldsymbol{\omega}$, improves the stability of the vMCG equations of motion.\cite{mendive2012towards} Direct-dynamics vMCG (DD-vMCG) was applied to study the photophysics and photochemistry of numerous systems in the gas phase, but also including the influence of a solvent using a polarizable continuum model.\cite{D1CP01843D,molecules26237247} For details on the method, the interested reader is referred to Ref.~\citenum{richings2015quantum}.

\paragraph{Multiconfiguration Ehrenfest and ab initio multiple cloning}
Another strategy, called multiconfiguration Ehrenfest (MCE), consists of propagating the TBFs (using a frozen width matrix) along Ehrenfest (mean-field) trajectories.\cite{shalashilin2009quantum,shalashilin2010nonadiabatic,saita2012fly} In its original version, MCE couples the TBFs via their complex amplitudes and their Ehrenfest equations of motion. A second version of the technique, more suitable for on-the-fly dynamics, removes the couplings between TBFs in their equations of motion.\cite{SHALASHILIN2024212} The mean-field nature of Ehrenfest trajectories sometimes challenges the description of branching at a CX, and a version of MCE, coined ab initio multiple cloning (AIMC), was developed to improve the description of a nuclear wavepacket splitting at a CX.\cite{makhov2014ab} The ab-initio version of MCE and the AIMC have been used to study the photodynamics of a broad range of molecular systems.\cite{doi:10.1021/acs.jctc.3c00583,Makhov_2022,doi:10.1021/acs.jpclett.1c00266,C8CP06359A,C9CP00039A} Additional information on these methods can be found in Refs.~\citenum{makhov2017ab,doi:https://doi.org/10.1002/9781119417774.ch15,SHALASHILIN2024212}

\paragraph{Full and ab initio multiple spawning}
Another famous method using the idea of TBFs is Full Multiple Spawning (FMS),\cite{martinez1996multi,martinez1997non,Ben-Nun1998,ben-nun00,hack2001comparison,toddaims,virshup2008photodynamics,curchod2018ab} where the TBFs have a frozen width matrix and follow classical trajectories. The key difference between FMS and MCE or vMCG is that the number of TBFs in the dynamics is not fixed but can vary over time. More specifically, the FMS technique offers a robust algorithm, called the spawning algorithm, that allows for a smooth expansion of the number of TBFs whenever a region of strong nonadiabaticity is encountered.\cite{yang2009optimal,levine2007isomerization,toddaims,curchod2018ab} A TBF evolving in a given electronic state approaching a region of strong nonadiabatic coupling can spawn, i.e., create a new TBF on the coupled state, and the two TBFs can exchange amplitude via the time-dependent Schr\"{o}dinger equation (Eq.~\eqref{eomec}). The FMS framework considers that the Hamiltonian matrix elements in Eq.~\eqref{eomec} are calculated exactly. However, such matrix elements can be approximated using a saddle point approximation, allowing for an on-the-fly propagation and coupling of the TBFs (see Ref.~\citenum{mignolet2018walk} for details and tests of this strategy). This strategy is often coined ab initio multiple spawning (AIMS) and can tackle small- to medium-size molecular systems to elucidate the photochemistry of numerous molecules (e.g., Refs.~\cite{tao2011ultrafast,mignolet2016rich,snyder2016gpu,pijeau2017excited,li2017ultrafast,glover2018excited,yang2018imaging,coates2018vacuum} for examples and Refs.~\cite{persico2014overview,curchod2018ab} for a more detailed list of applications) and was generalized to account for an external time-dependent electric field\cite{mignolet2016communication,mignolet2019excited} and intersystem crossings.\cite{curchod2016communication,fedorov2018predicting,fedorov2016ab,varganovGAIMS2021} Further information on the FMS formalism and the AIMS technique can be found in Refs.~\citenum{curchod2018ab,toddaims,aimschaptercurchod}. A typical drawback of the spawning algorithm is that a large number of TBFs can be created during an AIMS dynamics, leading to a computationally expensive determination of the coupling between all pairs of TBFs to form the Hamiltonian matrix. A series of techniques have been devised to detect when TBFs need to be coupled (in nonadiabatic regions), allowing us to decouple them otherwise with only a very limited loss of accuracy. These techniques, called stochastic-selection AIMS (SSAIMS)\cite{curchod2020ssaims,ibele2021ssaims} and AIMS with informed stochastic selections (AIMSWISS),\cite{lassmann2021aimswiss,lassmann2022extending} preserve the accuracy of an AIMS simulation while reducing their computational costs down to those typical of mixed quantum/classical methods that will be described in the following section.   

\subsubsection{Methods based on classical trajectories}
\label{sec:classtrajmethod}

Another strategy to perform nonadiabatic molecular dynamics involves the use of a swarm of classical trajectories to depict the dynamics of coupled nuclear probability densities. A commonality among all \textit{mixed quantum/classical} methods is the fact that the nuclear degrees of freedom are evolved classically while the electronic part of the problem (and, to some degree, the coupling between electronic states) is contained in a time-dependent electronic Schr\"{o}dinger equation where the electronic wavefunction is propagated on the support of the classical trajectory. Hence, mixed quantum/classical methods use a Newton equation for the propagation of the nuclei (for a trajectory $\alpha$):
\begin{align}
\mathbf F^{\alpha}(t) = -\boldsymbol\nabla_{\bs R} \varepsilon(\bs R)|_{\bs R = \bs R^{\alpha}(t)} \, .
\label{eqn: quantum-classical general}
\end{align}
where $\varepsilon(\bs R)$ is a (electronic) potential energy whose precise definition gives rise to different methods. $R^{\alpha}(t)$ is the nuclear configuration at time $t$ along trajectory $\alpha$. A time-dependent electronic wavefunction $\tilde{\Phi}(\bs r; \bs R^{\alpha}, t)$ is evolved on the support of each trajectory $\alpha$ according to a time-dependent Schr\"{o}dinger equation
\begin{equation}
i \hbar \frac{\partial \tilde{\Phi}(\bs r; \bs R^\alpha(t), t)}{\partial t} = \hat{\mathcal{H}}_{el}(\bs r, \bs R^\alpha(t)) \tilde{\Phi}(\bs r; \bs R^\alpha(t), t) \, .
\label{eq:tdese}
\end{equation}
To unravel the physics contained in the time-dependent electronic wavefunction, one could expand it in a basis of adiabatic electronic states,
given by 
\begin{align}
\tilde{\Phi}(\bs r; \bs R^{\alpha}, t) = \sum_I c_I^\alpha(t) \Phi_I(\bs r; \bs R^{\alpha}) \, ,
\label{eqn: surface-hopping BH expansion}
\end{align}
and insert this Ansatz into Eq.~\eqref{eq:tdese}. The usual algebra -- left multiplication by one of the electronic basis functions $\Phi_J^\ast(\bs r; \bs R^{\alpha})$ and integration over the electronic coordinates -- is used to reveal a set of coupled equations of motion for the complex electronic coefficients, 
\begin{align}
\dot c_{J}^{\alpha}(t) &= -i  E_J^{\text{el}}\left(\bs R^{\alpha}\right)c_J^{\alpha}(t)-\sum_I  \bra{\Phi_J(\bs R^{\alpha})}\frac{\partial}{\partial t}\ket{\Phi_I(\bs R^{\alpha})}_{\mathbf{r}}  c_I^{\alpha}(t) \, .
\label{tdese_coeff_intermed}
\end{align}

The term $\bra{\Phi_J(\bs R^{\alpha})} \frac{\partial}{\partial t}\ket{\Phi_I(\bs R^{\alpha})}_{\mathbf{r}}$ can be rewritten using the chain rule $\frac{\partial}{\partial t} = \frac{\partial}{\partial \bs R}\frac{\partial \bs R}{\partial t}$, leading to $\bra{\Phi_J(\bs R^{\alpha})}\frac{\partial}{\partial t}\ket{\Phi_I(\bs R^{\alpha})}_{\mathbf{r}} = \bra{\Phi_J(\bs R^{\alpha})}  \frac{\partial}{\partial \bs R}\frac{\partial \bs R}{\partial t}  \ket{\Phi_I(\bs R^{\alpha})}_{\mathbf{r}} = \bs d_{JI}\left(\bs R^{\alpha}\right)\cdot \dot{\bs R}^{\alpha}(t)$.\cite{hammes94} Hence, Eq.~\eqref{tdese_coeff_intermed} can be rewritten to unravel the presence of the NACVs:

\begin{align}
\dot c_{J}^{\alpha}(t) &= -i  E_J^{\text{el}}\left(\bs R^{\alpha}\right)c_J^{\alpha}(t)-\sum_I  \bs d_{JI}\left(\bs R^{\alpha}\right)\cdot \dot{\bs R}^{\alpha}(t) c_I^{\alpha}(t) \, .
\label{tdese_coeff}
\end{align}

Eq.~\eqref{tdese_coeff} shows that the time evolution of a given electronic coefficient along a trajectory $\alpha$ is composed of an adiabatic (first term on the right-hand side) and a nonadiabatic part (second term on the right-hand side) -- the latter clearly identified thanks to the presence of the NACVs $\bs d_{JI}\left(\bs R^{\alpha}\right)$.

\paragraph{Ehrenfest dynamics}
In Ehrenfest dynamics, the electronic potential used to drive the nuclear dynamics in Eq.~\eqref{eqn: quantum-classical general} is defined at time $t$ as the expectation value of the electronic Hamiltonian given by the time-dependent electronic wavefunction at time $t$, namely
\begin{align}
 \varepsilon(\bs R^{\alpha}(t)) := \langle \tilde{\Phi}(\bs R^\alpha,t) | \hat{H}_\text{el}(\bs R^\alpha) | \tilde{\Phi}(\bs R^\alpha,t) \rangle_{\bs r} \,.
\label{ehrenfest}
\end{align}
Hence, the classical nuclear dynamics in Ehrenfest dynamics is dictated by a mean-field potential, and the evolution of the time-dependent electronic wavefunction directly influences the nuclear dynamics. When the time-dependent wavefunction is described by a single adiabatic electronic state, the nuclear dynamics is adiabatic, that is, Born-Oppenheimer in nature, with respect to this electronic state. When nonadiabatic effects induce a mix of electronic adiabatic states, via the NACVs in Eq.~\eqref{tdese_coeff}, the nuclear dynamics is driven by a linear combination of electronic energies, \textit{i.e.}, a mean-field potential.  

\paragraph{Trajectory surface hopping}
Trajectory surface hopping (TSH), on the other hand, proposes to portray the nonadiabatic dynamics of nuclear wavepackets by using swarms of independent classical trajectories that can hop from one electronic state to the other in case of strong nonadiabaticity.\cite{bjerre1967energy,preston71} To achieve this goal, TSH uses the following definition for the potential energy driving the classical nuclear dynamics (Eq.~\eqref{eqn: quantum-classical general}):
\begin{align}
 \varepsilon(\bs R^{\alpha}(t)) := E_\ast^{\text{el}}(\bs R) \, ,
\label{ehrenfest2}
\end{align}
where the notation $E_\ast^{\text{el}}(\bs R)$ indicates that the classical force is obtained from a given adiabatic electronic state, but that the state label can change during the dynamics of a given trajectory as a result of a \textit{hop}, mimicking a nonadiabatic transition. In stark contrast to Ehrenfest dynamics, the classical nuclear dynamics in TSH is performed on adiabatic PESs -- not on a mean-field PES. The choice of the electronic state driving the dynamics is dictated by a recipe involving the time-dependent electronic wavefunction introduced earlier (Eq.~\eqref{tdese_coeff}). In the following, we introduce the recipe for hops suggested by Tully in 1990,\cite{tully90} coined fewest-switches, as this strategy is by far the most commonly used and implemented surface-hopping algorithm.  

Let us discover the working principle of the fewest-switches TSH by following a given TSH trajectory $\alpha$ of the swarm. As for all nonadiabatic molecular dynamics methods, a TSH run typically starts by defining a set of initial conditions (ICs; nuclear positions and momenta) for a given trajectory $\alpha$ and which electronic state will start driving the dynamics at time $t=0$ -- say $J$ in this case. Detailed information about these two processes will be provided below in Section~\ref{sec:stepsofnonaddyn} for all methods discussed in the current section. At time $t_0=0$, the complex electronic coefficients are set to $c_I^{\alpha}(t_0)=0+0i\, \forall I\neq J$ and $c_J^{\alpha}(t_0)=1+0i$. The TSH dynamics is initiated and the nuclei are propagated according to Eq.~\eqref{eqn: quantum-classical general} for a timestep $\Delta t$, with the electronic energy being $\varepsilon(\bs R^{\alpha}(t)) := E_J^{\text{el}}(\bs R)$. After each nuclear integration time step $\Delta t$, the following protocol is operated: (i) the electronic coefficients are propagated in time based on Eq.~\eqref{tdese_coeff} for the duration $\Delta t$ (using a smaller time step for this electronic propagation), (ii) at time $t+\Delta t$, the fewest-switches probability for the trajectory $\alpha$ to jump from the driving electronic state to another one is calculated, and (iii) a stochastic algorithm is used to determine whether the trajectory should change state. In the fewest switches algorithm,\cite{tully90} the probability to hop from state $J$ to any other electronic state $I$ between time $t$ and $t+\text{d}t$ is given by
\begin{align}
\mathcal P_{J\rightarrow I}^{\alpha} = \max\left[0, \frac{-2 \,\mathrm{d}t}{\left|c_J^{\alpha}(t)\right|^2} \,\Re\left[c_J^{\alpha \ast}(t)c_I^{\alpha}(t)\right] \dot{\bs R}^{\alpha}(t)\cdot \bs d_{IJ}\left(\bs R^{\alpha}\right)\right] \, .
\label{eq:tshproba}
\end{align}
A hop occurs from electronic state $J$ to another electronic state $I$ if
\begin{align}
\sum_K^{I-1} \mathcal P_{J\rightarrow K}^{\alpha} \le \zeta \le \sum_K^{I} \mathcal P_{J\rightarrow K}^{\alpha} \, ,
\end{align}
where $\zeta$ is a random number generated uniformly in the interval $[0:1]$. If a jump occurs, the nuclear kinetic energy needs to be rescaled to ensure that the total classical energy is conserved (see Ref.~\citenum{toldo2024velresc} for a recent discussion and caveat on this process). The trajectory $\alpha$ is propagated until the predetermined completion criterion -- often a given total propagation time. The user then needs to run a large swarm of (independent) TSH trajectories to ensure that the sampling of ICs \textit{and} the stochastic algorithm used for the nonadiabatic hops are converged.  

\begin{figure}[h!]
    \centering
    \includegraphics[width=1.0\linewidth]{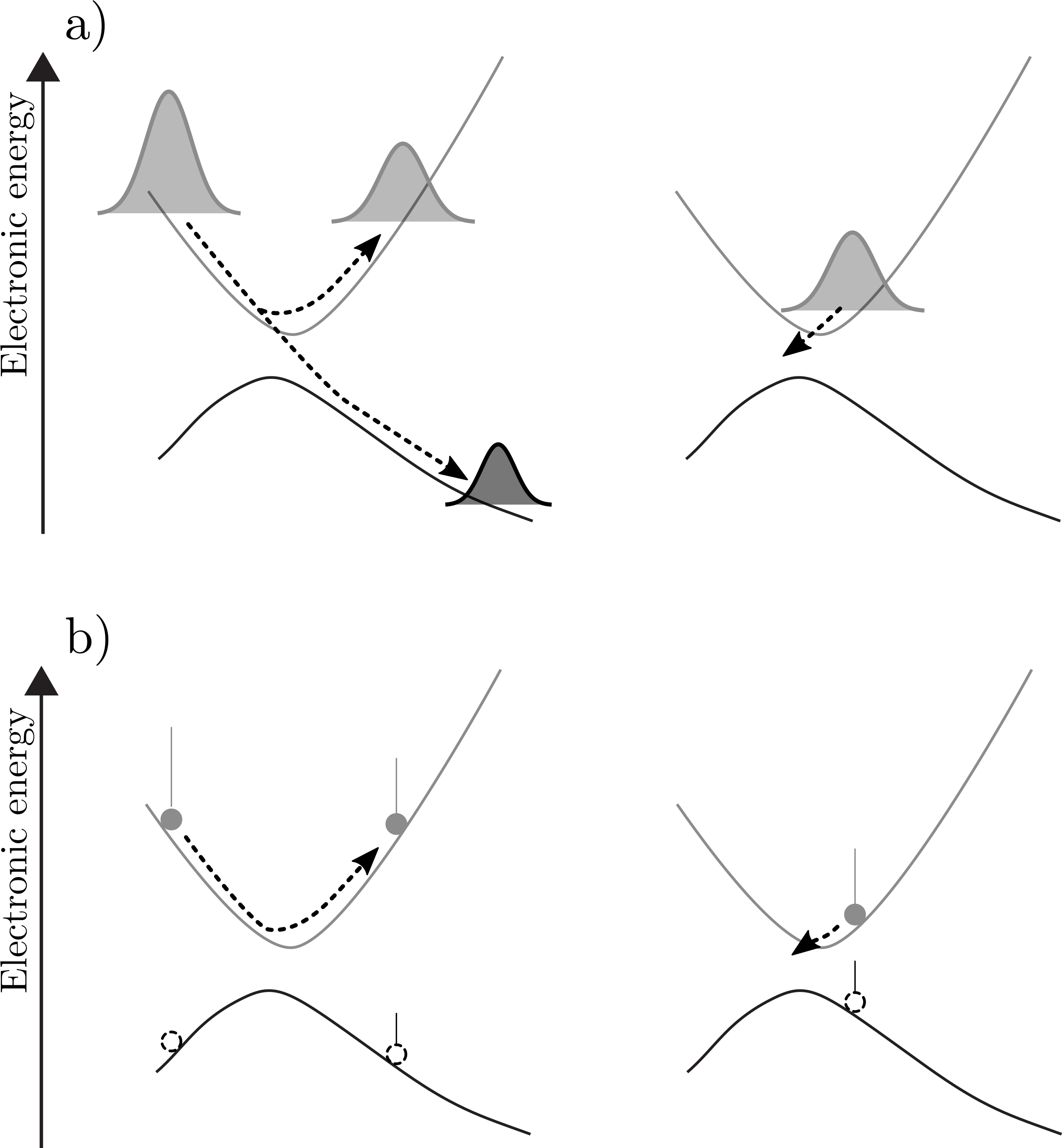}
    \caption{Schematic representation of a first branching of a nuclear wavepacket following a nonadiabatic region (left), followed by a second crossing event (right) in (a) quantum dynamics and (b) TSH. The right panel highlights the overcoherence issue of TSH in its original flavour. In panel (b), the trajectory that drives the TSH dynamics is given as a filled circle (the scenario here is that the driving trajectory remains in the upper electronic state). The dashed circle indicates a 'ghost' trajectory that follows the driving trajectory and is meant to highlight the dynamics of the electronic coefficients. The magnitude of the electronic coefficient in each electronic state is symbolized by the vertical bar on top of the circles: long bar for the upper electronic state at the beginning of the dynamics (the electronic coefficient is set to 1 for this state, 0 for the other), and smaller bars after the first crossing as electronic amplitude was shared between both electronic states due to nonadiabatic couplings. 
}
    \label{decoherence}
\end{figure}

\paragraph{A brief apart\'{e} on electronic decoherence}
When a nuclear wavepacket reaches a region of strong nonadiabaticity, the nonadiabatic couplings will transfer nuclear amplitude to the coupled electronic state. After the nonadiabatic region, each part of the nuclear wavepacket (in the original electronic state and in the coupled electronic state) will follow different PESs and move away from each other -- a process often referred to as branching of the nuclear wavepacket (see Fig.~\ref{decoherence}a, left, for a depiction of this process). If a component of the nuclear wavepacket comes back to the same region of nonadiabaticity after the first passage, it is expected that the other component (in the coupled state) would have left this region (see Fig.~\ref{decoherence}a, right). In other words, the overlap of the two nuclear wavepackets after the first passage would have reduced to zero -- a process called 'decoherence' in the nonadiabatic dynamics community. The picture is somewhat more complex in TSH. The original version of fewest-switches TSH, owing to its independent trajectory approximation, has been shown to suffer from \textit{overcoherence}. Overcoherence refers here to the fact that the electronic coefficients for a trajectory $\alpha$ (Eq.~\eqref{tdese_coeff}) are forced to evolve on the support of this single trajectory $\alpha$ and, as such, will interact coherently throughout the trajectory dynamics -- the independent trajectory approximation implies that the coefficients of a trajectory $\alpha$ remaining on the original state $J$ cannot apprehend the dynamics that the coefficients on another trajectory $\beta$, which jumped to the coupled state $I$, would undergo. Coming back to our scheme (Fig.~\ref{decoherence}b), a TSH trajectory evolving in the upper electronic state will pass through the region of strong nonadiabaticity and, imagining that no hop occurs, will carry on its dynamics on this state (Fig.~\ref{decoherence}b, left). Focusing on the electronic coefficients, the passage through the nonadiabatic region would have led to an (incomplete) transfer of electronic amplitude from the upper to the lower state (see the sticks on top of the trajectories in Fig.~\ref{decoherence}b). Both electronic coefficients (sticks) are, however, forced to follow the driving TSH trajectory -- decoherence could not take place. Hence, when this trajectory comes back in the nonadiabatic region (Fig.~\ref{decoherence}b, right), the electronic coefficient for the upper state will see a non-zero coefficient in the lower state, and they will interact coherently. This process is strikingly different from that expected in the exact quantum dynamics case (Fig.~\ref{decoherence}a, right), where the portion of the nuclear wavepacket on the upper state reaches the nonadiabatic region for a second time but does not interact anymore with the lower component (produced during the first event) thanks to decoherence. We note that AIMS would describe such branching naturally thanks to the spawning of coupled TBFs in the different electronic states (see Ref.~\citenum{aimschaptercurchod} for a discussion).

A plethora of works have focused on the overcoherence problem of TSH and suggested various patches for this method (e.g., Refs.~\citenum{Bittner1995,jasperJCP2005,granucci2010including,shenvi2011phase,shenvi2011simultaneous,shenvi2012achieving,subotnik2011new,subotnik2011decoherence}).
The most commonly employed strategy, used as a default in most TSH code, is coined the \textit{energy-based decoherence correction} (EDC),\cite{granucci2010including} and enforces \textit{internal consistency} of TSH,~\cite{fang99b,Granucci2007} that is, the fact that the \textit{fraction of trajectories} in an electronic state $I$ at time $t$, $P_I(t) = \frac{N_I(t)}{N_{\text{traj}}}$, should be equal to the \textit{averaged electronic population} in an electronic state $I$ at time $t$, $\langle|c_I(t)|^2\rangle=\frac{1}{N_{\text{traj}}}\sum_\alpha^{N_{\text{traj}}} |c_I^\alpha(t)|^2$, for all $I$ and $t$. To achieve this goal, EDC dampens the population of the inactive TSH states at each time step using a decoherence time that is deduced from the electronic energy difference between each inactive state and the active one based on earlier derivations in the context of mean-field methods.\cite{Zhu2004} More involved correction schemes have also been suggested, such as the (parameter-free) \textit{augmented fewest switches surface hopping} (A-FSSH) strategy\cite{subotnik2011new,subotnik2011fewest,subotnik2016understanding} or the decoherence correction inspired by the exact factorization.\cite{min2018decoh} 
Model systems were used to test the various decoherence-correction schemes for TSH in various scenarios of nonadiabatic transitions that would usually lead to a failure of uncorrected TSH (see for examples Refs.~\citenum{granucci2010including,subotnik2011new,subotnik2011fewest,Jain_2016,Granucci2007,subotnik2011decoherence,Gorshkov2013fk}). It is interesting to note that the widely used EDC appears to fail for a model of double crossings (Tully model 2).\cite{falkissueedc2014} A more limited number of works focused on the role and influence of decoherence corrections in the nonadiabatic dynamics of molecular systems (see, for example, Refs.~\citenum{granucci2010including,nelson2013nonadiabatic,plasser2019lvc,vindel-zandbergen2021decoh,ibele2020}).

Last but not least, we discussed in this section TSH in its fewest-switches flavour, but numerous other TSH schemes were developed, some also tested on molecular systems: surface hopping by consensus,\cite{martens2016surface} coupled-trajectory TSH based on the exact factorization,\cite{pieronicttsh2021} Zhu-Nakamura TSH,\cite{zhunakatsh2001,zhunakatsh2006} Landau-Zener TSH,\cite{slavicek2020} coherent switching with decay of mixing,\cite{Zhu2004,shu2020CSDMsharc} and the mapping approach to surface hopping (MASH).\cite{Mannouch2023,richardson2025mash}

\begin{figure*}[h!]
    \centering
    \includegraphics[width=0.7\linewidth]{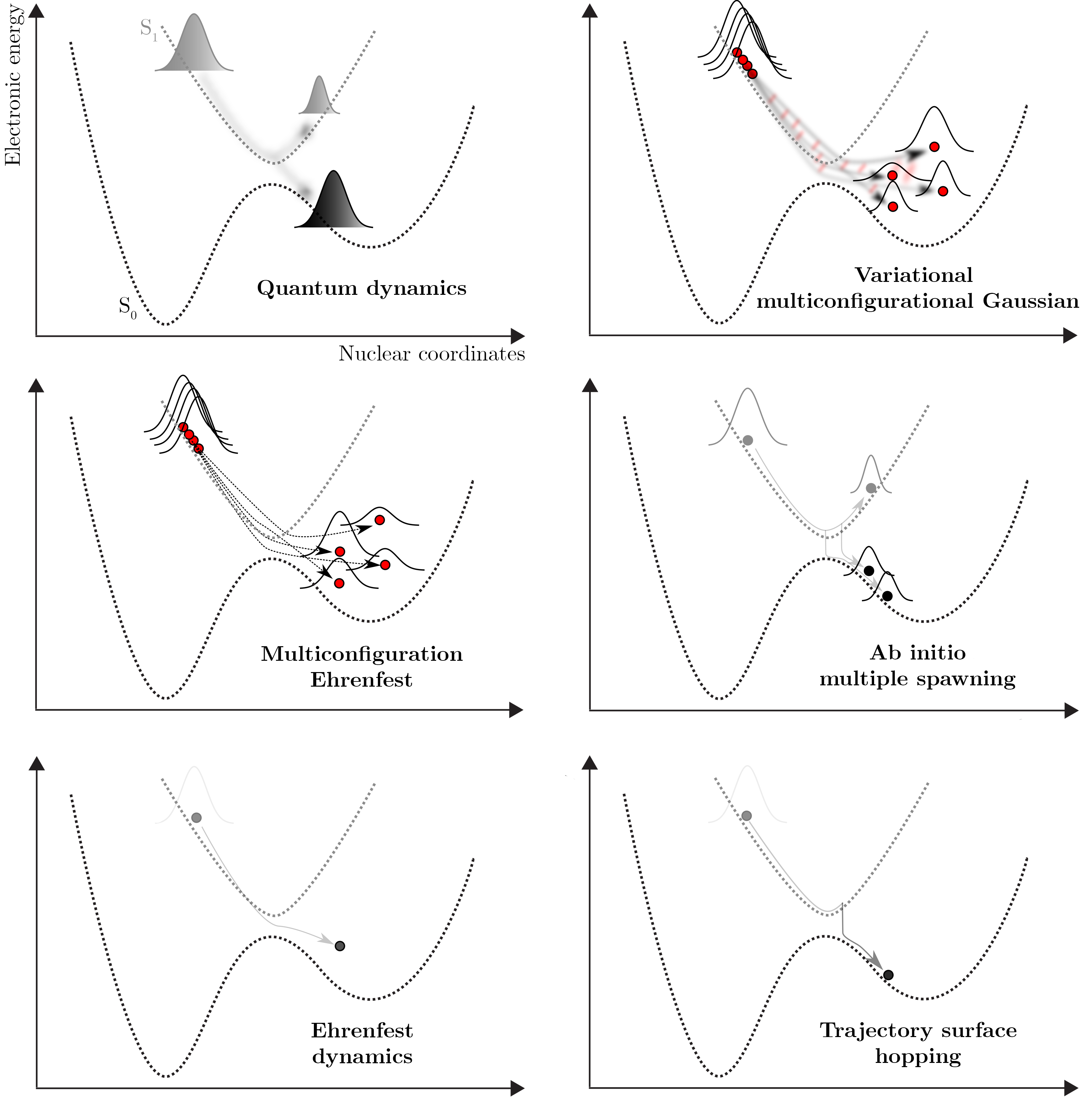}
    \caption{Schematic representation of the various families of methods for nonadiabatic molecular dynamics. }
    \label{fig:namdmethods}
\end{figure*}

\subsubsection{Hierarchy of methods for nonadiabatic molecular dynamics}

\begin{figure}[h!]
    \centering
    \includegraphics[width=1.0\linewidth]{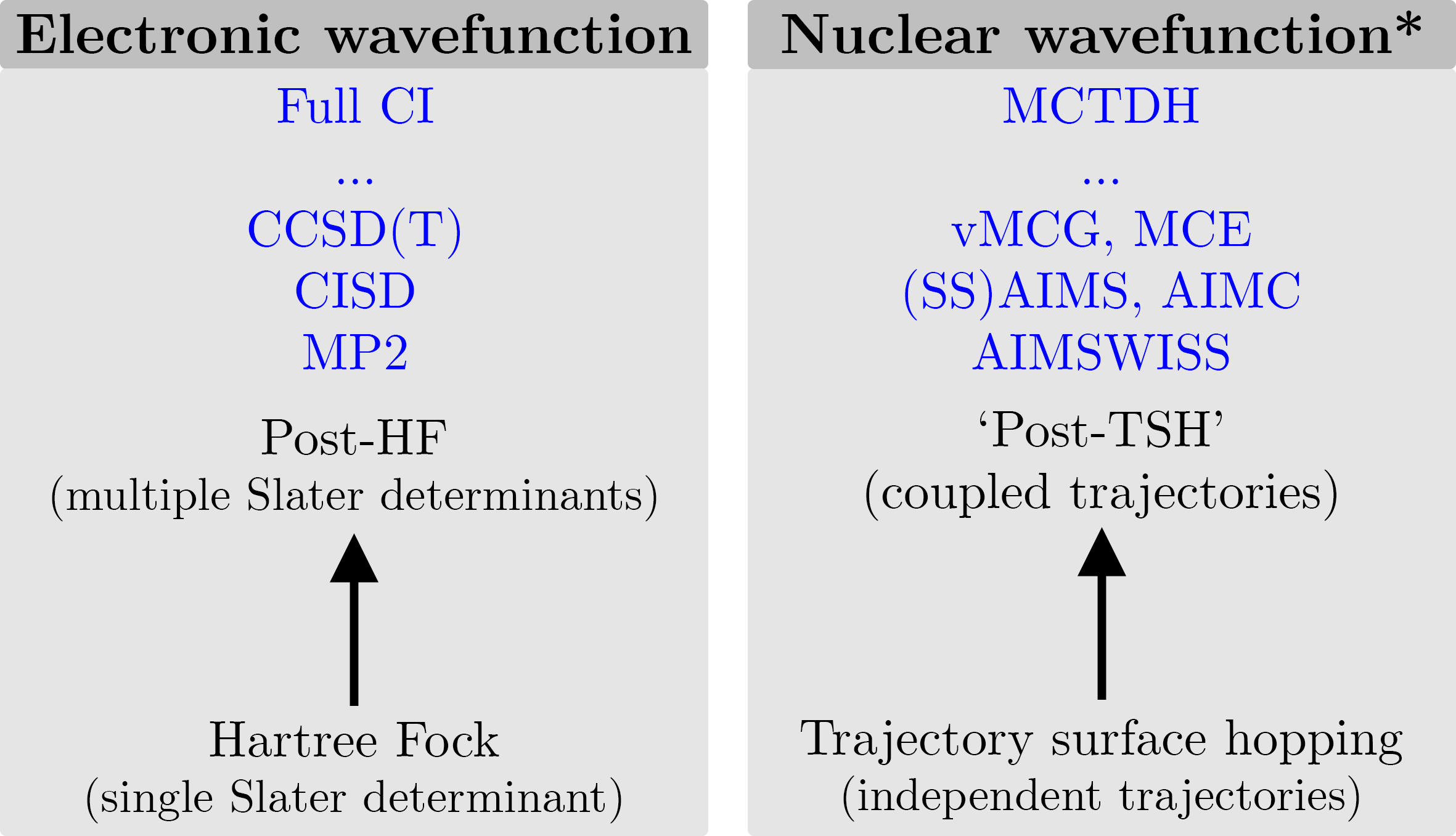}
    \caption{Analogy between quantum chemistry and nonadiabatic molecular dynamics -- the quest to uncover a hierarchy of methods to describe the electronic and nuclear wavefunction. The asterisk on 'nuclear wavefunction' is meant to indicate that mixed quantum-classical methods like TSH mimic the nuclear probability density in place of the actual nuclear wavefunction.}
    \label{fig:hierarchymethods}
\end{figure}

As a conclusion to this section on different methods for nonadiabatic molecular dynamics (see Figure~\ref{fig:namdmethods} for a schematic representation of all the methods discussed), an analogy with electronic-structure theory is tempting. In electronic-structure theory, one often takes HF, a method approximating the electronic wavefunction by a single Slater determinant composed of molecular orbitals, as a starting point for the build-up of a hierarchy toward a higher-level, more accurate (ground-state) electronic wavefunction (Fig.~\ref{fig:hierarchymethods}, left). Hence, one can offer a rationale for the expected improvement of the electronic wavefunction when starting from HF and climbing the ladder of the method hierarchy -- MP2 is seen as a first improvement, that can be further consolidated by higher order or by using coupled-cluster theory, up to the numerically-exact full configuration interaction (numerically, as the method is exact within a give basis set).  

Based on the presentation of the methods for nonadiabatic molecular dynamics above, it is tempting to offer a similar, yet looser, analogy for the approximations to the nuclear wavefunctions (Fig.~\ref{fig:hierarchymethods}, right).\cite{lassmann2025thesis} TSH offers a first approximation not to the nuclear wavefunctions \textit{per se} but their probability density, by using a swarm of independent classical trajectories. The next step 'post-TSH' to recover approximate nuclear wavefunctions would be AIMSWISS, which uses a basis of TBFs to describe the nuclear wavefunctions that are coupled only when strictly required (close to nonadiabatic regions). A next step in improving the quality of the nuclear wavefunctions would consist in describing better the coupling between TBFs, as done first by SSAIMS, then AIMS or AIMC. For an even more accurate depiction of the nuclear wavefunctions, one can use DD-vMCG or MCE in its original version, before resorting to the near numerically-exact multi-configuration time-dependent Hartree method (MCTDH).\cite{Beck2000,mctdhnonad,gatti2014molecular} As stated earlier, this analogy with electronic-structure theory should be taken with a grain of salt -- TSH is per se not used to build an AIMS or a vMCG wavefunction. Yet, this analogy allows to highlight how the description of the nuclear wavefunctions can be improved in nonadiabatic molecular dynamics, from only their probability densities being approximated by a swarm of independent classical trajectories (TSH) to the nuclear wavefunctions being depicted by coupled, traveling Gaussian basis functions. 

At this point, one can ask the question: why do we need improved nuclear wavefunctions? One important element of response can be found by considering the final results of a nonadiabatic molecular dynamics simulation. Often, the analysis of such dynamics is based on the decay of the electronic population -- mimicking how the excited nuclear wavepacket returns to the ground electronic state. The time traces of electronic populations are often fitted and compared to experimental proxies (evolving features in time-resolved spectra that are used as a measure of an electronic state). Electronic populations are also often used directly to compare methods for nonadiabatic molecular dynamics. The reason for this success is perhaps that electronic populations are usually a simpler quantity to calculate from a nonadiabatic dynamics simulation, even for mixed quantum/classical methods.\cite{mignolet2018walk} 
However, electronic populations are not observables -- they emanate from the Born-Huang representation of the molecular wavefunction -- and there is a growing need for the calculation of actual experimental observables with the emergence of advanced light sources with shorter (more intense) laser pulses. Some observables (as simple as a time-dependent dipole moment for example) require the evaluation of matrix elements directly involving the nuclear wavefunctions, which may become more challenging for TSH -- one often needs to use an \textit{ad hoc} prescription for the evaluation of such matrix elements. In any case, it becomes clear from this brief discussion that the ordering of nonadiabatic methods in the hierarchy above is likely to change depending on the type of quantities (observable or not) calculated. Dedicated benchmarking strategies for nonadiabatic methods have emerged and will help clarify which method should be used for which molecular case and observables.\cite{doi:10.1021/acs.jpclett.1c00266,ibele2020,cigrang2025roadmapNAMD}

\subsubsection{Books and reviews}
Additional information on nonadiabatic (molecular) dynamics can be found in the following selected books and book chapters\cite{gonzalez_quantum_2021,conicalintersection2004,gatti2014molecular,GONZALEZ20241,kutateladze2005computational,marxbook,tannor_book,persico2018photochemistry,tully1998modern,aimschaptercurchod} or reviews.\cite{crespo2018recent,curchod2018ab,agostini2019different,ibele2024EFreview,richardson2025mash,makhov2017ab,richings2015quantum,tully2012perspective,toddaims,WCMS:WCMS26,barbatti2011}

\section{The different steps of a nonadiabatic molecular dynamics simulation}
\label{sec:stepsofnonaddyn}

\begin{figure*}[h]
\centering
\includegraphics[width=0.654\textwidth]{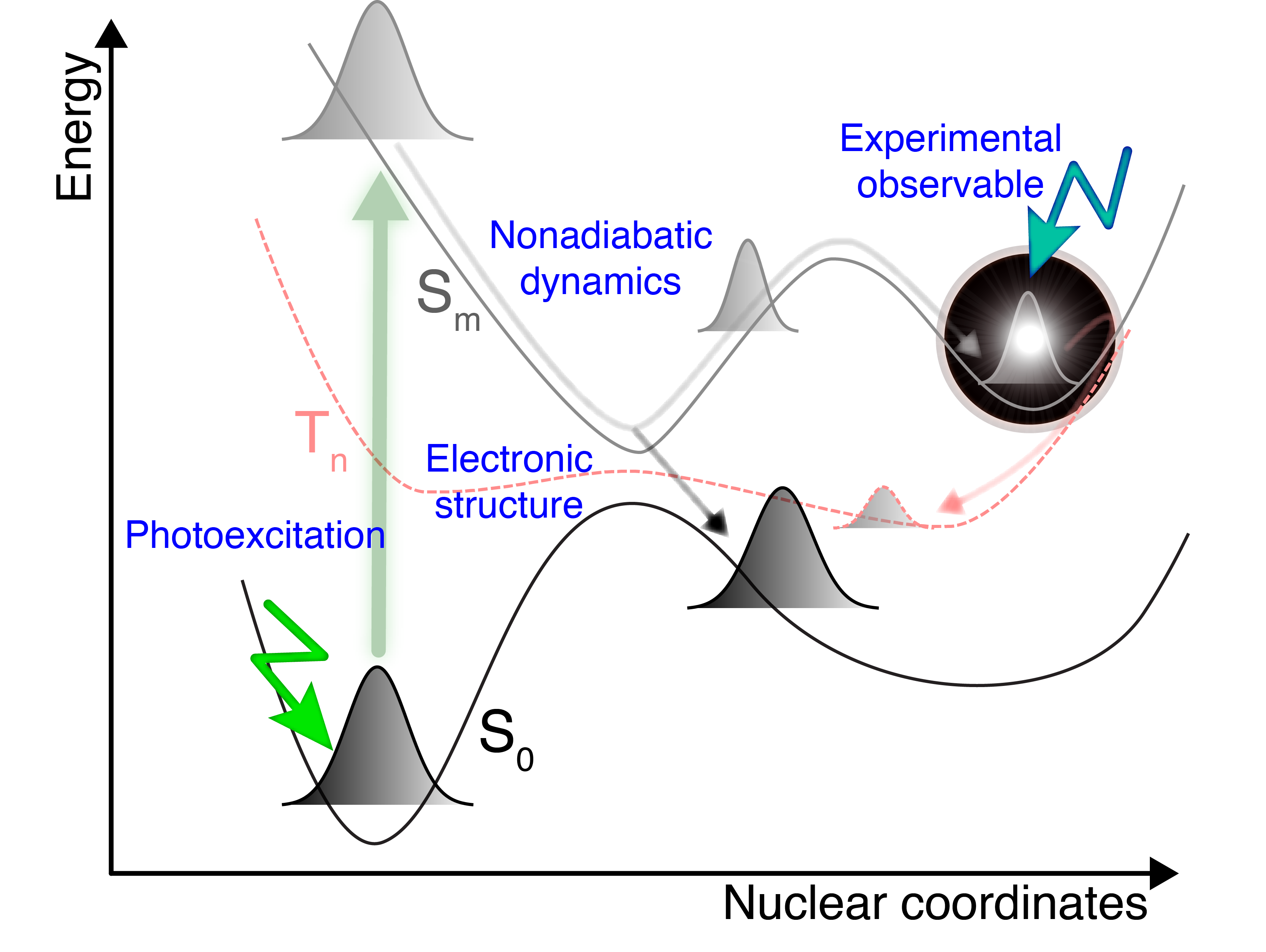}
\caption{Schematic representation of the four key ingredients to perform a nonadiabatic molecular dynamics simulation: photoexcitation process, electronic-structure method to provide electronic energies and other electronic properties, approach to treat the nonadiabatic dynamics, and the calculation of experimental observables.}
\label{fig:schemeglobal}
\end{figure*}

In this Section, we propose to dig deeper into the details of the various steps constituting a nonadiabatic molecular dynamics simulation (Fig.~\ref{fig:schemeglobal}). Mirroring the discussion above, we stress that a nonadiabatic molecular dynamics simulation is meaningless in itself if the various ingredients it requires, namely the choice of the electronic-structure method and the description of photoexcitation, have not been thoroughly benchmarked. We suggest here various protocols and strategies to build confidence in the several components necessary to perform a nonadiabatic molecular dynamics. We note that we will address more practical questions in subsequent Sections.

\subsection{Benchmarking electronic-structure methods}
\label{sec:banchmarkelstr}
Selecting an optimal electronic-structure method (hereby, listed in Sections~\ref{sec:wft} and ~\ref{sec:dftbased}) for propagating ab initio nonadiabatic molecular dynamics is a critical first step in ensuring a reliable simulation, and sometimes more important than the actual method used for the nonadiabatic dynamics, as identified in several recent works.\cite{janos2023controls,papineau2024,crespo2018recent} These works illustrate how the choice of electronic structure significantly influences the calculated lifetimes and quantum yields for the photochemistry of model molecules. The question surrounding the choice of electronic-structure methods is an old concern within the nonadiabatic dynamics community. Comprehensive benchmark studies addressing this issue remain scarce, but recent community efforts such as the prediction challenge on the photochemistry of cyclobutanone have provided a prototypical system exhibiting the dramatic impact of the electronic-structure method on experimental observables.\cite{predictionchallenge} A recent work also proposed a sensitivity analysis of nonadiabatic molecular dynamics with the photodynamics of cis-stilbene.\cite{tomas2024}

In general, a good electronic-structure method is one that accurately captures the photophysics/photochemistry of a given system. Nevertheless, the meaning of `accuracy' can vary depending on the context. Sometimes, a qualitative understanding suffices, while in other cases, a quantitative calculation of observables is necessary for comparison with experimental data (regardless of whether experimental measurements are available or anticipated). It is important to note that a computationally inexpensive electronic structure method with low accuracy does not guarantee even a qualitatively correct outcome. Therefore, a thorough benchmark of electronic structure methods is imperative before starting with any nonadiabatic simulation.

Even prior to any preliminary calculation, it is advisable to review the existing literature on the system of interest, gaining insight into the anticipated excited-state processes and identifying appropriate methods. For instance, if bond breaking is an expected excited-state behaviour, fine-tuning multireference methods capable of accurately describing bond dissociation are essential. Conversely, for larger systems undergoing a photodynamical process over long timescales, the focus should be on identifying cost-effective (potentially single-reference) methods that can qualitatively capture the dynamics. For systems undergoing intersystem crossing processes, an electronic-structure method giving access to spin-orbit coupling matrix elements and providing a balanced treatment of singlets and triplets is needed.

Depending on the type of processes under study and the size of the molecular system, nonadiabatic simulations can require substantial computational resources, making the choice of electronic-structure method a crucial consideration. Due to their size and complexity, many experimentally relevant molecules and materials may only become computationally feasible over time. At this point, it is pivotal to realize that many photophysical processes, which (as stated in the introduction) often do not involve substantial nuclear reorganization or any bond breaking, may tolerate a slightly more approximate level of theory. However, photochemical reactions are often highly challenging from an electronic-structure perspective, and any attempt to simulate such processes with low-quality electronic-structure methods (to save computational effort, for example) may become a futile task. Fortunately, ongoing advances in electronic structure methods, with promising results from machine-learning potentials, aim to meet the computational demands of large and complex systems, with dynamics potentially occurring on long timescales. Nevertheless, most nonadiabatic dynamics applications to photochemistry have, to date, primarily focused on small to medium-sized molecular systems.

\subsubsection{Tests in the Franck-Condon region}\label{sec:fctest}
Initial calculations typically begin with an assessment of vertical excitation energies at the optimized ground-state geometry, i.e., the Franck-Condon (FC) geometry. The number of excited states of interest varies depending on the experimental conditions being simulated -- the laser excitation wavelength or the excitation range of an external source (such as sunlight), as discussed in further detail in Section~\ref{sec:initcond}. During the nonadiabatic dynamics, many excited electronic states that were initially high in energy at the FC point may subsequently stabilize and influence the results of the simulation. Therefore, the best practice is to calculate more than the minimal required number of transitions in the FC region.

\begin{figure*}[h!]
    \centering
    \includegraphics[width=0.7\linewidth]{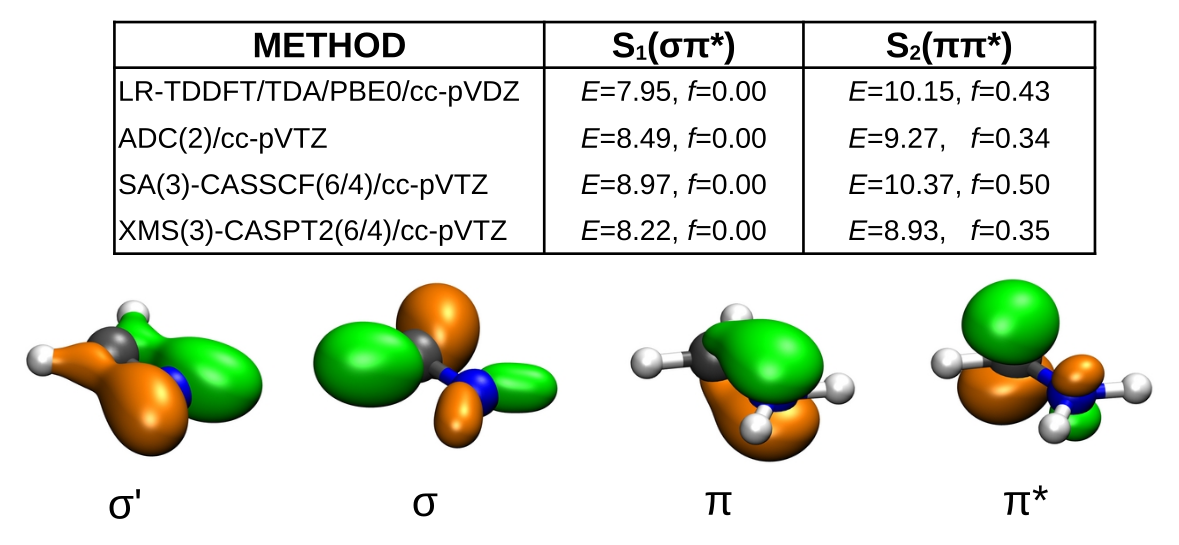}
    \caption{Excitation energies (in eV) of the two lowest singlet excited electronic states of protonated formaldimine. Active space orbitals used in the multiconfigurational calculations are listed. Computational details follow those reported in Ref.~\citenum{taylor2023description}.}
    \label{fig:formaldimine_exc}
\end{figure*}

In addition to excitation energies, molecular excited states are characterized by their electronic properties. Transition dipole moments ($\boldsymbol{\mu}_{IJ}$) and the associated oscillator strengths ($f_{IJ}=\frac{4 \pi m_e }{3 \hbar e^2} (E_J^{\text{el}}-E_I^{\text{el}}) |\boldsymbol{\mu}_{IJ}|^2$) indicate the probability of absorption/emission, typically between the ground ($I$) and excited state ($J$) (transition dipole moments between excited electronic states are also available with some specific electronic-structure methods\cite{jacquemin2025betweenes}). Understanding the character of calculated excited electronic states in terms of hole (i.e., donating) and particle (i.e., accepting) orbitals is particularly useful in organic photochemistry (e.g., $\pi\pi^\ast$, $n\pi^\ast$, $\pi\sigma^\ast$, $n\sigma^\ast$ states, etc.). Excited-state transitions may also exhibit charge-transfer or diffuse Rydberg characters, which can be analyzed through canonical or, preferably, natural transition orbital analysis. Post-processing wavefunction analysis tools such as TheoDORE\cite{plasser2020theodore} and Multiwfn\cite{lu2024comprehensive} help quantify excited-state characters through measures like electron-hole correlation plots, charge transfer numbers, and exciton size. Assessing the character of the different excited electronic states of interest is critical to compare their energy ordering with different electronic-structure methods.
Figure~\ref{fig:formaldimine_exc} presents a comparative illustration of excitation energies for protonated formaldimine calculated using four different electronic-structure methods: LR-TDDFT, ADC(2), SA-CASSCF, and XMS-CASPT2. For the latter two methods, the active space orbitals are explicitly listed. The two lowest singlet excited states, labeled $\sigma\pi^\ast$ and $\pi\pi^\ast$, can be readily distinguished based on their oscillator strengths and the associated orbitals. This assignment can be further supported by comparing the natural transition orbitals (of LR-TDDFT and ADC(2)) with the active space orbitals involved in two excitations. Therefore, in this relatively simple case, further analysis of state character appears unnecessary. XMS-CASPT2 and ADC(2) show good agreement in excitation energies and oscillator strengths within the expected error margins. Although the ADC(2) energies are slightly higher than those of XMS-CASPT2, they maintain a similar interstate energy gap. In contrast, the results from LR-TDDFT and SA-CASSCF appear less balanced. The accuracy of LR-TDDFT depends on the underlying approximations (see Section~\ref{sec:dftbased}) and the choice of xc-functional, whereas (SA-)CASSCF lacks dynamic correlation, which limits its predictive reliability (see Section~\ref{sec:wft}).

Thorough characterization of the excited electronic states for a given molecule allows for a strict comparison of different electronic-structure methods. It is always advisable to compare predictions from DFT-based and wavefunction-based methods, as well as to assess single-reference vs. multireference approaches for a given molecular geometry (often, the FC geometry), extending the test to different basis sets.  While small basis sets enable fast dynamics, they can significantly reduce the accuracy of the calculations depending on the character of the electronic states of interest. Achieving full basis set convergence for excited states is generally difficult, but convergence trends should always be explored by comparing several basis sets of varying sizes, including those with diffuse functions. Ref.~\citenum{hait2024prediction} highlighted, for example, the sensitivity of the electronic energy of a Rydberg state (S$_2$) of cyclobutanone to the precise type of diffuse basis set used: 6-31+G$^\ast$ or def2-SVPD were found to overestimate the energy of the Rydberg state in comparison to aug-cc-pVDZ. 

Minimal benchmarking typically involves comparing vertical excitation energies as well as oscillator strengths. When comparing electronic-structure methods, it is important to recognize that the accuracy of excited-state methods does not adhere to the same error expectations as for the ground-state thermochemistry (e.g.,  1 kcal/mol) -- larger deviations between methods are often anticipated. Discrepancies in excitation energies of less than 0.2 eV ($\sim 4.6$ kcal/mol) are generally considered small, particularly when they are consistent across the spectrum of excited electronic states (and, as we shall see later, when the molecule is pushed away from the FC region).

The results obtained with the widely used LR-TDDFT (within its conventional approximations) should always be compared with those of more rigorous wavefunction-based methods (e.g., ADC(2), CC2, EOM-CCSD, or CC3 if affordable), while cross-comparisons of results among 'higher-level' methods are also encouraged. Multireference electronic structure should be preferred whenever the system size allows, as they often more effectively address the diverse challenges encountered in excited-state dynamics. A common protocol is performing a sequence of SA-CASSCF and (X)MS-CASPT2 calculations. SA-CASSCF alone does not always provide high accuracy, but it can highlight potential issues with a single-reference description of excited states -- e.g., when excited states with antibonding character induce photochemical bond cleavage, or doubly excited states play a significant role in dynamics (note that double excitations are not an inherent problem of a single-reference description, but most single-reference methods happen to have issues with their description). Multireference methods are also well-suited for describing the topology and topography of CXs -- tasks that are not reliably handled by conventional single-reference methods, especially in the case of S$_{1}$/S$_{0}$ CXs.\cite{tuna2015assessment,taylor2023description}

(X)MS-CASPT2 offers a more quantitative description of excited states by incorporating dynamic-correlation effects in addition to the static correlation handled by SA-CASSCF. However, the accuracy of both SA-CASSCF and (X)MS-CASPT2 methods depends largely on the choice of the active space, which has to include all relevant orbitals for an accurate description of the electronic states of interest. Designing a suitable active space is often a labor-intensive task that also requires careful consideration of numerical stability -- we discuss various strategies to select an active space in Section~\ref{sec:activespace} and highlight different stability issues in Section~\ref{sec:practicalcons}, as well as the need to perform exploratory nonadiabatic trajectories to test the electronic structure. It is crucial to always report the active-space orbitals used in a multiconfigurational or multireference calculation -- at least as a figure of the orbitals in a manuscript, but even better as a file containing the orbitals themselves (for example, using a molden format). Mentioning only the number of electrons and orbitals (and their character) is not enough to ensure that results can be reproduced. 

When available, experimental data can serve as a reference for calculated FC excitations. However, one needs to be cautious when comparing vertical excitation energies with experimental absorption band maxima, as these values may not always align. Band maxima are often red-shifted relative to vertical excitation energies due to frequency changes between the ground and excited states.\cite{bai2020origin} The shift can vary for different electronic states: shifts between 0.1 and 0.2 eV are typically reported for conjugated organic molecules, but shifts as large as 0.4 eV can be observed.\cite{daday2012full} `Correcting' experimental maxima with calculated shifts indirectly allows to obtain `experimental vertical excitations'.\cite{crespo2014spectrum} Nevertheless, as vertical excitation energies are not considered experimental observables, their values for a given electronic-structure method are ideally benchmarked against more accurate approaches as described above. In contrast, 0-0 transitions are proper observables, allowing a direct theory-to-experiment comparison.\cite{jacquemin20150,winter2013benchmarks}

To test electronic-structure methods beyond the FC point, i.e., beyond the equilibrium geometry, it is instructive to calculate absorption (and emission) spectra, observables that can be directly compared to experimental measurements. An elegant method for simulating spectral shapes is the nuclear ensemble approach (NEA),\cite{crespo2014spectrum} a numerical realization of the reflection principle,\cite{schinkebook} which is intrinsically linked to sampling ICs for nonadiabatic dynamics. A detailed description of the NEA will follow after we introduce methods for sampling ICs for nonadiabatic dynamics -- we propose here only a brief mention in the context of calculating absorption spectra. The NEA involves sampling an ensemble of molecular geometries representing the nuclear probability density of the initial electronic state (for absorption, typically the ground state), calculating electronic transitions for each geometry from the ensemble, and applying line broadening to obtain a photoabsorption cross-section directly comparable to experiments. While spectra calculated with NEA lack vibronic features, the method can predict the correct widths of bands. Alternatively, if a spectrum with fine vibronic structure is required, calculations of Franck-Condon Herzberg-Teller factors are also possible, though usually limited to rigid harmonic molecules with bound electronic states.\cite{santoro2016beyondvert} The FCClasses code, for example, offers a broad toolbox of strategies to calculate properties beyond vertical transitions.\cite{cerezo2023fcclasses} We stress, nevertheless, that comparing theoretical and experimental absorption spectra may not be a sufficient benchmark for electronic-structure methods -- dark states (i.e., electronic states with small oscillator strength) are likely to be obscured in absorption spectra but often play a crucial role in nonadiabatic dynamics. 

\subsubsection{Tests beyond the Franck-Condon region}\label{sec:beyondfstest}
It is entirely possible for an electronic-structure method to show excellent agreement with a reference method for excitation energies at the FC point, but suffer significant deviation when the molecular geometry is distorted from the optimal ground-state minimum -- departing from the FC geometry is indeed what will happen in practice during the nonadiabatic molecular dynamics. There are multiple reports in the literature of such an inhomogeneity in the quality of an electronic-structure method as a function of nuclear coordinates (see, for example, Refs.~\citenum{wiggins09,tuna2015assessment,Prljphotolysis2020}), and this issue is common for molecular systems with excited electronic states of antibonding character leading to bond breaking -- a process challenging for single-reference methods. Mapping PESs beyond the FC region with a given electronic-structure method and comparing its result with a reference approach helps anticipate potential relaxation pathways in nonadiabatic dynamics and provides a more rigorous benchmark of its performance. Such a benchmark beyond the FC region also permits assessing the numerical stability of the electronic-structure method for different molecular distortions or relevant reaction coordinates. For instance, PES scans can help with constructing a stable active space for multireference methods and fine-tuning relevant parameters (e.g., real and imaginary shifts in (X)MS-CASPT2 are examples of parameters used to avoid numerical issues due to the appearance of intruder states), as discussed in further detail in Section~\ref{sec:activespace}.
 
The ultimate goal of a benchmark beyond the FC region is to select an electronic-structure method that is sufficiently accurate, computationally efficient, and numerically stable to be used in the subsequent nonadiabatic molecular dynamics. Yet, a thorough mapping of PESs for molecules with many degrees of freedom can be tedious. One strategy to alleviate this burden is to start by locating key critical molecular geometries for each electronic state considered. Excited-state minima can be routinely located with most electronic structure codes, followed by frequency calculations to confirm whether a true minimum has been obtained. Optimizing CXs is more challenging, as they do not represent a single geometry but rather an intersection seam of geometries (see Section~\ref{ch2_cx}).\cite{pieri2021namdreactor} Identifying the most relevant CXs for the system under study sometimes also requires running a few exploratory nonadiabatic trajectories in advance (as we also suggest in Section~\ref{sec:practicalcons}). Typically, one optimizes a minimum-energy CX (MECX), although it has been suggested that calculating minimal distance CXs -- i.e., points on the intersection seam closest to the FC geometry or excited-state minimum -- may often be more meaningful.\cite{levine2008optimizing} Furthermore, many single-reference methods, such as LR-TDDFT and ADC(2), are known to predict an incorrect dimensionality of the S$_1$/S$_0$ intersections -- their intersections do not appear to have a correct conical shape when plotted against branching space coordinates.\cite{tuna2015assessment,taylor2023description} Intersections between excited states are typically less problematic, at least in terms of their topology. Despite these theoretical peculiarities, optimizing crossings between electronic states is often meaningful, even when single-reference methods are employed.\cite{taylor2023description} General-purpose algorithms, such as CIopt,\cite{levine2008optimizing} enable the optimization of CXs without derivative coupling vectors and can thus be coupled with any electronic-structure method (see Ref.~\citenum{keal_comparison_2007} for a detailed discussion on the various algorithms to locate MECXs). Multireference methods, in principle, describe CXs accurately as long as they use an adequate active space and often serve as a benchmark for other methods. A commonly employed strategy in computational photochemistry is to balance the computational cost by using a multireference methods based on (multistate) CASPT2 to perform single-point calculations on a SA-CASSCF-optimized geometries (an approach often denoted 'CASPT2//CASSCF').\cite{sanjuan2012reviewcasscfpt2,gonzalez2012progress} Yet, this strategy may often suffer from shortcomings as solely correcting the electronic energy with second-order perturbation theory can be insufficient to provide reliable photochemical paths, and dynamic correlation is often required also at optimization stage for the different critical geometries on the PESs, in particular near CXs (an issue coined 'differential correlation problem'\cite{gonzalez2012progress} in this context) and for an adequate description of states involving a ionic/covalent electronic character. We note that this statement then extend to nonadiabatic molecular dynamics (e.g., Ref.~\citenum{liu2016dyncorrel}).

\begin{figure*}[h!]
    \centering
    \includegraphics[width=0.8\linewidth]{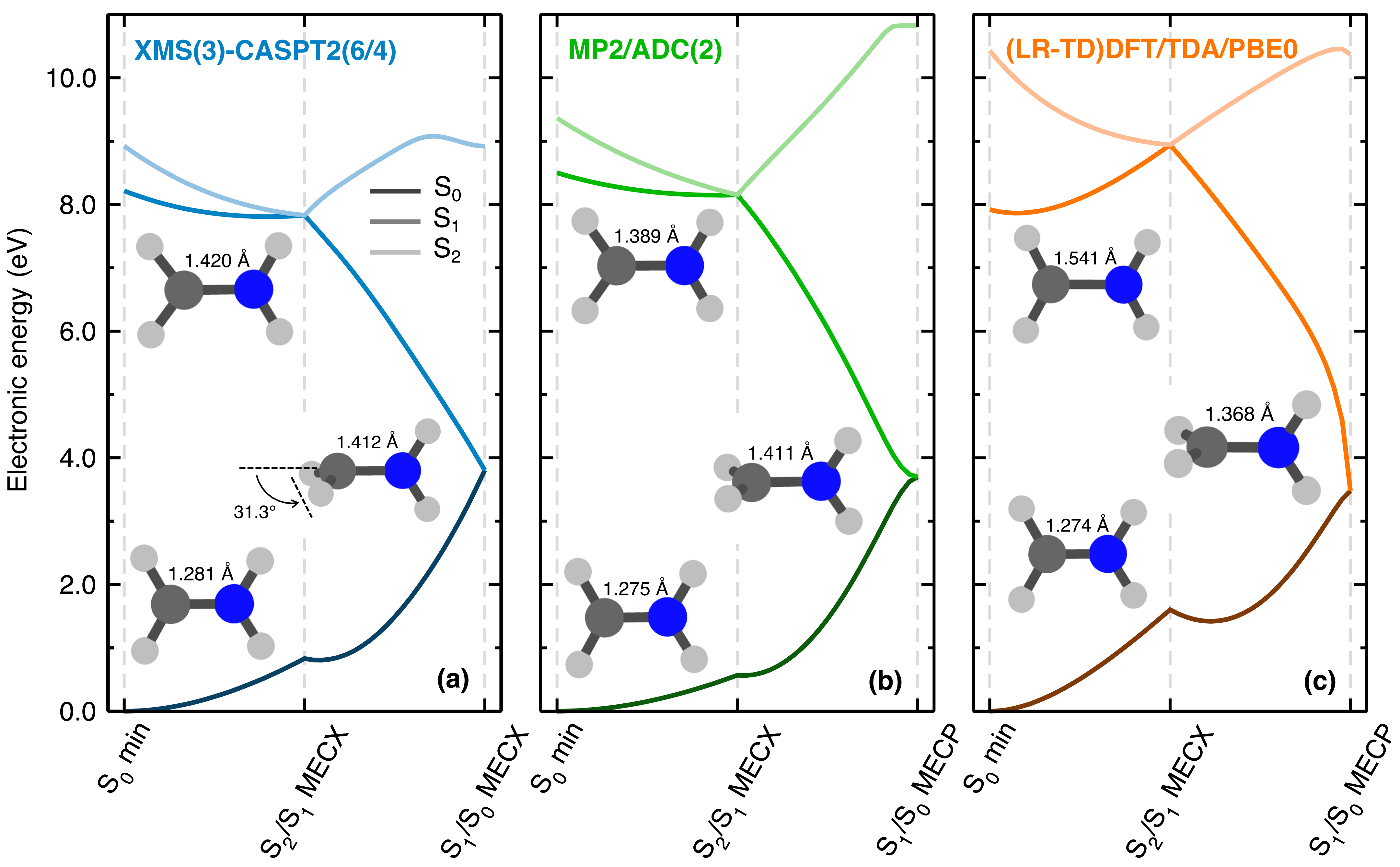}
    \caption{LIIC pathways connecting the critical geometries optimized at three different levels of theory for ground and two lowest singlet excited states of protonated formaldimine. Reproduced from Ref.~\citenum{taylor2023description}.}
    \label{fig:liic_jack}
\end{figure*}

Identifying critical geometries for the different electronic states of interest is not the final goal of the benchmark approach. Once these geometries (e.g., excited-state minima, MECXs) have been identified, one can construct simple interpolation pathways representing a reaction coordinate. Linear interpolation in internal coordinates (LIIC) pathways are a popular and cost-effective alternative to computationally demanding calculations of minimum-energy paths (which, in addition, become difficult to implement when multiple electronic states are important in the process of interest).\cite{hudock_ab_2007} LIICs allow one to obtain a map of the PESs by interpolating geometries through incremental displacements along internal coordinates (a common Z-matrix is used for all geometries). Fig.~\ref{fig:liic_jack} shows different electronic-structure methods along their LIIC paths. For large and complex systems, constructing Z-matrices becomes more intricate -- LIICs can vary significantly due to the non-unique choice of internal coordinates. In such cases, it is advisable to use geodesic interpolation.\cite{zhu2019geodesic} Once the LIIC of interest is obtained, it becomes simple to test and compare the electronic energies obtained with various electronic-structure methods on the support of the same LIIC pathway (i.e., the same set of nuclear geometries). In case of large discrepancies between methods, the critical geometries used to produce the LIIC can be reoptimized with different electronic-structure methods, and the resulting LIICs can be compared (as done for example in Fig.~\ref{fig:liic_jack}). 

In addition to interpolations between critical geometries, it is also convenient to scan surfaces along relevant nuclear degrees of freedom for the excited-state dynamics. When an excited electronic state of interest exhibits an antibonding character (the corresponding PES typically lacks a stable minimum), one can perform a rigid scan along a bond-breaking coordinate. If significant geometry relaxation is expected along the dissociation path, a relaxed scan would be more appropriate -- all degrees of freedom are being optimized except the dissociation coordinate. Relaxed scans can reveal the influence of specific degrees of freedom on the photochemical process, whether it be bond distance, angle, or torsion. Such analyses are convenient not only for benchmarking but also for interpreting the results of nonadiabatic dynamics and providing a mechanistic understanding of the photoreactivity of interest.

\subsubsection{Choosing an active space for a multiconfigurational/multireference method to be used for nonadiabatic dynamics}
\label{sec:activespace}
In Section~\ref{sec:wft}, we introduced single-reference electronic structure methods, which can generally be applied in a somewhat more 'grey-box' fashion, contrasting them with the class of multiconfigurational/multireference methods that require the user to define an active space.
Experience shows that the 'chemical intuition' used to construct an active space for the ground electronic state does not apply anymore to the determination of an active space for multiple electronic states, in particular if one is interested in exploring a photochemical process. In addition, the strategy to determine an active space for excited electronic states may vary depending on whether one uses a multiconfigurational (e.g., SA-CASSCF) or a multireference (e.g., XMS-CASPT2) method. In the following, we discuss a potential protocol that may help to determine an active space for excited electronic states, but the reader should keep in mind that it is purely grounded in (mostly painful) experiences and its applicability may vary depending on the molecule of interest. As such, the tone of the following paragraphs will be somewhat different and should be understood as a narrative of a 'path towards an active space'.

For the ground-state optimized geometry of the molecule of interest, perform a series of electronic-structure calculations with various sets of electronic-structure methods and a medium-size basis set: LR-TDDFT/TDA, ADC(2), CC2, EOM-CCSD, and, if affordable, CC3. Compare the various electronic energies and oscillator strengths, and check the influence of the basis set on these results, possibly detecting the importance of Rydberg states (keeping in mind that LR-TDDFT may not describe them adequately). More importantly, check the electronic character of the range of excited electronic states of interest and compare the different electronic-structure methods to see where they agree and where they disagree (for example, for the ordering of electronic states). Based on the electronic character of the states (and the excitation energies and oscillator strengths), and using the literature or any available experimental data on the system, identify the similarities between methods. In the FC region and for closed-shell molecules, it is likely that most methods will show some form of agreement, with a possible exception being the ordering of some electronic states if close in energy to each other. Notorious issues will appear with electronic states showing a doubly-excited character, but we will come back to this point later. At this stage, the goal is not to find the best possible single-reference electronic-structure method, but merely to identify the various electronic-state characters, i.e., the important occupied and virtual orbitals, for the electronic states of interest. We are now at a turning point, depending on whether you want to identify an active space for SA-CASSCF  or XMS-CASPT2. 

Let us start with an active space for XMS-CASPT2. Bring into the active space the smallest possible number of occupied/virtual orbitals that were identified earlier (use natural orbitals from a CIS calculation as an initial guess for the orbitals, or even Kohn-Sham orbitals, but avoid HF orbitals -- virtual HF orbitals are often more delocalized), using the default value for any shift parameters. Set the state-averaging/multistate value to the number of electronic states of interest in the first place (these values are likely to change later, in particular for photochemical processes). Perform a first single-point XMS-CASPT2 calculation and check your orbitals (CYO) that are in your active space at the end of the calculation -- detect in particular any orbital rotations and the nature of the newly entered/departed orbitals from your active space. Compare the XMS-CASPT2 electronic energies, oscillator strengths, and electronic character (be careful to look at the XMS-CASPT2 result and not accidentally use the electronic character for the underlying SA-CASSCF electronic states), and determine whether they match your best theoretical estimate from the single-reference methods. Check for the presence of electronic states with a doubly-excited character. It is very likely that you will then need to include (or exclude) further orbitals to adequately describe the electronic states of interest. To do so, always restart from your last XMS-CASPT2 calculation and develop the active space from there. Sometimes, voluntarily setting a larger active space or increasing the number of electronic states in the state-averaging/multistate procedure further than needed may help catch 'shy' orbitals in your active space. After each XMS-CASPT2 test, CYO. Evolving the active space and using the single-reference calculations as an anchor, together with a lot of patience and perseverance, should eventually lead to an active space that will describe all the electronic states of interest in the FC region. This process is tedious but also allows you to learn about the electronic states of your molecules (and their interplay). This is only the first step of the process, as it is very likely that the found active space is not the final one -- moving away from the FC region will unravel other possible orbitals that may be needed to fully describe the photochemical/photophysical process of interest.  

The next step will be to validate the active space away from the FC region. To achieve this, one can locate critical points -- minima, MECXs -- along a path that would mimic the potential nonradiative decay. It is key to note at this point that we are here exploring the nuclear configuration space of our molecule to find a stable active space -- the pathway will need to be recalculated eventually with the final stable active space for interpretation purposes. Starting from the excited electronic state that one would expect to be dominantly populated by the photoexcitation process, one can first conduct a geometry optimization in this electronic state with XMS-CASPT2 and the active space defined earlier. While the active space might be stable during this process, its composition will likely vary. In any case, at the end of the geometry optimization, and independently on whether the calculation crashed or succeeded, CYO! Try to identify whether the active space was preserved or if orbital rotations occurred. If new orbitals appear in your active space, try to include them as a starting point in the FC region -- perform a new XMS-CASPT2 single-point calculation, CYO, and check your results against the single-reference methods. If the orbitals do not stay in your active space in the FC region, you can include them at the step of your geometry optimization. You may want to question the number of electronic states considered at this point, too.
Before restarting the geometry optimization, make sure that the active space was the only reason for the problem: if the crash occurred solely because of an electronic-structure problem, the above-mentioned protocol may have helped and the calculation can be restarted; if the geometry optimization process itself failed, you may want to check the energy difference between your driving electronic state and the surrounding electronic states and determine whether you should instead locate a MECX for this state. This overall process may need to be repeated several times until you find a critical geometry. At this point, CYO and compare the orbitals of the active space with those of the FC region. 
The critical point here is to determine whether you have the same active space at the critical geometry as in the FC region. A powerful trick here to confirm this match is to use LIICs or, more conveniently, geodesic interpolation between the two critical points (as discussed in Section~\ref{elecstructproblem}). Determine this path between the FC geometry and your located critical point, and calculate the XMS-CASPT2 electronic energies for all electronic states considered along this path, starting from your FC point. If your active space was (mostly) stable, you should see no discontinuities along your interpolation pathways, focusing in particular on the electronic state of interest. If you see a discontinuity, CYO in your active space before and after the discontinuity occurred in your path. You can also calculate the electronic energy along your interpolation pathway backward and check for a hysteresis in electronic energy, indicating the divergence of the active space. Check also that there might not be a new electronic state appearing along your interpolation pathway -- this appearance is often identified by a cusp (and not a discontinuity \textit{per se}) in the energy of one or more electronic states along the path, and may require an extension (or reduction) of the state-averaging/multistate value for your XMS-CASPT2 calculation. In some extreme cases, dynamical weighting can also be considered (see Ref.~\citenum{glover2014dw} for an example with DWS-CASSCF). Hence, the interpolation pathway offers a stringent test for the stability of the active space and a powerful tool to identify important orbitals (or unimportant orbitals) as well as more/less electronic states to be considered in the XMS-CASPT2 calculation. Interpolations paths are the best 'mimics' of a nonadiabatic dynamics trajectory, and any discontinuity in your interpolated path means that a nonadiabatic dynamics simulation with this active space will lead to instabilities and non-conservation of your total energy.\footnote{Instabilities in an active space used for the electronic-structure ingredients of a nonadiabatic dynamics simulation would be as if you were to use LR-TDDFT and swap xc-functionals randomly along the dynamics -- a trajectory evolving based on different 'electronic-structure realities'.} Once a stable active space is obtained for the interpolation pathway, you can proceed with the identification of further critical points, repeating the full process of the interpolation pathway and the study of the active space along the extended path. As always, CYO! You can repeat this until you reach the last representative critical geometry for the process of interest. Once you find a stable active space for the entire path, you can reoptimize each critical point -- and at this stage, also the FC geometry -- to this final XMS-CASPT2 level of theory. This final overall interpolated pathway can also be used at this stage to check the impact of the basis set in different regions of the nuclear configuration space, as well as to obtain some information about the timing of the XMS-CASPT2 calculation for an electronic energy calculation -- a key information to evaluate the feasibility of the subsequent nonadiabatic molecular dynamics. 

The protocol would be rather similar if you want to identify an active space for SA-CASSCF, but keep in mind that this method is not expected to give absolute values for excitation energies in agreement with single-reference methods (with a few exceptions for specific electronic characters) due to its lack of dynamic correlation. Hence, the objective of the process here is to make sure that the active space for the SA-CASSCF calculations reproduces the proper ordering of the electronic states (and their energy separation) in the FC region, and that the overall interpolated pathway eventually obtained reproduces the topography of the pathway obtained with XMS-CASPT2 (if this level of theory is achievable just for the interpolated pathway). We note, however, that higher excitation energies in the FC region would result in trajectories with a higher internal energy when they reach the ground electronic state via nonadiabatic transitions, and great care would then be required when analyzing and interpreting the ensuing ground-state dynamics. Possibly more important is to realize that, with SA-CASSCF, 'bigger is not always better' as stated in the supporting information of Ref.~\citenum{levine2009}. Indeed, active spaces for SA-CASSCF can often be unintuitive in comparison to the active spaces used with XMS-CASPT2. Numerous examples from the literature highlight this fact: e.g., butadiene\cite{levine2009} with a (4/3) active space instead of a (4/4), cyclohexadiene\cite{tao2011first} with a (6/4) active space instead of the expected (6/6), benzene\cite{thompson2011} with a (6/5) active space instead of a (6/6), or provitamin D\cite{snyder2016gpu} with a (6/4) active space instead of a (6/6).

It is finally important to stress that, once the orbitals composing an adequate active space are identified and the stability/quality of the active space is validated, the precious orbitals selected should always be used as an initial guess for subsequent electronic-structure calculations (photoabsorption cross-sections or nonadiabatic molecular dynamics trajectories).

\subsection{Describing the photoexcitation process}
\label{sec:initconds}

In the following, we discuss the critical steps required to initiate a (trajectory-based) nonadiabatic molecular dynamics simulation. As discussed below, this process is often split into two subsequent steps: the definition of a ground-state nuclear distribution for the molecular system of interest, sampled to obtain multiple pairs of nuclear positions + momenta (often called 'initial conditions' - ICs). The precise nature of the light source employed to excite the molecule then informs how one should select these ICs in subsequent excited-state dynamics.

\subsubsection{Sampling the ground-state nuclear density to obtain initial conditions}
\label{sec:gssampling}

In trajectory-based methods (introduced in Section~\ref{nonadiabaticdynproblem}), ICs refer to a set of molecular coordinates and momenta that define the starting point of individual excited-state trajectories. ICs should accurately represent the initial state of the system, as they directly influence its time evolution and the resulting calculated observables (e.g., excited-state lifetimes, quantum yields, etc). Strategies to acquire ICs for excited-state dynamics will be outlined in the following paragraphs. For a more detailed discussion, the reader is referred to Ref.~\citenum{janos2025selecting}. 

A common approach to sample the ground-state nuclear density is to perform long Born-Oppenheimer ground-state dynamics and collect (uncorrelated) snapshots of molecular geometries and momenta. This is known as Boltzmann (thermal) sampling. However, Boltzmann sampling does not capture the quantum delocalization of nuclei or zero-point vibrational effects present in the ground state.\cite{barbatti2016effects} Neglecting these effects can pose a challenge when investigating the photodynamics of many molecules. Some nuclear quantum effects can be incorporated by using the Wigner distribution,\cite{tannor_book,barbatti2016effects} a more rigorous way of mapping quantum nuclear densities onto classical phase-space quantities -- positions and momenta. Sampling ICs from a Wigner distribution is commonly used in conjunction with nonadiabatic dynamics methods, such as TSH and multiple spawning.\footnote{Note that multiple spawning appears to be less sensitive to the choice of ICs compared to TSH.\cite{ben2007continuous}} In practice, the Wigner distribution is typically applied to harmonic and uncoupled normal modes, requiring only ground-state optimization and harmonic frequency calculations as input. However, this implementation of Wigner sampling has a number of limitations.\cite{persico2014overview,mccoy2014role,suchan2018importance,mai2018novel,prlj2023deciphering} While effective for rigid, harmonic molecules with well-defined structures represented by one or a few ground-state minima, a harmonic Wigner sampling becomes impractical for flexible systems featuring many interconverting local minima or significant anharmonic behavior. Moreover, rectilinear normal modes inadequately capture torsional degrees of freedom (low-frequency motions), potentially introducing artifacts when ICs are generated via a harmonic Wigner sampling.\cite{mccoy2014role,mai2018novel,persico2014overview,prlj2023deciphering} 

Quantum thermostat (QT) \cite{ceriotti2009nuclear,ceriotti2010colored,Finocchi2022} (or quantum thermal bath) is an alternative sampling technique based on Born-Oppenheimer molecular dynamics. Unlike Boltzmann sampling, QT incorporates basic nuclear quantum effects arising from zero-point delocalization. QT is based on a generalized Langevin equation (GLE) thermostat, which maintains the normal modes of a molecule at frequency-dependent temperatures. As quantum vibrational energies are intrinsically linked to mode frequencies (recall the energy expression for the quantum harmonic oscillator), QT can accurately reproduce zero-point vibrational energies by effectively thermalizing each mode accordingly. Temperature effects can also be incorporated into QT dynamics. QT is applicable to both harmonic and moderately anharmonic systems\cite{suchan2018importance,Vuilleumier2013}, making it a viable alternative to Wigner sampling when the latter’s applicability is limited. For a detailed discussion of QT (and GLE thermostat), the reader is referred to Ref.~\citenum{ceriottigle}. Notably, QT often yields a distribution of geometries consistent with more rigorous path integral dynamics simulations, but with significantly reduced computational cost.\cite{prlj2023deciphering} However, standard path integral dynamics is not suitable for sampling nuclear phase space, as this approach does not provide both nuclear coordinates and momenta simultaneously, due to the inherent nature of quantum uncertainty.

How to approach the sampling of ICs for a real molecular system?  Wigner sampling with uncoupled harmonic normal modes is often a reasonable first choice (assuming the system is harmonic), due to its straightforward implementation and availability in nonadiabatic dynamics codes. Temperature effects can also be easily incorporated in the Wigner distribution, although sampling at 0K is also common (corresponding to the ground vibrational level of the ground electronic state). In cases where multiple local minima have significant ground-state populations, different sets of ICs can be calculated, with their contributions weighted by the appropriate Boltzmann factors.\cite{marsili2022theoretical} Caution is advised when dealing with systems exhibiting anharmonicity or modes with low harmonic frequencies (typically below 500 cm$^{-1}$). Such modes can be removed from the sampling if problematic,\cite{persico2014overview,favero2013dynamics} but one should pay great care that the modes are not photoactive.\cite{prlj2023deciphering} Torsional or out-of-plane modes, in particular, can be difficult to represent using linear normal modes.\cite{mccoy2014role,C8FD90051E} Another central point to consider when using the harmonic Wigner sampling is the importance of locating all possible conformers (minima) in the ground electronic state that could be Boltzmann populated at the temperature of interest. A bilirubin subunit offered a striking example of the importance of conformers in photochemical processes. A photocyclization of bilirubin was observed experimentally with a very small quantum yield. Yet, none of the identified main conformers could undergo such a cyclization, except for a higher-energy conformer with a Boltzmann weight of only 0.5\%, which has a 45\% quantum yield for photocyclization.\cite{janos2020bilirubin} This example shows that a low photochemical yield may be rooted not only in the mechanisms for nonradiative decays, but also in the photoactivity of a specific conformer. A careful search for potential conformers extends to sampling strategies based on dynamics, too. 

QT sampling offers a generally more reliable alternative to select ICs. A recent case study\cite{prlj2023deciphering}  demonstrated that QT yields results similar to Wigner sampling for stiff harmonic molecules, but outperforms Wigner for molecules featuring low-frequency torsion and weakly interacting hydrogen-bonded systems (involving both low-frequency and large-amplitude motions). QT may be advantageous for complex systems such as solute-solvent clusters, for which preparing ICs with a Wigner distribution requires a more complicated protocol \cite{ruckenbauer2010azomethane} (for a more recent discussion on ICs for solvated systems, see Ref.~\citenum{curtis2024preparing,avagliano2025aimsxtb,sheik2025investigating}). As long as issues with zero-point energy leakage do not significantly affect the dynamics, QT provides broad applicability for ICs sampling in molecular systems. Energy leakage from high-frequency to low-frequency modes is a known issue in classical trajectories, and several extensions to QT dynamics have been proposed to mitigate this problem. System-bath coupling typically helps reduce the leakage issue.\cite{Brieuc2016, Mangaud2019} (We note that the zero-point energy leakage described here is different from that observed in the context of nonadibabatic molecular dynamics, see Section~\ref{sec:faq} for a brief discussion.) On a practical note, QT dynamics is available in free software packages such as ABIN\cite{abin} and i-PI\cite{kapil2019pi}, while the GLE4MD website\cite{GLE4MDwebsite} provides additional details on input generation.

\subsubsection{Calculating a photoabsorption cross-section}
\label{sec:photoabscrosssec}
The nuclear ensemble approach (NEA)\cite{crespo2014spectrum,crespo2018recent} is a convenient method for calculating photoabsorption cross-sections using the same set of sampled ICs (in this case, only the nuclear geometries) that will be used to initiate the nonadiabatic dynamics. In the NEA, electronic transitions are calculated on the support of hundreds of IC geometries, which represent the ground-state nuclear density distribution of our molecule. Subsequently, individual transitions are broadened using appropriate shape functions (typically Lorentzian or Gaussian) to yield a convoluted photoabsorption cross-section, $\sigma (\lambda)$.\cite{crespo2014spectrum,crespo2018recent} The cross-section can be calculated using the equation:
\begin{equation}
\sigma(E)=\frac{\pi e^2 \hbar}{2m_e c \epsilon_0 c E} \sum_{J=1}^{N_s} \frac{1}{N_p} \sum_n^{N_p}\Delta E_{0J}(\mathbf{R}_n)f_{0J}(\mathbf{R}_n)w_s[E-\Delta E_{0J}(\mathbf{R}_n),\delta] \, .
\end{equation}
Here, $E$ represents the photon energy, while $e$, $m_e$, $\epsilon_0$, and $c$ denote fundamental constants for the electron charge, electron mass, vacuum permittivity, and the speed of light, respectively. $\Delta E_{0J}(\mathbf{R}_n)$ is the vertical transition energy between the ground state $0$ and the excited electronic state $J$, for the $n^\text{th}$ sampled molecular geometry with the corresponding nuclear geometry $\mathbf{R}_n$. $f_{0J}(\mathbf{R}_n)$ represents the oscillator strength at the geometry $\mathbf{R}_n$. $w_s[E-\Delta E_{0J}(\mathbf{R}_n),\delta]$ denotes a normalized lineshape centered at energy $\Delta E_{0J}(\mathbf{R}_n)$ with a phenomenological width $\delta$. Individual line shape widths are typically set to be much narrower than the overall absorption band widths.

\begin{figure*}[h!]
    \centering
    \includegraphics[width=0.65\linewidth]{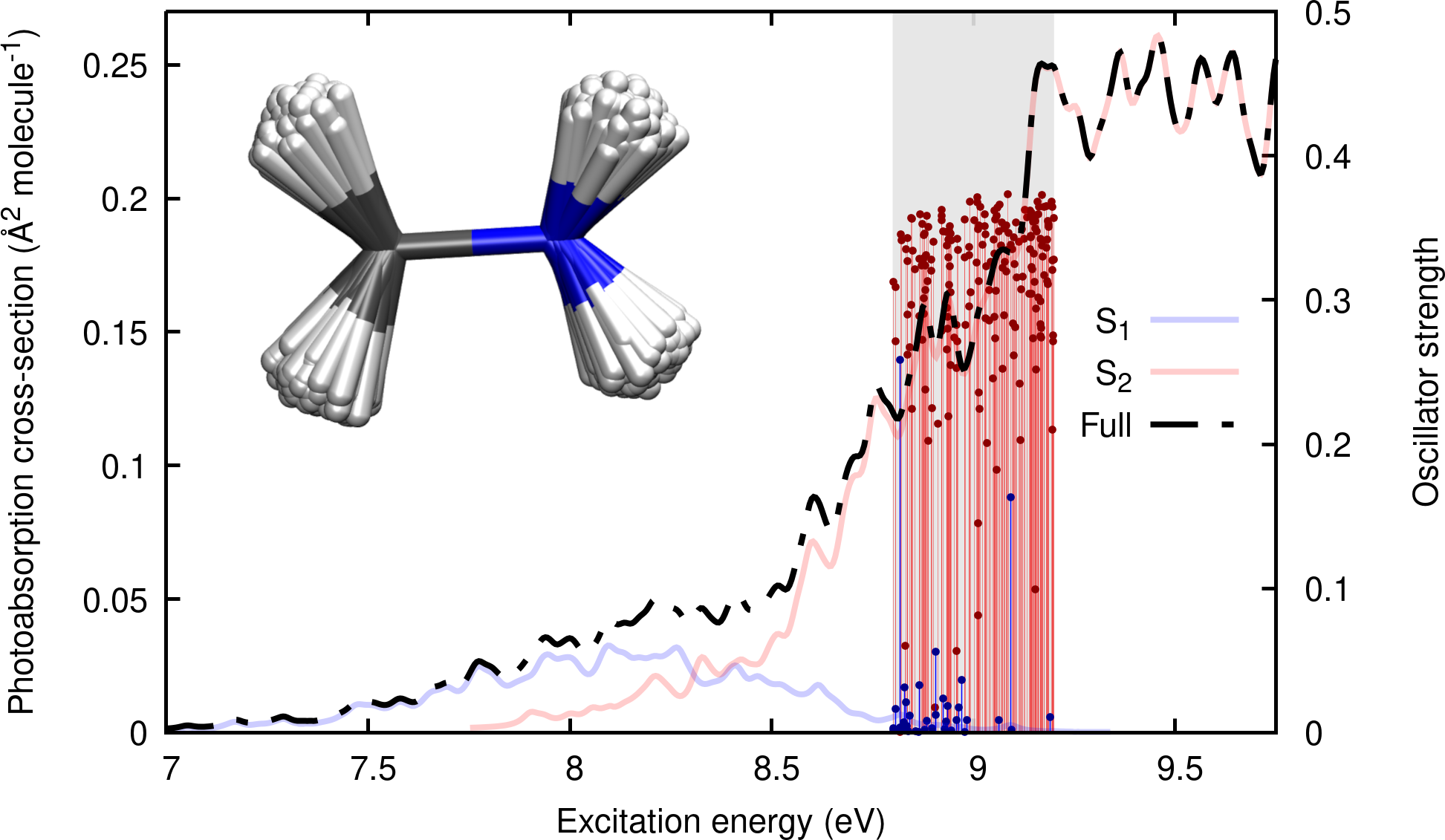}
    \caption{Photoabsorption cross-section of protonated formaldimine decomposed into S$_1$ and S$_2$ contributions. The inset shows 1000 structures sampled from a Wigner distribution, while the narrow energy window includes transitions to both excited states. ADC(2)/def2-SVP level of theory.}
    \label{fig:ic_sampling}
\end{figure*}

Absorption spectra estimated using the NEA account for non-Condon effects by considering the dependence of transition dipole moments on molecular geometries. In addition, the NEA can predict the overall widths, heights, and positions of molecular absorption bands. However, the NEA does not describe the vibronic structure, as it lacks information about the nuclear wavefunctions of the excited states.\cite{crespo2014spectrum,crespo2018recent}

Although more rigorous methods exist for calculating photoabsorption cross-sections,\cite{prlj2021calculating,santoro2016beyondvert} the NEA offers a convenient alternative, enabling the estimation of absorption spectra without incurring additional computational cost beyond that required for sampling the ICs (see previous section). The calculated NEA spectra can be directly compared to experimental spectra, providing another useful means to evaluate the suitability of the chosen electronic structure method (at least for bright transitions). The NEA is also particularly interesting for the simulation of photoabsorption cross-sections for molecules exhibiting dark states, given its ability to recover non-Condon effects and produce absolute values for the cross-sections, often in close agreement with experiment.\cite{prlj2021calculating} An illustrative example of a photoabsorption cross-section calculated using the NEA is shown in Fig.~\ref{fig:ic_sampling}.

\subsubsection{Selecting the adequate initial conditions to mimic a photoexcitation}
\label{sec:initcond}

The selection of appropriate ICs depends not only on the nuclear density of the initial state (e.g., the ground state) but also on the nature of the light source. In this context, we will primarily focus on scenarios where the molecular system is excited by an external ultrashort laser pulse (from around one to a few hundreds femtoseconds). However, other setups are also possible. Barbatti\cite{barbatti2020simulation} proposed a protocol for simulating nonadiabatic molecular dynamics following photoexcitation by continuous thermal light (e.g., solar radiation). Continuous radiation can be conceptualized as an ensemble of coherent short pulses, allowing for the propagation of conventional nonadiabatic dynamics with varying initial times for individual trajectories, corresponding to distinct field realizations. Suchan et al.\cite{suchan2018importance} discussed sampling for dynamics initiated by a continuous-wave laser field using the concept of importance sampling. Curchod and coworkers\cite{janos2025selecting} provided a detailed account about the selection of ICs for nonadiabatic molecular dynamics, from the sampling of the ground-state distribution to the description of the photoexcitation process in itself. 

For a system excited by an ultrashort laser pulse, ICs are typically selected under the assumption of a sudden vertical electronic excitation. Sudden excitation maps the ground-state distribution directly onto a given excited-state PES to initiate the dynamics -- such a sudden excitation is, in principle, justified by first-order time-dependent perturbation theory for an infinitely short laser pulse.\cite{tannor_book} However, an infinitely short laser pulse has an infinitely wide energy spectrum and should excite the molecule to all electronic excited states. In contrast to that, nonadiabatic trajectories are frequently initiated from only a single excited state (e.g., the $S_x$ adiabatic state) regardless of the pulse energy spectrum. Although this may be the easy way to generate an ensemble of trajectories (with no selection criteria apart from the index of the initially excited state), it is not justified when the molecular system interacts with a (not infinitely) short laser pulse. In such a case, a narrow excitation energy window may be defined, and ICs selected based on excitation probabilities within this window. Moreover, the initial population may, in principle, be distributed among multiple states that fit within the excitation energy window.

By applying excitation energy windowing, results from nonadiabatic dynamics can be more directly compared to experiments. A simple selection algorithm that biases the ICs towards those with a high absorption probability (as reflected by transition oscillator strengths) has been proposed.\cite{barbatti2010non, barbatti2022newton} For every transition that fits within the excitation energy window $\varepsilon \pm \delta \varepsilon$, a selection probability can be calculated by dividing its transition oscillator strength by the maximum value among all transitions in the window ($p_i = f_i / f_{max}$). Note that one IC can have multiple transitions within the window. The IC (along with the index of the initial transition) is selected if its normalized probability $p_i$ is greater than a random number from a uniform distribution in the [0,1] interval, thus favouring ICs associated with brighter transitions.

Recent work has discussed the benefits of biasing the selection of ICs.\cite{prlj2023deciphering} Uniform selection, where ICs are randomly chosen without applying any filters, was compared to $f$-biased selection based on the aforementioned algorithm. A case study using TSH dynamics revealed that biasing the selection can significantly affect observables, particularly when relevant electronic states have very different oscillator strengths (e.g., bright and dark states). The $f$-biased selection was found superior when compared to available experimental measurements. However, oscillator strengths used for $f$-biased selection should be sufficiently accurate and balanced across the excitation spectrum, highlighting the importance of electronic structure benchmarking.

More recently, a strategy coined the promoted density approach (PDA) was devised to incorporate the effect of a laser pulse at the level of the ICs.\cite{janos2024including} The PDA will select given nuclear momenta-positions pairs based on their coupling with the provided parameters of the laser pulse (frequency, pulse shape, full width at half-maximum -- parameters provided to mimic the experiment). Hence, a given IC within the PDA will incorporate an initial time (different from $t_0=0$ and dependent on the characteristics of the laser pulse) and the excited electronic state reached by this IC (again, determined by the characteristics of the laser pulse). When analyzing the swarm of (TSH, AIMS, AIMC, MCEv2) trajectories initiated from the PDA, any observables will naturally encode the time broadening (and energy constraints) set by the initial laser pulse that serves for the photoexcitation.   
The PDA is rigorously obtained from perturbation theory and can also be used to derive, as a practical approximation, the use of a windowing approach as discussed above, but with a theoretical justification. The windowing strategy that can be derived from PDA, coined PDAW, assigns weights to the selected trajectories based on the transition dipole moment of its IC. PDAW is compatible with arbitrary pulse envelopes (giving the same result as PDA for an unchirped Gaussian pulse), and outperforms other windowing strategies described above. A simple code, \texttt{promdens},\cite{promdens} is available to generate the ICs for a given laser pulse automatically, using as input solely quantities that are usually determined to calculate a photoabsorption cross-section (see Section~\ref{sec:photoabscrosssec}), namely the excitation energies and transition dipole moments for each nuclear momenta/positions pair sampled from a given ground-state distribution.

\begin{figure}[h!]
    \centering
    \includegraphics[width=1.0\linewidth]{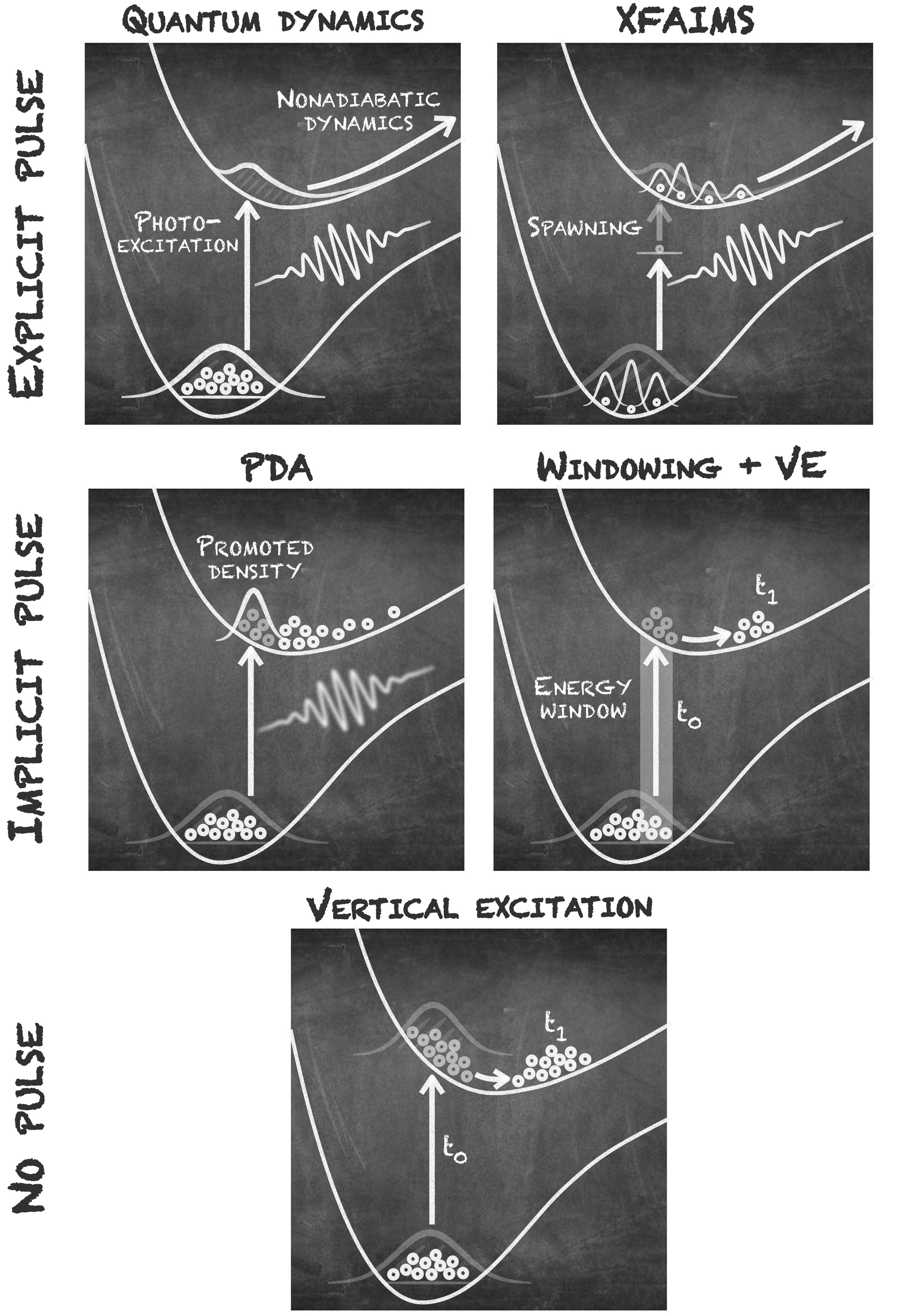}
    \caption{Schemes of the different approaches to describe photoexcitation (here with a laser pulse) in nonadiabatic molecular dynamics. Upper panel: approaches describing explicitly the interaction between a laser pulse and a molecule. Middle panel: approaches encoding the interaction between a laser pulse and a molecule implicitly (VE stands for vertical excitation). Lower panel: approach where the precise details of the photoexcitation are omitted (a pulse with a broad bandwidth is assumed, projecting perfectly the initial nuclear wavefunction in the ground state to the selected excited electronic state). See text for additional details.}
    \label{fig:pulses}
\end{figure}

Photoexcitation can also be described explicitly by including in the simulation the coupling between the molecule and the time-dependent electric field of a laser pulse, for example. While this statement sounds trivial in theory, it often leads to multiple challenges in practice, particularly for trajectory-based nonadiabatic molecular dynamics methods. The central idea is to describe the interaction of the laser pulse and the molecule by using a classical light-matter interaction term, coupling the time-dependent electric field of the laser pulse with the dipole moment operator for the molecule. The trajectory-based dynamics can then start in the ground electronic state from the ICs sampled from an approximate nuclear probability density (using any methods described in Section~\ref{sec:gssampling}), and the laser pulse promotes these ICs in one (or more) excited electronic state(s) based on its central frequency and bandwidth. 
The inclusion of a light-matter interaction term is rather straightforward in methods based on TBFs like FMS/AIMS and vMCG.\cite{mignolet2016communication,penfold2019excited} Nevertheless, the number of TBFs generated to describe the nuclear amplitude transfer to excited electronic states rapidly becomes intractable for long laser pulses. External field in AIMS (XFAIMS) nevertheless remains a method of choice to study the formation of (transient) electronic wavepackets upon photoexcitation by a ultrashort few-cycle (attosecond) laser pulse.\cite{mignolet2016communication,mignolet2018walk,mignolet2019steering,mignolet2019sub} Including an explicit laser pulse in mixed quantum/classical methods like TSH has been suggested,\cite{mitric2009c,tavernelli2010mixed,richter2011sharc} but this extension appears to push the limits of their implicit approximation (like the independent trajectory approximation in TSH) for any pulses but ultrashort ones.\cite{bajo2014interplay,mignolet2019excited} Hence, while the explicit description of photoexcitation is in principle feasible for trajectory-based nonadiabatic molecular dynamics, its use in practice is limited to specific cases and remains rather niche due to the limitations listed above. We note that molecular electronic states can also be directly mixed with quantum states of light, see e.g. Refs.~\citenum{luk2017multiscale,fregoni2020photochemistry,rana2024aimspolarito}. Figure~\ref{fig:pulses} offers a pictorial summary of the different approaches describes in this Section to incorporate photoexcitation in nonadiabatic dynamics.

\subsection{Performing the nonadiabatic molecular dynamics simulation}
\label{sec:performingNAMD}
In this Section, we discuss various aspects to consider when performing the nonadiabatic molecular dynamics simulations. We begin this Section with some general yet important considerations and elements to keep in mind when performing nonadiabatic molecular dynamics. We then present an example of a typical dynamics simulation using TSH (Section~\ref{sec:classtrajmethod}) and AIMS (Section~\ref{sec:gaussmethod}).

\subsubsection{Practical (and numerical) considerations} \label{sec:practicalcons}
In the following, we highlight some main considerations and concepts that are pivotal when performing nonadiabatic molecular dynamics. 

\paragraph{Running a few trajectories first}
As discussed above, nonadiabatic dynamics methods like TSH and AIMS use a swarm of trajectories (uncoupled in TSH and coupled in AIMS) to depict the nuclear wavepacket dynamics. Focusing on TSH, a certain number of ICs is required to adequately sample the ground-state distribution of the molecule of interest and, in principle, each IC should be used for multiple TSH dynamics (each initiated with a different seed for the random number generator) to adequately sample the stochastic nature of the nonadiabatic process in this method.\cite{Granucci2007,ibele2020} In practice, this process is often substituted by a larger number of ICs, each used to run a single TSH trajectory. AIMS only requires an adequate sampling of the ground-state distribution thanks to its spawning algorithm (except when a stochastic-selection version of the spawning is used). As a result, AIMS is expected to converge faster than TSH with respect to the total number of ICs\cite{Toniolo2005,Mukherjeelongtime2025} (but care is needed to ensure that the ground-state distribution is adequately sampled\cite{10.1063/5.0248950}). In any case, a recommendation for simulating the swarm of TSH or AIMS trajectories is to first start with a small subset of ICs ($\sim$ 10) and carefully analyze the resulting trajectories, focusing in particular on the behavior of the electronic-structure method and its stability (see below) along the trajectories. This initial swarm can also be used to deduce what the timescale of the propagation should be to reach the desired outcome (e.g., multiple fs or ps) and the associated computational cost. These trajectories can also be used to identify unexpected regions of the configuration space visited during the propagation, which may need additional attention to confirm that the electronic-structure method chosen describes them adequately. Some trajectories could also crash, and identification of the reasons for this instability is key (see also the FAQ in Section~\ref{sec:faq} for a discussion about discarding crashed trajectories). Once this first test-swarm is analyzed and all indicators are green, another set of trajectories can be propagated, but we would advocate to always propagate them by small bunch -- this strategy allows (i) for a more careful analysis and an earlier detection of any issues with the electronic structure chosen and (ii) for a better control of the convergence of the quantities of interest.

\paragraph{Conservation of total energy}
Conservation of total energy is a key consideration for nonadiabatic molecular dynamics given that these simulations are performed in the microcanonical ensemble (as required by the equations of motion introduced in Section~\ref{setting-the-scene}). Conservation of the total (classical) energy along each individual trajectory, for methods employing classical trajectories like TSH and AIMS, also constitutes a stringent test for the stability of the electronic-structure method used for their propagation -- in particular when discontinuities in the total energy are observed. While different rules of thumb have been used to judge the severity of a total energy discontinuity (e.g., energy discontinuity should be less than 10\% of nuclear kinetic energy), a detailed analysis of the cause (and potential consequences) of such discontinuities is mandatory.

Discontinuities in total energy along classical nonadiabatic trajectories can occur for different reasons from an electronic-structure perspective. Let us focus here on multiconfigurational and multireference methods, i.e., electronic-structure methods using an active space and a state-averaging process, as they often leads to the most problematic issues with total energy conservation. A first reason for the observation of total energy discontinuities is that, over the course of a nuclear time step, the electronic-structure calculation involves the rotation of orbitals in and out of the active space. In other words, one molecular orbital (or more) originally in the active space is replaced by another molecular orbital (or more). This exchange of orbitals means that the electronic structure calculated for this time step is no longer consistent with that of the previous steps, leading to an abrupt change of the electronic energy and, as a result, a mismatch in the sum of potential (electronic) energy and kinetic (nuclear) energy between time steps -- a discontinuity. This type of total energy non-conservation -- coined here 'orbital-rotation discontinuity' -- frequently happens when studying a photochemical process involving bond breaking/forming and requires a careful analysis of the reasons for this orbital rotation -- questioning whether these newly observed orbitals may be required in the active space or not. We discussed strategies to devise an active space in Section~\ref{sec:activespace}. 

\begin{figure}[h!]
    \centering
    \includegraphics[width=1.0\linewidth]{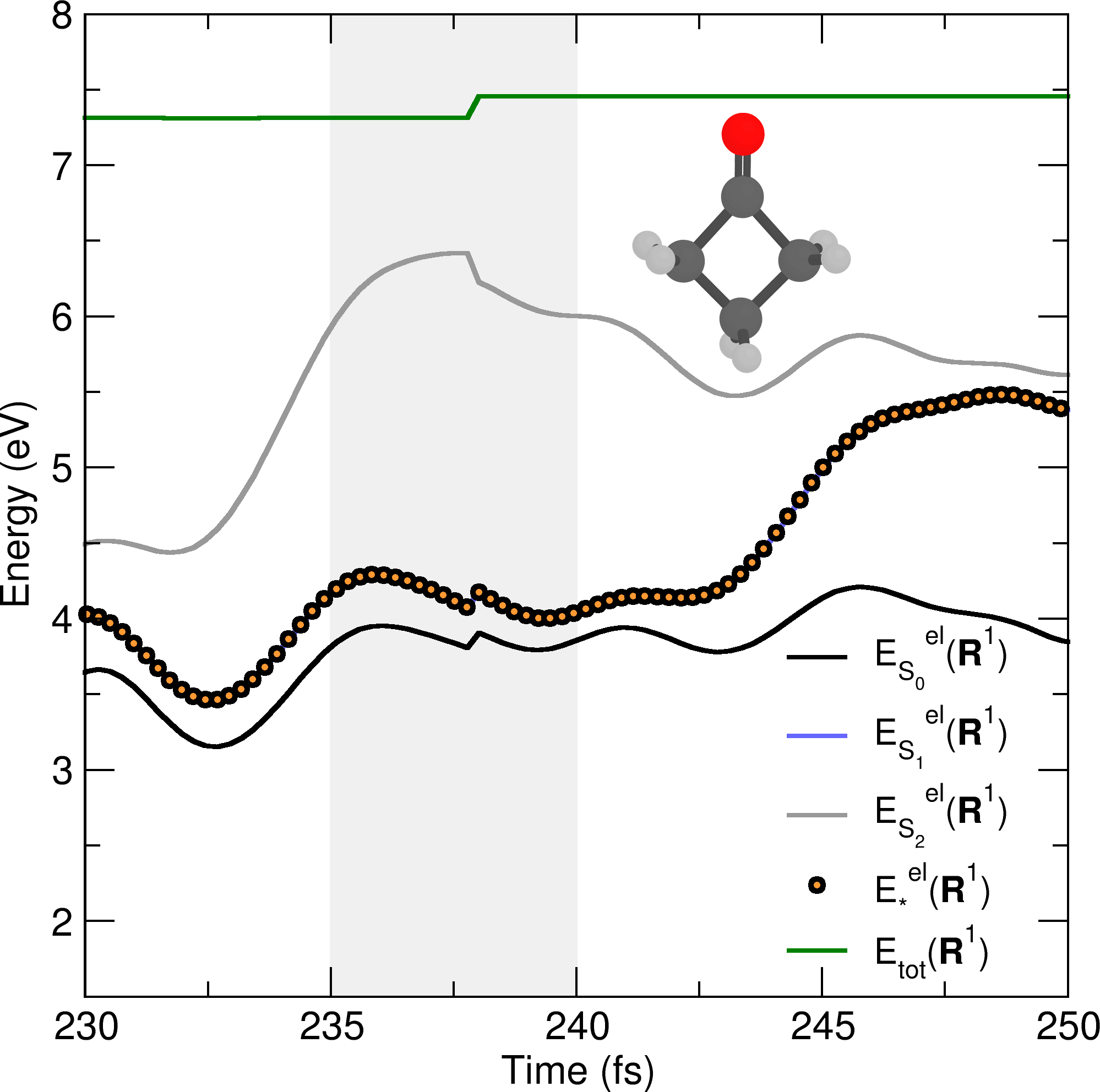}
    \caption{Exemplary TSH trajectory ($\alpha=1$) for cyclobutanone (inset) exhibiting an orbital-rotation discontinuity of total energy (grey area).}
    \label{fig:etot_orbrot}
\end{figure}
An example of such an orbital-rotation discontinuity in total energy is provided in Figure~\ref{fig:etot_orbrot} for an exemplary trajectory simulating the nonadiabatic molecular dynamics of cyclobutanone using TSH and XMS(3)-CASPT2(8/8). The TSH trajectory evolves in the first excited state (S$_1$) and a discontinuity in all electronic energies can be observed at around $t=237$ fs, caused by the rotation of an orbital out of the active space. As a result, this orbital rotation impacts the total classical energy of the TSH trajectory (discontinuity of $\sim 0.15$ eV). The impact of this specific orbital rotation on the electronic energies (and total energy) is mild, as the orbital that rotated out of the active space was not a strong contributor, at that time of the dynamics, to the different electronic characters of the underlying adiabatic electronic states.

Another cause for a discontinuity in the total energy along a classical nonadiabatic trajectory is due to the sudden importance of initially neglected excited electronic states for the nonadiabatic dynamics. A subset of electronic states needs to be selected for state-averaged (multi-state) methods (see Section~\ref{sec:activespace}), but in the course of a trajectory some other electronic states may come close in energy to the driving state. This scenario, that we could coin 'state-averaged discontinuity', is often observed when a bond is dissociated in the dynamics, leading to different electronic states becoming nearly degenerate in this region of configuration space. 

\begin{figure}[h!]
    \centering
    \includegraphics[width=1.0\linewidth]{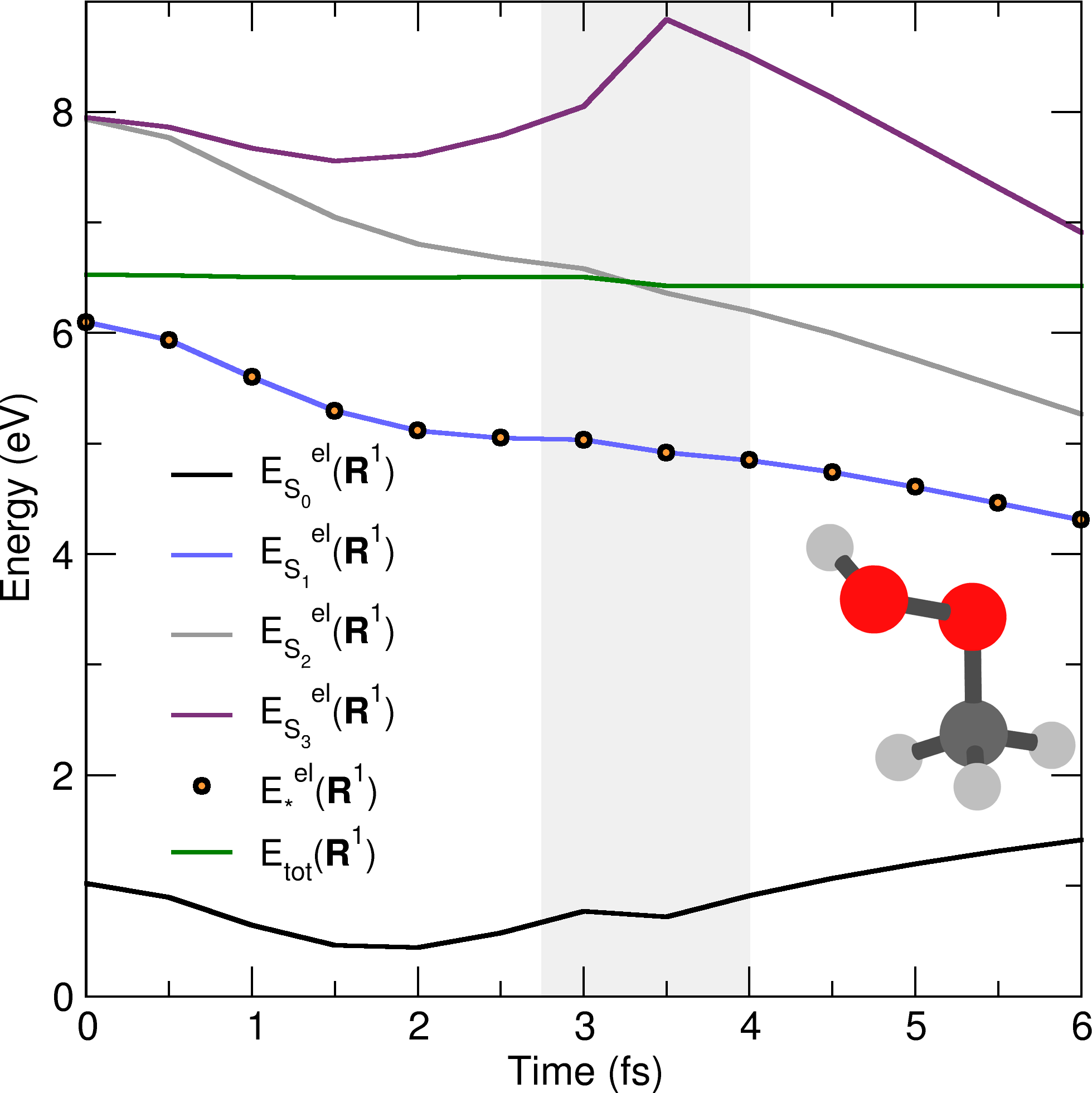}
    \caption{Exemplary TSH trajectory ($\alpha=1$) for methylhydroperoxide (inset) exhibiting a state-averaged discontinuity of total energy (grey area).}
    \label{fig:etot_extrastate}
\end{figure}
An example of such a state-averaged discontinuity is provided in Figure~\ref{fig:etot_extrastate}, where electronic energies for the lowest four electronic states are plotted along a TSH trajectory for methylhydroperoxide, together with the total energy and the energy of the TSH driving state. (XMS(4)-CASPT2(8/6)) is used for the electronic structure, see Ref.~\citenum{prlj2023deciphering} for details.) The excited-state dynamics of MHP is here initiated in S$_1$ and this state drives the dynamics during the trajectory segment presented here (6 fs). In the time interval $2.75$ fs $< t < 4$ fs, the energy of the third excited state (S$_3$) increases sharply before abruptly decreasing, exhibiting a discontinuity at 3.5 fs. This discontinuity in the electronic energy of S$_3$ attests to a sudden change of electronic character for this state (calculating an additional excited electronic states would unravel an avoided crossing with S$_4$ near the discontinuity), caused by the stretch of the \ce{O-OH} bond of methylhydroperoxide in the early times of the dynamics and the lowering in energy of an electronic state with a $n\sigma^\ast(\text{O-O})$ character (the character of S$_3$ after 3.5 fs). This missed avoided crossing (as one of the electronic state is not included in the state-averaging process) is in stark contrast with the orbital-rotation discontinuity presented above (Figure~\ref{fig:etot_orbrot}), where electronic energies evolve smoothly along the TSH trajectory before experiencing a sudden jump due to orbital rotation. The abrupt change of electronic character leads to a small discontinuity in the total energy ($\sim 0.08$ eV) and also impacts the behavior of the electronic energy of the other states (e.g., S$_0$). It is important to stress here that the orbitals composing the active space are strictly preserved along the TSH trajectory presented in Figure~\ref{fig:etot_extrastate} -- the discontinuity in total energy is purely the result of a new electronic-state character entering the state-averaging process. 

Sometimes, state-averaged discontinuities may be less severe than the orbital-rotation discontinuity if the character of the incoming states is encompassed by the orbitals already present in the active space (as exemplified by the case presented in Figure~\ref{fig:etot_extrastate}), and incorporating these electronic states leads to stable trajectories (see Ref.~\citenum{Prljphotolysis2020} for another example). Increasing the number of electronic states considered in the state-averaging procedure is a solution in principle, but in practice, this strategy often fails as the electronic states that may become important in the dissociation limits are often high in energy in the FC region and hard to capture. Strategies like dynamically-weighted state-average methods were defined to alleviate this issue\cite{glover2019analytical,battaglia2020extended,li2019dynamically} and started to appear in combination with nonadiabatic molecular dynamics (even though dynamically-weighted approaches rely on additional parameters and may suffer from additional instabilities).\cite{ibele2024aimspt2}

When one is interested in the formation of photoproducts in the ground electronic state, the active space defined for the nonadiabatic dynamics may not be sufficient to describe the diversity of electronic structures for the products. A strategy consists of switching the nonadiabatic dynamics for a regular adiabatic Born-Oppenheimer dynamics, together with a switch to an unrestricted ground-state electronic structure method (with a small time step). This switching strategy requires a strict benchmark to validate the change in the electronic-structure method, but can provide insight into the formation of ground-state photoproducts (see Refs.~\citenum{mignolet2016rich,pathak2020tracking,sanchez2021dasa,janos2024prediction,nunes2024ued} for examples of this approach). 

From a numerical perspective, the time step used for the propagation of nonadiabatic trajectories can also lead to issues with total energy conservation, for example, when the molecule reaches the ground electronic state and its nuclei exhibit a high kinetic energy. Drifts (and not sudden jumps) in total energy can be observed in adiabatic Born-Oppenheimer molecular dynamics for various reasons, and the reader is referred to Ref.~\citenum{marxbook} for a discussion on this topic. Strategies for adaptive time steps were devised for AIMS.\cite{levine2008implementation} The situation is trickier for TSH, given that the time step features in the definition of the hopping probability in its fewest-switches version, see Ref.~\citenum{parker2020} for a detailed discussion of this issue. Similar issues can be observed for nonadiabatic dynamics employing LR-TDDFT/TDA in regions of the nuclear configuration space near an intersection seam between the ground and first excited electronic states, where the topography of the PESs is known to exhibit sharp energy variations.\cite{levine06,tapavicza08,tavernelli09b,taylor2023description} 

We also note that conservation of the total quantum energy is achievable for Gaussian-based methods only in the limit of a complete number of traveling Gaussian functions\cite{habershon2012} or if all parameters are propagated variationally as in vMCG\cite{richings2015quantum}, even though further practical approximations for this method can lead to deviations from total energy conservation.\cite{christopoulou2021developing} Various strategies were devised to improve the total (quantum) energy conservation for these methods.\cite{izmaylov2022}

\paragraph{Nonadiabatic couplings and nonadiabatic coupling vectors} 

A first numerical consideration emerges when observing the nonadiabatic coupling term in the equations of motion for the electronic coefficients of TSH or the coupling between TBFs in AIMS (see Section \ref{setting-the-scene} for the definition of nonadiabatic couplings and Section \ref{nonadiabaticdynproblem} for their role in the equations of motion). The NACVs are projected onto the nuclear velocity of the trajectory (in TSH) or the nuclear momentum for the centroid position between TBFs (in AIMS). Focusing on TSH and its equations of motion for electronic coefficients (Eq.~\eqref{tdese_coeff}), this projection appears as $\sigma_{JI}(\bs R^{\alpha}(t))=\bs d_{JI}\left(\bs R^{\alpha}\right)\cdot \dot{\bs R}^{\alpha}(t) = \bra{\Phi_J(\bs R^{\alpha})}\frac{\partial}{\partial t}\ket{\Phi_I(\bs R^{\alpha})}_{\mathbf{r}}$.\cite{hammes94} As discussed above, the last term is, in fact, the term that appears first upon derivation of the TSH equations of motion (see Eq.~\eqref{tdese_coeff_intermed}), but it is often rewritten as a projection of the NACVs onto the nuclear velocities using the chain rule $\frac{\partial}{\partial t} = \frac{\partial}{\partial \bs R}\frac{\partial \bs R}{\partial t}$. This projection is often referred to as the 'nonadiabatic couplings' (NACs) and is also found in the definition of the fewest-switches hopping probability (Eq.~\eqref{eq:tshproba}). More than a curiosity, the appearance of NACs has allowed, from the early times of TSH simulation, to determine the nonadiabaticity along a trajectory without requiring an explicit evaluation of the NACVs, using finite differences to get the derivative of the electronic wavefunctions with respect to time.\cite{hammes94} Given that it requires the overlap of electronic wavefunctions at different time steps of the dynamics, this strategy is nowadays often referred to as the 'wavefunction overlap approach' and allows to perform TSH (and AIMS) dynamics with an electronic-structure method for which NACVs are either unavailable (e.g., ADC(2)) or computationally demanding to calculate. Different approaches were developed to calculate wavefunction overlaps\cite{tapavicza07,mitric2008nonadiabatic,PITTNER2009147,meek2014tdc,Ryabinkin2015,plasser2016,Xiaorui2023} or approximate them,\cite{baeck-an2017} together with a comparison of their performance.\cite{merritt2023} Similar approaches were suggested for AIMS.\cite{levine2008implementation,tao2009ab} Last but not least, a local diabatization strategy has also been proposed to ensure a smoother numerical propagation of the electronic coefficients without the explicit need for the NACs, which might develop strong peaks near intersection regions.\cite{plasser2012} 

Another potential danger of the numerical integration of the equations of motion in nonadiabatic molecular dynamics: the possibility of jumping over a coupling region between two consecutive integration time steps (for the classical propagation of a TSH trajectory or a TBF in AIMS). The impact of this problem was made clear by studies involving electron transfer processes,\cite{fernandezalberti2012tuc} where the diabatic coupling between diabatic electronic states describing the electronic transfer might be small, leading (according to our discussion in Section~\ref{sec_adiabatic_diabatic}) to very localized and strong NACs in the vicinity of the state degeneracy between the corresponding adiabatic electronic states. Such couplings are so localized that nuclear motion can easily jump over within a typical time step used for the numerical integration of the classical equations of motion in TSH (or the TBFs in AIMS). A TSH trajectory should, in principle, jump from one electronic state to another with nearly complete probability to ensure the preservation of its diabatic character. Yet, if the numerical trajectory misses this narrow region of coupling, it will continue on the same adiabatic state, meaning that the trajectory abruptly changed its electronic character -- a clear artifact.\cite{fernandezalberti2012tuc,NELSON2013208,MEEK2015117} Such trivially unavoided crossings have been triggering the development of various strategies for TSH and AIMS to better follow the electronic character of the electronic states,\cite{NELSON2013208,meek2014tdc,MEEK2015117,Lee2019,temen2021tuc,qiu2023practical} most of them available in current codes proposing TSH or AIMS dynamics. The current implementation of AIMS in \texttt{FMS90} does follow the electronic character of each electronic state considered in the dynamics (as part of the calculation of couplings between TBFs) and will reject a time step if it leads to a switch of electronic character between electronic states (a marker that the TBF may have jumped over a coupling region), as part of an adaptive time step strategy.\cite{levine2008implementation} Finally, the interested reader may also consult Ref.~\citenum{nelson2012nonadiabatic} for various numerical tests on the stability of a TSH dynamics. 

\paragraph{Nonadiabatic transitions} 

As stated in Section~\ref{elecstructproblem}, electronic-structure methods like LR-TDDFT and ADC(2) do not adequately describe the topography and topology of intersection regions between S$_1$ and S$_0$, either due to their practical approximations or their formalism. Various strategies have been suggested to mitigate these particular issues. A pragmatic solution consists of stopping the nonadiabatic dynamics whenever trajectories reach a region of nuclear configuration space where the energy gap between S$_1$ and S$_0$ falls below a certain value (for example, 0.1 eV).\cite{crespo-otero2011,plasser2014surface}. This approach, even if perhaps the safest, prevents the explicit description of the nonadiabatic transfer to the ground electronic state. Other schemes were suggested to either approximate the NACs in this critical region (Baeck-An\cite{baeck-an2017,do2022fewest}) or model the nonadiabatic transitions with a different strategy than the fewest-switches approach to TSH (Landau-Zener\cite{belyaev2011nonadiabatic,slavicek2020}). Inducing artificially a hop from S$_1$ to S$_0$ constitutes another brute-force alternative (see Ref.~\citenum{papineau2024} for an example and discussion of this approach). The strategies mentioned in this paragraph should be restricted to exploratory dynamics, and a proper description of the nonadiabatic dynamics between S$_1$ and S$_0$ requires a higher level of electronic-structure theory. 

\paragraph{Velocity rescaling}

The previous paragraph naturally brings the discussion to another potential issue, in particular for TSH, related to rescaling the nuclear velocities after a hop. The rescaling of velocities has been a matter of debate in the community, also connected to the issue of frustrated hops (hops to a higher electronic state that cannot occur due to insufficient nuclear kinetic energy, along the NACVs, required to compensate for the increase in electronic energy).\cite{fang99b,huang2023,subotnik2016understanding} A recent work has provided strong evidence (validated by other approaches\cite{Mannouch2023}) that the rescaling should be performed by along the NACVs (or, if not available, the gradient difference) and, possibly more importantly, that an isotropic rescaling of velocities should be avoided.\cite{toldo2024velresc} AIMS is less sensitive to the strategy employed for the rescaling of velocities, due in part to the nature of the spawning algorithm.\cite{ibele2020} 

\paragraph{Final considerations}

The complexity of a nonadiabatic dynamics simulation and the number of ingredients it depends on (in comparison to regular Born-Oppenheimer molecular dynamics) means that great care should be paid when restarting a trajectory. If using a multiconfigurational or multireference method for the electronic structure, the precise electronic energies should be carefully checked upon restart to ensure that the very same level of electronic-structure theory was recovered. The sign of the NACs and/or NACVs should be continuous throughout the dynamics and checked after a restart.

We should stress again the critical importance that the electronic-structure method can play in the outcome of a given nonadiabatic dynamics, as identified in a series of recent articles and highlighting the crucial role of benchmarking electronic-structure methods before any dynamics.\cite{janos2023controls,papineau2024,tomas2024}

\subsubsection{Outcome of a typical TSH dynamics}
\label{sec:tshoutcome}

\begin{figure*}[h!]
    \centering
    \includegraphics[width=1.0\linewidth]{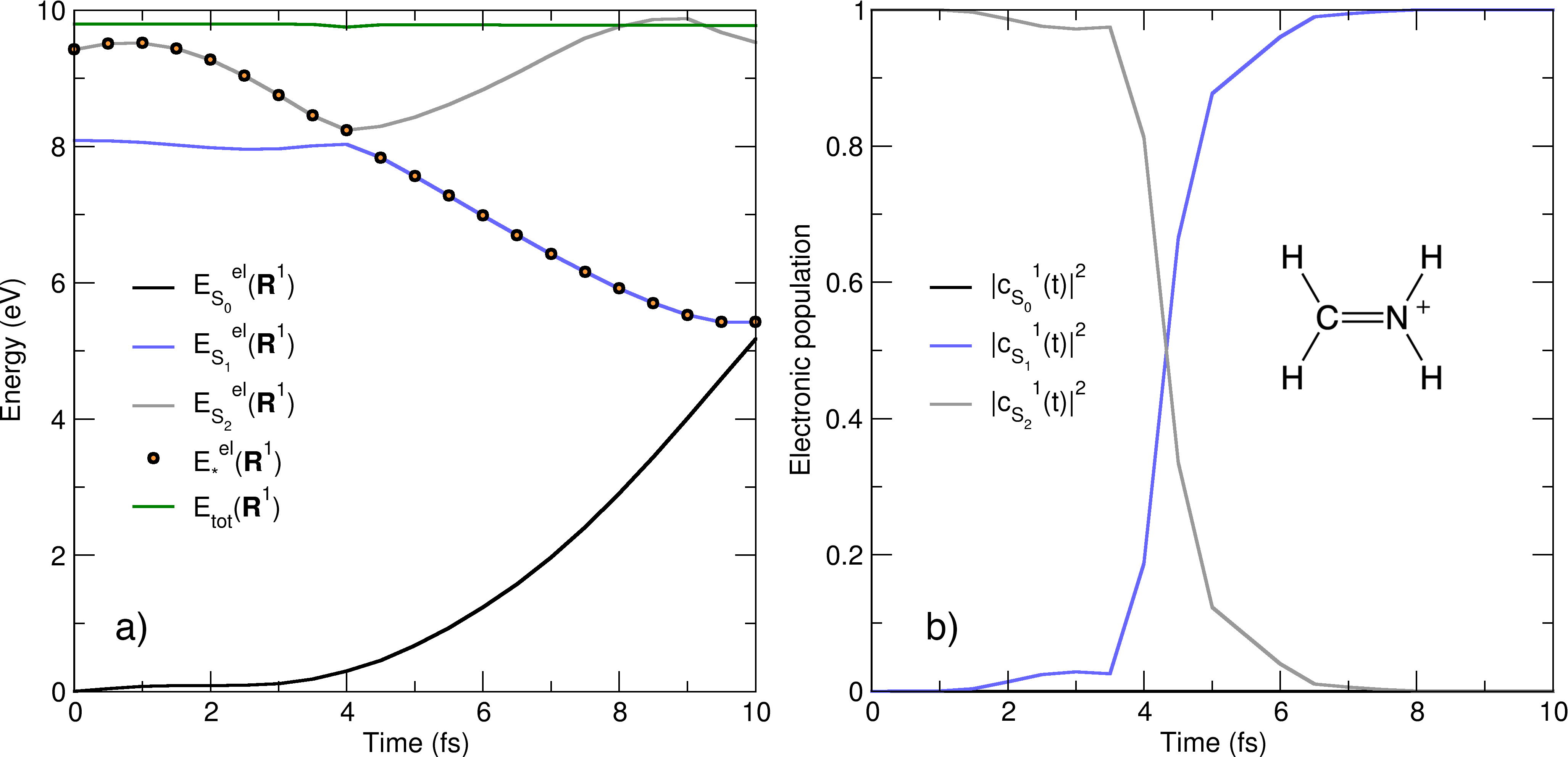}
    \caption{A single TSH trajectory for the photodynamics of protonated formaldimine. (a) Electronic energies along the TSH trajectory $\alpha$ (here $\alpha=1$), labeled as $E_I^{el}(\bs R^1)$. The electronic state driving the TSH trajectory is indicated by filled circles ($E_\ast^{el}(\bs R^1)$). (b) The electronic coefficients along the trajectory $\alpha=1$ for the three electronic states considered. The molecular representation of protonated formaldimine is given as an inset.}
    \label{fig:tshrun}
\end{figure*}

We discuss here the outcome of a nonadiabatic molecular dynamics simulation conducted with TSH, using the photodynamics of protonated formaldimine (molecular structure given in the inset of Fig.~\ref{fig:tshrun}b) as an example. We consider a photoexcitation of protonated formaldimine to its second excited electronic state (S$_2$) and observe the nonadiabatic decay of the created nuclear wavepacket. ADC(2) is employed for the electronic structure (along with def2-SVP basis set), meaning that we limit our study to the nonadiabatic dynamics between S$_2$ and S$_1$. As discussed in Section~\ref{elecstructproblem}, ADC(2) offers an adequate description of the S$_2$/S$_1$ seam of intersection in comparison to XMS-CASPT2. The ICs were sampled from a harmonic Wigner distribution for the ground state of protonated formaldimine. 

\paragraph{Monitoring a single TSH trajectory}

We begin by monitoring the fate of a single TSH trajectory, labeled $\alpha=1$ in the following (following our earlier notation for TSH trajectories). The TSH trajectory is initiated in the electronic state S$_2$ and will be propagated classically in this electronic state following this electronic state until a region of nonadiabaticity can induce a hop to a different electronic state. Figure~\ref{fig:tshrun}a shows the electronic energy of TSH trajectory 1 during the dynamics (red circle), evolving initially in S$_2$ before it reaches a region of strong nonadiabaticity (after 4fs), triggering a jump to electronic state S$_1$. The nonadiabatic transition is clearly explained by looking at the electronic coefficients for this TSH trajectory (Figure~\ref{fig:tshrun}b). The nonadiabatic region encountered after around 4 fs of dynamics leads to an almost complete transfer of electronic population from S$_2$ to S$_1$, ensuring a high probability for the TSH trajectory to hop from S$_2$ to S$_1$. The TSH trajectory then rapidly decays toward the S$_0$ state. Due to the limitations of ADC(2) in describing the S$_1$/S$_0$ intersection seam, the TSH trajectory is stopped when the electronic energy gap between S$_1$ and S$_0$ reaches a small value (here $<0.3$ eV). As discussed earlier, a critical quantity to monitor for each TSH trajectory is its total classical energy, in particular its conservation along the full dynamics (see green curve in Fig.~\ref{fig:tshrun}a). For the present case, the total energy is conserved satisfactorily, with small fluctuations but no clear drifts or jumps observed. 

\paragraph{Analysis of the ensemble of TSH trajectories}

\begin{figure}[h!]
    \centering
    \includegraphics[width=1.0\linewidth]{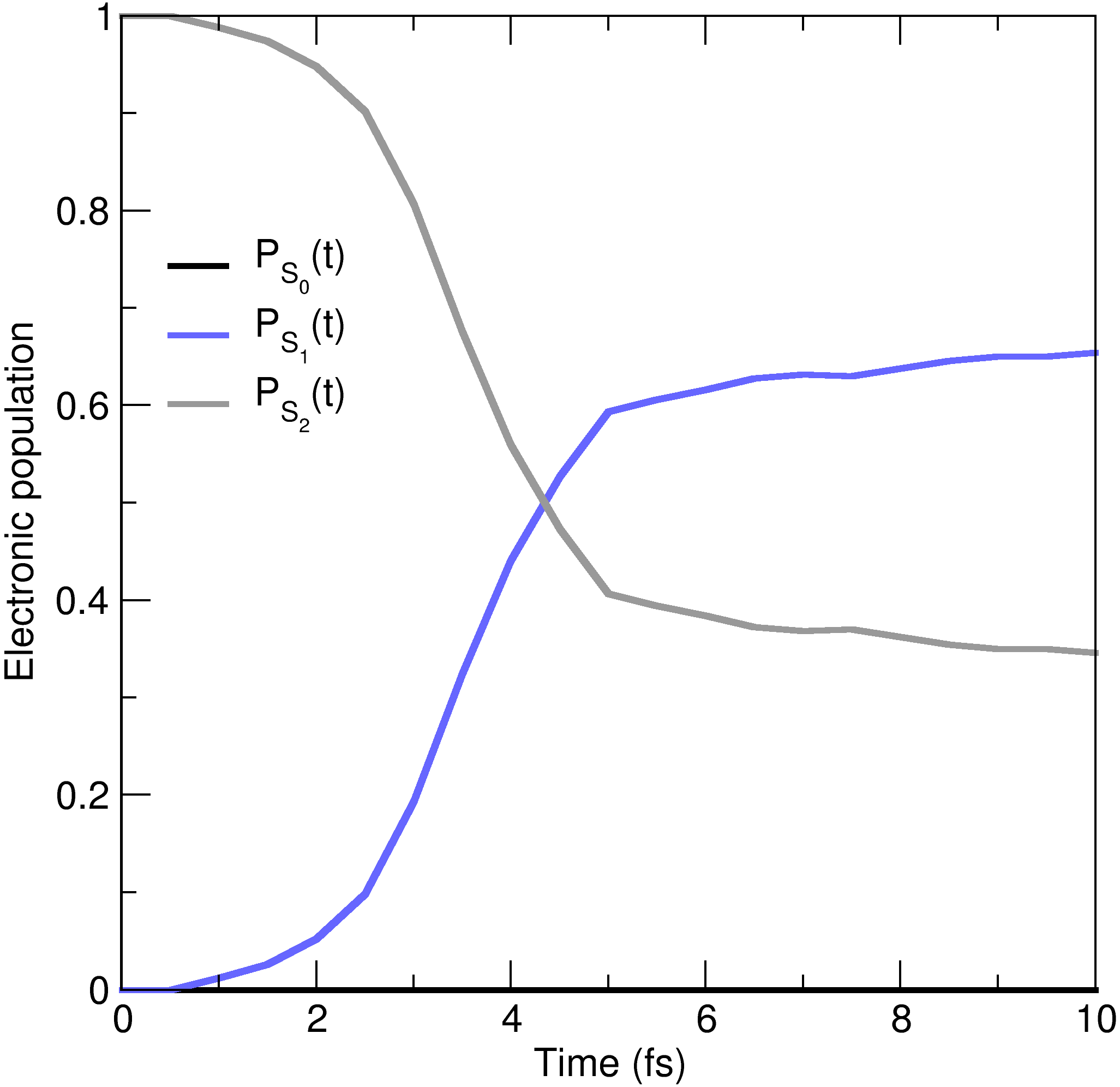}
    \caption{Full TSH dynamics for the nonadiabatic molecular dynamics of protonated formaldimine. The electronic populations (fractions of trajectories, $P_I(t)$) deduced from 497 independent TSH trajectories are represented with thick lines, colored by electronic state.}
    \label{fig:tshrun2}
\end{figure}

A nonadiabatic molecular dynamics simulation using TSH relies on the sampling of a large number of trajectories (to offer a proper description of the ICs mimicking the initial nuclear wavepacket and to converge the stochastic algorithm responsible for the nonadiabatic transitions). We focus here on the populations obtained from an entire swarm of TSH trajectories (497 in the present case). As discussed in Section~\ref{nonadiabaticdynproblem}, the population of a given electronic state $I$ in TSH can be described in two different ways: the fraction of TSH trajectories in this particular state, $P_I(t)$, or the average electronic population for this state, $\langle|c_I(t)|^2\rangle$ (see Section~\ref{sec:classtrajmethod}). For the TSH algorithm to be considered consistent, both definitions of the electronic population should provide very similar trends.\cite{Granucci2007} The populations depicted in Fig.~\ref{fig:tshrun2} offer an approximation to the branching of the nuclear wavepacket -- the photoexcited nuclear wavepacket in the S$_2$ electronic state relaxes significantly toward S$_1$ within the short timescale presented here. The ground electronic state would, in principle, reduce the S$_1$ population, but this sink of population is not accounted for in these calculations employing ADC(2) for the electronic structure. We stress again that a single TSH trajectory (as depicted earlier) does not carry much information, and only the swarm of trajectories should be used to extract meaningful quantities like populations or variations of geometric parameters. TSH, being a mixed quantum/classical method, it does not give access to nuclear (or molecular) wavefunctions, meaning that the determination of observables relies on a given 'recipe' (extracting a given quantity of interest along each TSH trajectory and incoherently averaging it over the entire swarm) more than an actual well-defined mathematical protocol (i.e., expectation values of the molecular wavefunction). For a given property $\mathcal{O}$, an incoherent average in TSH would be given by $\mathcal{O}(t) = \frac{1}{N_{\text{traj}}}\sum_\alpha^{N_{\textit{traj}}}\mathcal{O}_\alpha(t)$. More information will be given on the determination of observables in Section~\ref{sec:analysisandobservables}.

\subsubsection{Outcome of a typical AIMS dynamics}
\label{sec:aimsoutcome}

\begin{figure*}[h!]
    \centering
    \includegraphics[width=1.0\linewidth]{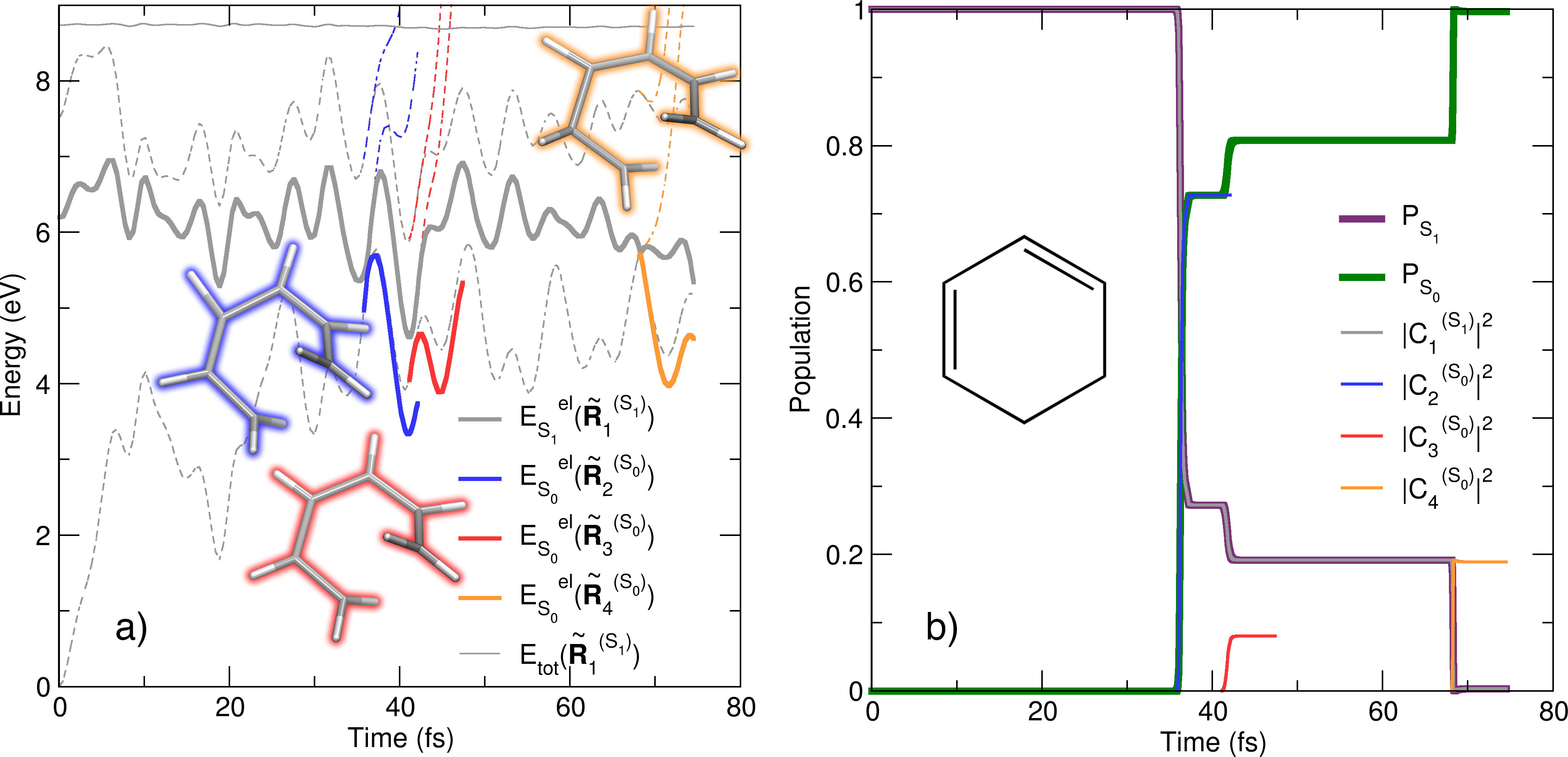}
    \caption{A single AIMS run for the photodynamics of cyclohexadiene. The parent TBF in S$_1$ ($\tilde{\chi}_{1}^{(S_1)}$) spawns three child TBFs in S$_0$ ($\tilde{\chi}_{2}^{(S_0)}$, $\tilde{\chi}_{3}^{(S_0)}$, and $\tilde{\chi}_{4}^{(S_0)}$). Their respective electronic energy over time is labeled as $E_I^{el}(\tilde{\bs R}_k^{(I)})$.  a) Electronic energies along the four TBFs, with plain lines indicating the electronic energy driving each TBF and dashed lines (with corresponding colors) highlighting the electronic energy of the other electronic states at the current nuclear position of each TBF ($\tilde{\bs R}_k^{(I)}(t)$). The three molecular structures represent the nuclear geometries at the spawning time of each child TBF. b) Electronic-state populations during the AIMS run, with the overall S$_1$ and S$_0$ population given by a thick palatinate and green line, respectively. The population for each TBF is depicted by a thin line (color code is the same as in panel (a)). Adapted from Ref.~\citenum{aimschaptercurchod} with permission.}
    \label{fig:aimsrun}
\end{figure*}

We provide here an example of an AIMS simulation and the quantities that can be monitored. We focus here on the nonadiabatic molecular dynamics of cyclohexadiene (CHD, molecular structure given in the inset of Fig.~\ref{fig:aimsrun}b). When photoexcited to its first excited electronic state (S$_1$), CHD undergoes an excited-state dynamics that can potentially result in its ring-opening to form hexatriene. The electronic structure employed for the following calculations is SA(3)-CASSCF(4/3), combined with a 6-31G$^\ast$ basis set (this level of electronic-structure theory was validated in previous work\cite{kim2011control,kim2015ab,snyder2016gpu}). The ICs were sampled from a harmonic Wigner distribution for the ground state of CHD, and the AIMS dynamics were initiated in the first excited state, S$_1$. We note that the example presented below for the photodynamics of CHD is adapted from a detailed discussion of the practical use of AIMS that can be found for further information in Ref.~\citenum{aimschaptercurchod}.

\paragraph{Monitoring a single AIMS run}

An AIMS simulation is composed of multiple independent runs (or 'branches', using the vocabulary of Refs.~\citenum{curchod2018ab,aimschaptercurchod}), each starting with a single TBF, called the parent TBF, propagated from a selected set of ICs. The complex amplitude (see Eq.~\eqref{eomec}) for the parent TBF $k$ attributed to state $I$ at the beginning of the dynamics is $C_{k}^{(I)}(t=0)=1.0+0.0i$. An AIMS run consists of propagating the parent TBF $k$ until new child TBFs are created (spawned) in regions of strong nonadiabaticity. The AIMS run terminates when a certain criterion is met (e.g., when most of the nuclear amplitude is in the ground electronic state). 

Figure~\ref{fig:aimsrun} shows a typical AIMS run for CHD, with a parent TBF being initially propagated and leading to the spawning of three child TBFs to ensure a population transfer from the first excited to the ground electronic state. Only the parent TBF is present at $t=0$, labeled with $k=1$ and evolving in the electronic state S$_1$ ($\tilde{\chi}_{1}^{(S_1)}$). The thick gray line in Fig.~\ref{fig:aimsrun}a depicts the electronic energy as a function of time for the parent TBF during this AIMS run, while the time trace of its population ($|C_{1}^{(S_1)}(t)|^2$) is shown by a similar line in Fig.~\ref{fig:aimsrun}b. As TBFs follow classical trajectories, their total classical energy should be conserved (see thin full line in gray in Fig.~\ref{fig:aimsrun}a) -- this classical energy should not be confused with the total quantum energy for the system, which would be expressed as an expectation value of the molecular Hamiltonian over the full swarm of TBFs (see below for a discussion on expectation values in AIMS). In the early times, the parent TBF evolves adiabatically in S$_1$ until it reaches a region of strong nonadiabaticity with S$_0$ (lower dashed gray line in Fig.~\ref{fig:aimsrun}a) after 30 fs of dynamics. The parent TBF reaches a region of the nuclear configuration space where the two electronic states become nearly degenerate at $t\sim36$ fs. In this region of strong nonadiabaticity, the parent TBF spawns a new TBF ($\tilde{\chi}_{2}^{(S_0)}$) in the coupled electronic state S$_0$ (whose electronic energy is depicted by a thick blue curve in Fig.~\ref{fig:aimsrun}a). The spawning geometry for $\tilde{\chi}_{2}^{(S_0)}$ is given as an inset in Fig.~\ref{fig:aimsrun}a (with a blue shadow). The two electronic traces (for the parent and the child TBFs) become nearly degenerate at the time of the spawning. The fact that the blue curve in Fig.~\ref{fig:aimsrun}a appears before the intersection region is due to the spawning algorithm, which ensures that any new child TBF is included in the AIMS dynamics before it reaches the region of strong nonadiabaticity to ensure a proper transfer of nuclear amplitude -- see Refs.~\citenum{curchod2018ab,aimschaptercurchod} for details on the spawning algorithm. The child TBF $\tilde{\chi}_{2}^{(S_0)}$ rapidly relaxes in the ground electronic state S$_0$ -- its corresponding S$_0$ energy drops shortly after the nonadiabatic region. A large transfer of nuclear population is observed between the parent and first child TBF (large variations in the gray and blue curves in Fig.~\ref{fig:aimsrun}b at around 36 fs). When a TBF evolves in a 'target state', here the electronic ground state, and is no longer coupled to any other TBFs, its dynamics can be stopped (as seen for the curves corresponding to $\tilde{\chi}_{2}^{(S_0)}$ after 42 fs).\footnote{Stopping TBFs when they reach the ground electronic states may prevent numerical instabilities caused by the large amount of kinetic energy gained by these TBFs when they relax in S$_0$, often necessitating a smaller time step. If one is interested in the overall dynamics of the nuclear wavepacket (and not just the corresponding electronic populations), the TBFs in S$_0$ can be propagated individually with a smaller time step and included in the overall analysis of the AIMS dynamics.} After meeting this first nonadiabatic region, the parent TBF continues its evolution in S$_1$ until it again hits the S$_1$/S$_0$ intersection seam, spawning a second child TBF, $\tilde{\chi}_{3}^{(S_0)}$ (given by red lines in Fig.~\ref{fig:aimsrun}). Only a small amount of population transfer results from the nonadiabatic interaction between the parent and second child TBFs ($\sim$ 8\%). After 68 fs of dynamics, the parent TBF hits the intersection seam a final time, leading to an almost complete transfer of its nuclear amplitude to the third child TBF $\tilde{\chi}_{4}^{(S_0)}$ (orange lines).

In stark contrast with TSH, the `TBF population' mentioned above and given by $|C_{k}^{(I)}(t)|^2$ does not represent the actual population of a given electronic state as one needs to account for the population of all the TBFs within the run and their interaction (as we are using a nonorthogonal Gaussian basis set). Hence, the population in electronic state $L$ for a given AIMS run is determined by calculating the expectation value of the projector $\hat{\mathcal{P}}_L = | \Phi_L \rangle\langle \Phi_L |$ using the AIMS molecular wavefunction for a given run:
\begin{align}
P_L(t) &  = \sum_{IJ}^\infty \Big[ \sum_k^{N_I(t)}\sum_{k'}^{N_J(t)}
\left(C_{k}^{(I)}(t)\right)^\ast C_{k'}^{(J)}(t)  \langle   \Phi_I \tilde{\chi}_{k}^{(I)} |  \hat{\mathcal{P}}_L  | \tilde{\chi}_{k'}^{(J)} \Phi_J \rangle_{\bs r, \bs R}
\Big] \notag \\
& =  \sum_{kk'}^{N_L(t)}\left(C_{k}^{(L)}(t)\right)^\ast C_{k'}^{(L)}(t)  \langle \tilde{\chi}_{k}^{(L)}  | \tilde{\chi}_{k'}^{(L)}  \rangle_{\bs R}  \notag \\
&=   \sum_{kk'}^{N_L(t)} \left(C_{k}^{(L)}(t)\right)^\ast C_{k'}^{(L)}(t)  (\bs S)^{LL}_{k,k'} \, .
\label{eqpop}
\end{align}
Eq.~\eqref{eqpop} highlights another key difference between AIMS and TSH: by propagating nuclear wavefunctions, AIMS can be used to approximate a molecular wavefunction from which one can express expectation values. This equation also shows the influence of the Gaussian interference terms in the calculation of expectation values in AIMS. Coming back to our example of an AIMS run for CHD, the population trace for the S$_1$ and S$_0$ electronic states -- $P_{L}(t)$, with $L=\text{S}_1,\text{S}_0$ -- is depicted in Fig.~\ref{fig:aimsrun}b with thick lines. 

\paragraph{Analysis of the full ensemble of AIMS TBFs}

An AIMS nonadiabatic molecular dynamics simulation requires more than a single run (or single parent TBF) to describe properly the dynamics of the initial nuclear wavepacket. In practice, one needs to initiate a certain number of parent TBFs, each of them having different ICs, sampled based on the recommendation made in Section~\ref{sec:initconds} above. The total number of parent TBFs depends on the nonadiabatic molecular dynamics simulated but typically ranges from 20 to more than 100 (see Ref.~\citenum{lassmann2023} for a brief discussion on the convergence of AIMS with respect to the number of initial parent TBF). To recover the final AIMS population from a swarm of $N_{ini}$ independent AIMS runs (a given AIMS run being labeled by $\beta$), one would simply average incoherently (that is, without accounting for the interferences of TBFs between branches) over the AIMS runs according to
\begin{equation}
\tilde{P}_L(t) \approx \frac{1}{N_{ini}} \sum_{\beta}^{N_{ini}} \left[   \sum_{kk'}^{N_L^\beta(t)} \left(C_{k\beta}^{(L)}(t)\right)^\ast C_{k'\beta}^{(L)}(t)  (\bs S)^{LL}_{k\beta,k'\beta} \right] \, ,
\label{eqifgacoherent2}
\end{equation}
which, considering that the total AIMS molecular wavefunction is normalized, corresponds to an average of Eq.~\eqref{eqpop} over the $N_{ini}$ branches.
Additional details about the calculation of expectation values in AIMS and FMS can be obtained from Ref.~\citenum{curchod2018ab}.

\begin{figure}[h!]
    \centering
    \includegraphics[width=1.0\linewidth]{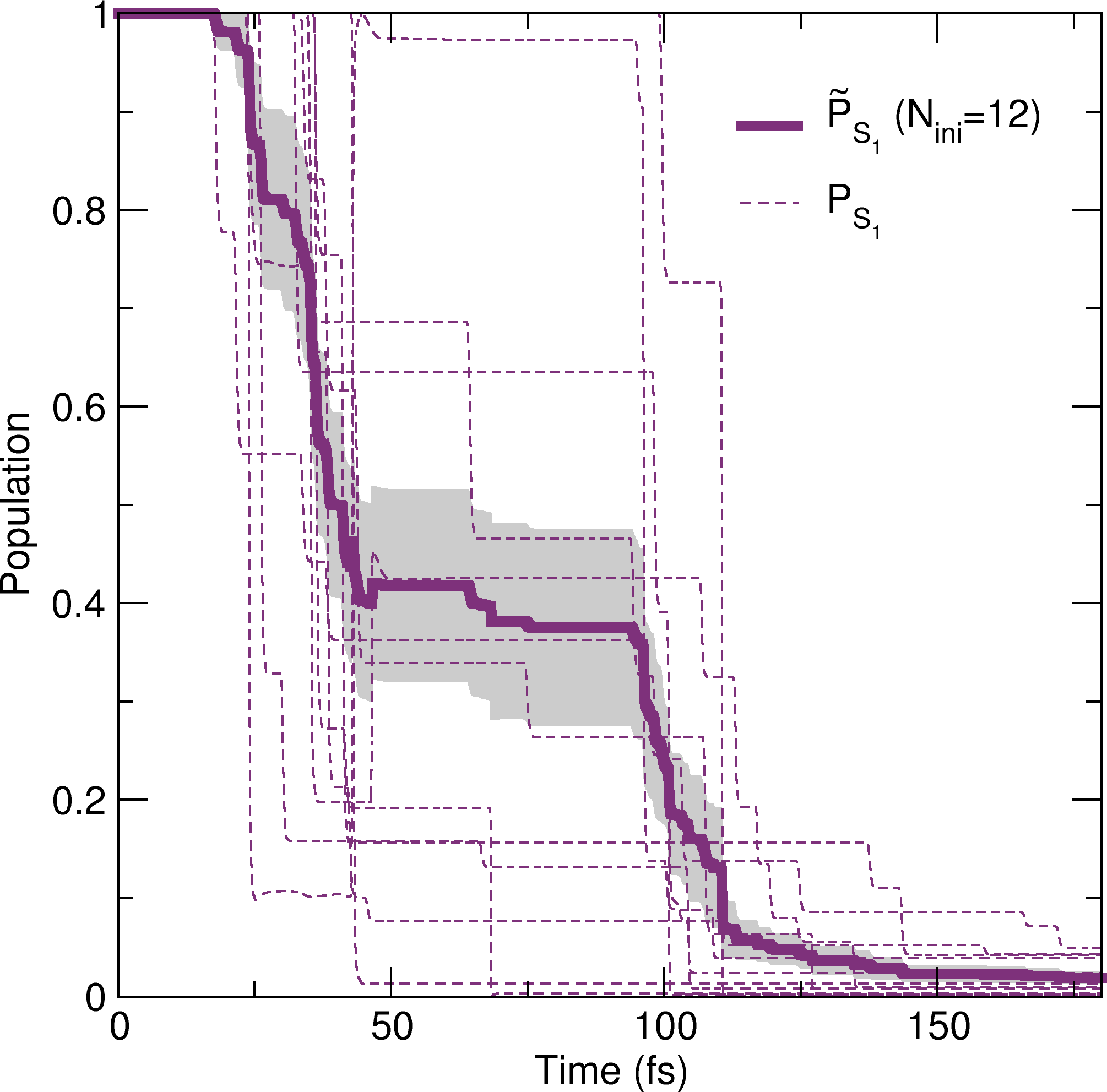}
    \caption{Nonadiabatic molecular dynamics of cyclohexadiene with AIMS using 12 runs (or 12 initial parent TBFs or ICs), generating 72 TBFs in total due to the spawning algorithm. The overall population of the S$_1$ state, $\tilde{P}_{S_1}(t)$ (thick palatinate line, gray area indicates the standard error), was obtained by averaging the 12 $P_{S_1}(t)$. The S$_1$ population time trace coming from each of the 12 AIMS runs, $P_{S_1}(t)$, is indicated by dashed lines.}
    \label{fig:tbfallic}
\end{figure}

Coming back to the nonadiabatic dynamics of CHD, one needs to perform more AIMS runs to converge the expectation value given in Eq.~\eqref{eqifgacoherent2}. For the current example of CHD, we used 12 different initial parent TBFs ($N_{ini}=12$) (one would need more initial TBFs to fully converge the simulation). The time trace for the population of electronic state S$_1$ ($\tilde{P}_{S_1}$(t), as defined in Eq.~\eqref{eqifgacoherent2} for 12 AIMS runs) is given in Figure~\ref{fig:tbfallic}. This representation allows us to talk about the branching of the full nuclear wavepacket (originally in S$_1$) among the different electronic states considered in the simulation. In the case of CHD, the initial nuclear wavepacket decays to the ground electronic state in less than 180 fs (thick line in Fig.~\ref{fig:tbfallic}). Dashed lines in Fig.~\ref{fig:tbfallic} symbolize the S$_1$ population of each AIMS run ($P_{S_1}(t)$, as individually obtained from Eq.~\eqref{eqpop}). It is important to note that, while the AIMS dynamics was initiated with 12 parent TBFs, the spawning algorithm triggered the generation of new TBFs to adequately describe the nonadiabatic transition between the first and the ground electronic state, meaning that 72 TBFs are present by the end of this nonadiabatic dynamics. 

\subsection{Calculation of experimental observables}
\label{sec:analysisandobservables}

Calculating the actual signal that was measured experimentally is perhaps the best strategy to compare the outcome of a nonadiabatic dynamics to a spectroscopic experiment and provide insight. This direct comparison avoids the potential pitfalls of trying to match an experimental signal or observable with a proxy quantity obtained from the simulation. For example, the time trace of electronic populations, which is a representation-dependent quantity and \textit{per se} not an experimental observable, is often used to determine electronic-state lifetimes, then compared with lifetimes obtained from an experimental signal. 

However, it is important to realize that calculating an experimental signal or observable often comes with a new layer of approximations on top of those deployed for the nonadiabatic molecular dynamics. The calculation of certain observables indeed requires additional electronic-structure quantities: for example, one could cite the need for transition dipole moments between excited states to determine a transient absorption signal, or for ionization energies and intensities to obtain a time-resolved photoelectron spectroscopy signal, or the recent development of electronic-structure methods to simulate X-ray absorption or photoemission spectra.\cite{vidal2020equation,seidu2019simulation,leyser2024general} The calculation of certain observables requires information about the nuclear wavefunctions too (molecular time-dependent dipole moments), while others can use the swarm of trajectories as a proxy (X-ray or electron diffraction). 
In the following, we list a series of experimental observables that were calculated on the support of nonadiabatic molecular dynamics simulations, highlighting the potential of such approaches to offer insights into complex experiments. We focus on gas-phase experiments only, and we invite the reader to consult the references provided for the details of such calculations. We expect that future iterations of this living review will fill out this list and highlight the splendid diversity of applications of nonadiabatic molecular dynamics.

We begin this survey with time-resolved spectroscopies, where a molecule is electronically excited by a pump laser and its resulting dynamics monitored by a probe at various time delays following the pump pulse. Such pump-probe experiments differ in name depending on the nature of the probe pulse used to monitor the molecular dynamics. If the probe pulse is a laser pulse with an energy large enough to ionize the molecule, one can follow the experimental binding energy of the electron as a function of time delay and the anisotropy of the electron angular distribution collected on a detector, informing on the electronic character and the structure of the molecule during its nonradiative pathway.\cite{schuurman2018reviewCI} Nonadiabatic molecular dynamics has been vastly employed to simulate such a time-resolved photoelectron spectroscopy (TRPES) signal -- see Refs.~\citenum{tao2011ultrafast,doslic2021TRPES,matsika2021trpes} and Ref.~\citenum{suzuki2015ethylenepredict} for a successful prediction of a TRPES signal using AIMS. Using a laser probe pulse in the X-ray regime opens doors to investigating the nonadiabatic molecular dynamics using other probing processes. For example, time-resolved X-ray absorption spectroscopy (TRXAS) uses a probe laser pulse that will excite a core electron from a specific atom of the molecule to populate a frontier (empty) orbital, offering an element-specific probe for the change of electronic character (and possibly molecular structure) during the excited-state dynamics, as the core electron can populate a molecular orbital involved in the excited electronic state the molecule is in -- see Refs.~\citenum{schuurman2022trxas,kjonstad2024photoinduced} for recent examples. The X-ray probe pulse can also lead to an ionization of the core electron targeted, resulting this time in time-resolved X-ray photoelectron spectroscopy (TRXPS), used for example in combination with AIMS in Ref.~\citenum{schuurman2023aimscs2trxps}. Ionizing a core electron can also lead to a secondary process called Auger decay, which takes place when a valence electron of a core-ionized molecule relaxes to fill in the core hole created (within a few fs to a few tens of fs) leading to the simultaneous emission of another electron (see Ref.~\citenum{li2017ultrafast} for an example of nonadiabatic molecular dynamics used to account for Auger decay, resulting in a Coulomb explosion). Coming back to the world of electronic excitations, the probe laser pulse could also be in the visible region to trigger an electronic excitation of the (already excited) molecule of interest. This strategy, called transient absorption spectroscopy (TAS), can be performed easily in a laboratory environment (in contrast with other experiments that might require a dedicated facility) but often causes challenges for nonadiabatic molecular dynamics due to the need for accurate electronic energies and transition dipole moments between excited electronic states. Examples of nonadiabatic molecular dynamics employed for TAS can be found in Ref.~\citenum{levine2024tasaims} or Ref.~\citenum{gelin2021TAS}, the latter using the doorway–window representation.\cite{yan1989doorwaywindow,yan1990pumpprobe} A more recent flavour of TAS employed attosecond X-ray pump/probe pulses to trigger the nonadiabatic dynamics.\cite{schuurman2021ATAS} Nonadiabatic molecular dynamics can also provide information about the dynamics of electronic wavepackets -- a coherent superposition of nuclear wavepackets on different electronic states -- generated in the context of attosecond spectroscopy. Such processes often require a more accurate description of the nuclear wavefunctions, and TBFs-based methods are often required. For example, DD-vMCG was used to investigate the electronic relaxation upon photoionization,\cite{vacher2017electron} AIMS was employed to simulate the time-dependent dipole moment of transient electronic wavepackets,\cite{mignolet2018walk} and a semiclassical thawed Gaussian approximation scheme could resolve the coupled electron-nuclear dynamics resulting from a core-valence attosecond TAS experiment.\cite{golubevPRL2021}  The probe pulse can also be scattered by the excited molecule to recover information about its molecular structure at a specific time delay (and sometimes also its electronic character). In time-resolved ultrafast electron diffraction (TRUED), a bunch of near-relativistic (MeV) electrons is sent to probe the excited molecule, and the collected diffraction pattern contains information about pairwise atom distances. The resulting signal of this probing strategy is rather simple to determine from the result of a nonadiabatic molecular dynamics simulation (with some caveats), allowing for the investigation of various photophysical processes\cite{yang2018imaging,wolf2019photochemical,PhysRevLett.131.143001} and the formation of photoproducts.\cite{nunes2024ued} TRUED was recently used in the context of a prediction challenge for the nonadiabatic dynamics community.\cite{predictionchallenge} The goal of the challenge was to predict the TRUED signals resulting from the photoexcitation of cyclobutanone at 200nm. This excitation wavelength leads to a very diverse photochemistry of cyclobutanone: the molecule undergoes an adiabatic dynamics in the second excited electronic states, evolving its electronic character from the initially formed 3s Rydberg state to a valence excited state; this adiabatic change of electronic character leads to a ring opening and nonadiabatic transitions to the ground electronic state, where further the parent molecule fully dissociates into photoproducts. This challenge highlighted, once again, the sensitivity of the nonadiabatic molecular dynamics to the underlying electronic-structure theory. Hard X-ray scattering can be used to obtain information about electronic rearrangement during photodynamics.\cite{gabalski2025imaging}
X-ray induced Coulomb explosion imaging is another experimental approach that can use nonadiabatic molecular dynamics for the interpretation of the produced signal (e.g., Ref.~\citenum{unwin2023x,schnorr2023direct}). 

Nonadiabatic molecular dynamics can also be deployed to obtain information about time-independent experimental observables. The yield of photoproducts is a typical quantity that can be extracted from a nonadiabatic molecular dynamics solution. A caveat is that, in the gas phase, some photoproducts can be formed in the ground electronic state\cite{mignolet2016rich} and their population can continue to evolve for numerous picoseconds following the nonradiative decay (see Ref.~\citenum{pathak2020tracking} for an example where photoproducts evolve over tens of picoseconds). As such, yields of photoproducts may be seen as time-dependent in the early times of the dynamics, and care is required when comparing these quantities with experiment. Nonadiabatic dynamics can be extended in the ground electronic state by switching to Born-Oppenheimer molecular dynamics if the molecule leaves the nonadiabatic region between the ground and excited electronic state(s), but note that this is not always possible (in particular in the case of near degeneracies). This switch of approaches was discussed above (Section~\ref{sec:practicalcons}). It is important to stress here that long-time simulations can be challenging for nonadiabatic dynamics\cite{Mukherjeelongtime2025} and lead to issues with zero-point energy leakage.\cite{PersicoZPE2023} 

Photolysis processes are of prime importance for atmospheric photochemistry, where sunlight can excite (transient) atmospheric molecules leading to photochemical pathways that can compete with typical removal chemical reactions. The characterization of atmospheric photochemical reactions requires the determination of wavelength-dependent quantum yields for the various possible photolysis channels, a quantity often challenging to obtain experimentally\cite{Curchodatmophotochem2024} but which can be approached computationally.\cite{Prljphotolysis2020} Nonadiabatic molecular dynamics simulations was used to unravel the formation of photoproducts for molecules in the troposphere (e.g., \ce{Hg(II)} species,\cite{rocasanjuan2020mercury}, volatile organic compounds,\cite{marsili2022theoretical,HuttonPA2022}), the upper atmosphere (\ce{HOSO}\cite{carmona-garcia2021hoso}), and the interstellar medium (methanol\cite{cigrang2024methanol}). Another example of energy-resolved quantities reachable by nonadiabatic molecular dynamics simulations are stationary photoionizaton and photoabsorption spectra -- see Refs.~\cite{toddaims,taylor2018phenol,parkes2024thiophene} for typical examples. Last but not least, translational kinetic energy maps can be obtained from nonadiabatic molecular dynamics simulations and support the interpretation of various experiments like velocity map imaging (for selected examples, see Refs.~\citenum{lodriguito2009tshmethane,prlj2023deciphering,janos2023cf3cocl,qu2005ozone}).

\section{FAQ}
\label{sec:faq}
In the following, we try to address (in a more informal tone) some typical questions related to the practical use of nonadiabatic molecular dynamics. 
\begin{itemize}

\item{\textit{Do I really need nonadiabatic molecular dynamics?} \\
Not always! There are cases where the general photochemistry of a molecule can be deduced from simpler calculations, like LIICs between critical points. Experimental observables can also be calculated along these pathways and already provide key information. 
}
\item{\textit{I want to use a certain level of electronic-structure theory, but a single-point calculation (just for electronic energies) at a single geometry of a LIIC takes already 2 minutes on my workstation -- is nonadiabatic dynamics doable at all with this method?} \\
Let us consider that the goal of the nonadiabatic molecular dynamics is to obtain some information on the early times of the dynamics, say the first 100 fs, and no minor channels are expected (requiring more trajectories for convergence). As a very rough back-of-the-envelope calculation, let us multiply by two the time of the electronic-energy calculation to account for the need in dynamics to obtain nuclear gradients (and possibly NACVs) for each time step. Using a 0.5 fs time step for a 100 fs trajectory, and considering a swarm of 100 TSH trajectories, the estimated time for this nonadiabatic dynamics simulation would already be $\sim56$ days -- \textit{à bon entendeur}! For a more precise estimate of the computational effort of running a TSH calculation, see Ref.~\citenum{barbatiblogtime} 
}
\item{\textit{Can I simulate intersystem crossing processes -- for example, nonadiabatic transfer between a singlet and a triplet state -- with nonadiabatic molecular dynamics?} \\
The vast majority of nonadiabatic methods have been generalized for the description of intersystem crossings. The influence of spin-orbit coupling on excited-state dynamics can be implemented either in a spin-diabatic basis -- electronic energies are those obtained from solving the usual electronic Schr\"{o}dinger equation and are spin pure, the coupling between electronic states of different spin multiplicity is obtained by calculating spin-orbit matrix elements of a spin-orbit coupling Hamiltonian (as done for diabatic couplings in the diabatic representation discussed in Section~\ref{sec_adiabatic_diabatic}) -- or in a spin-adiabatic basis -- electronic states are obtained upon diagonalization of the spin-diabatic electronic Hamiltonian containing spin-orbit coupling, the spin-adiabatic electronic states do not have a proper spin multiplicity and are coupled solely via the nonadiabatic coupling terms.\cite{michl,mai2015IJQC,granucci:22A501} The reader interested in learning more about the various approaches available is referred to Ref.~\citenum{penfold2018spin,mai2015IJQC}. From a practical aspect, performing nonadiabatic dynamics that includes intersystem crossing requires additional benchmarks to validate (1) the description and ordering of electronic states with a different spin multiplicity (e.g., LR-TDDFT is known to underestimate the energy of triplet states\cite{peach2011influence}) and (2) the calculation of spin-orbit coupling matrix elements (e.g., use of one- or two-electron spin-orbit coupling Hamiltonian vs effective one-electron operators, impact of the basis set, in particular when using the ZORA Hamiltonian). For trajectory-based approaches like surface hopping, additional considerations would be: (i) energy conservation when using a spin-adiabatic basis (as nuclear gradients for spin-adiabatic states often neglect a contribution coming from the spin-orbit coupling contribution to the Hamiltonian,\cite{mai2015IJQC}) and (ii) certain surface hopping schemes using a spin-diabatic basis may lead to issues with rotational invariance.\cite{granucci:22A501} 
Generalized AIMS allows for the description of internal conversion and intersystem crossing processes in a spin-diabatic representation, but comes at a high computational cost given the number of TBFs spawned to describe the spin-orbit coupling interaction.\cite{curchod2016communication,varganovGAIMS2021}
The exact-factorization-based method coupled-trajectory mixed quantum/classical was also extended to include spin-orbit coupling.\cite{talotta2020soc} The algorithm was formulated both in the spin-adiabatic and the spin-diabatic basis, and it was shown, based on numerical tests, that the results slightly depend on the electronic representation -- the spin-adiabatic being the most suitable. Last but not least, the timescales for intersystem crossing are often significantly longer than those of internal conversion, meaning long-timescale nonadiabatic dynamics simulations\cite{favero2013dynamics} with possible limitations for some methodologies.\cite{Mukherjeelongtime2025}}  

\item{\textit{If a trajectory crashes, shall I simply discard it?} \\
The simple answer is: no! A crashed trajectory may tell us something important about the limitations of our model. A trajectory often crashes due to an issue with the electronic-structure convergence -- electronic energy convergence, but also sometimes calculation of the nuclear gradients or the NACVs -- and the precise reason for this behavior should be investigated. An instability in the electronic-structure calculation may indicate that the trajectory visited a specific (possibly yet unexplored or unexpected) region of the nuclear configuration space, for which the level of electronic-structure theory fails (due to an active space problem, or the need for more electronic states to describe this region adequately). Hence, a careful examination of the molecular geometry at the time of the crash and its electronic structure is required. If the convergence issue is not related to any specific photochemical or photophysical pathway but is most likely due to another factor considered not important for the process under study, one may discard the trajectory. However, one should always report the fraction of discarded trajectories in the computational details for transparency. As a note, AIMS may be more prone to numerical instabilities in the electronic structure because TBFs remain throughout the propagation in their assigned electronic states (unlike TSH, where a trajectory will hop between electronic states). Hence, there is a larger probability that a TBF may find a region of nuclear configuration space where the electronic structure may become unstable. If the TBF that crashed carries a very low nuclear amplitude, it may, in principle, be discarded, but the coupled nature of TBFs requires great care in this process. SSAIMS offers a strategy to perform a selection of the important TBFs and dramatically reduces this issue for AIMS simulations.\cite{curchod2020ssaims,ibele2021ssaims}
}

\item{\textit{How do I estimate the error bar for the result of my simulation, e.g., electronic populations?} \\
A common practice is to employ the multinomial distribution for estimating errors from trajectories. Taking the example of electronic populations, the electronic population of state $J$ reads $P^\mathrm{el}_J = \frac{N_J(t)}{N_\mathrm{traj}}$, where $N_J$ is the number of trajectories in state $J$ at time $t$ and $N_\mathrm{traj}$ is the total number of trajectories. The standard deviation (error) of the population from the statistical sample is by multinomial distribution defined as $\sigma_J = Z\sqrt{\frac{P^\mathrm{el}_J (1-P^\mathrm{el}_J)}{N_\mathrm{traj}}}$ where Z depends on the desired confidence interval ($Z=1.96$ corresponds to 95\% confidence interval). The same approach can be applied to the error estimate of quantum yields. A more robust approach to estimating errors without any assumptions about the distribution is bootstrapping. Other strategies for trajectory-based methods are discussed in Ref.~\citenum{persico2014overview}.  

At this point, we also need to comment on the common issue of failed trajectories in the sample and how to deal with them when evaluating observables such as electronic populations. First, one should consider whether the trajectories can be artificially prolonged. For example, if the trajectory failed in the ground state and we know that the molecule never hops back to the excited state, we can consider the trajectory in the ground state until the final time. Yet, this assumption is not possible for trajectories failing in the excited state. These trajectories can be either discarded from the sample or considered in the statistics only up to the time they reach. Neither of these two approaches can be claimed to be better, and one should always try both of them. If the populations calculated by both approaches do not significantly differ, we can safely choose one of them. If there is a significant difference, one should always report on that and investigate the implications.
}

\item{\textit{Shall I use a thermostat with nonadiabatic molecular dynamics to mimic the temperature effect?} \\
No! The equations of motion used in approximate nonadiabatic molecular dynamics methods are designed to mimic quantum equations of motion, which conserve the total energy of a closed system in the absence of time-dependence within the Hamiltonian (e.g., from external time-dependent fields). Therefore, conventional nonadiabatic simulations should be carried out in the microcanonical ensemble, where total energy is conserved. Temperature effects are typically incorporated through sampling of ICs. In this context, we do not discuss systems interacting with an environment, where Langevin thermostats are sometimes employed ad hoc to mimic environmental effects by introducing random kicks to the molecules not explicitly present in the simulation. 
}

\item{\textit{How many trajectories do I need in TSH (or AIMS)?} \\
As advocated in Section~\ref{sec:practicalcons}, one should start with a few trajectories first ($\sim10$) to make sure all the practical and numerical aspects of the simulation are under control. The number of trajectories required highly depends on the quantities of interest from the simulation, as discussed thoroughly in Ref.~\citenum{persico2014overview}. An electronic population trace can be well described by a few tens of trajectories, but may require much more to be fully converged. The determination of a quantum yield for a minor channel can require hundreds of thousands of trajectories -- 169200 trajectories were required to simulate the (weak) chemiluminescence caused by \ce{Al + H2O} collisions.\cite{alvarezbarcia2013chemiluTSH}
An element to keep in mind is that a mixed quantum/classical method like TSH does not reach an exact solution (of the time-dependent molecular Schr\"{o}dinger equation) with a large number of trajectories, but a numerically-converged TSH result within all its underlying approximations. See Ref.~\citenum{barbattinumberoftrajs} for further discussion on this topic.
}

\item{\textit{How important are the initial nuclear momenta in a trajectory-based nonadiabatic molecular dynamics?} \\
Initial nuclear momenta may be important on certain types of photo-triggered processes that take place on a very short (less than 100fs) timescale. Some of the fastest chemical reactions are associated with proton transfer. The vibrational period of the OH bond, $\sim10$ fs, represents the ultimate time limit for a chemical reaction. However, in photoacids, proton transfer typically occurs over a longer timescale, as the slower motion of heavier atoms governs the overall process. One of the fastest known reactions is proton transfer following the ionization of liquid water: \ce{H2O^{+.} + H2O -> H3O+ + OH^.}, with a lifetime of around 40 fs.\cite{schnorr2023direct} An even more rapid proton transfer is observed after the ionization of a water core electron, occurring in less than 10 fs.\cite{stia2010theoretical} This reaction can be effectively studied using a method called 'core-hole clock', where autoionization during the Auger decay (with a time constant of 4 fs) acts as a strict halt in the process. This type of ultrafast process is predominantly driven by the initial nuclear momentum distribution rather than the nuclear position distribution. A simple one-dimensional model appears revealing in this case. Let us consider the time-evolution for the position $x$ of a classical particle with mass $m$:
\begin{equation*}
x(t)=x(0) + \frac{p}{m}t - \frac{1}{2} \frac{F}{m} t^2 \, .
\end{equation*}
At very short times, the linear term, controlled by the initial momentum $p$, dominates the dynamics. The influence of the term containing the force $F$ becomes significant only later in the process. In some sense, the initial part of the reaction can then be conceptualized as the dispersion of a free wavepacket, meaning that an accurate sampling of the initial nuclear momentum space is crucial for correctly describing such extremely fast reactions. 
}

\item{\textit{When should I worry about zero-point energy leakage in trajectory-based nonadiabatic molecular dynamics?} \\
Zero-point energy leakage may be problematic for slower photochemical reactions that occur on picosecond timescales -- timescales where intramolecular vibrational relaxation (IVR) becomes a critical factor. When a molecule is promoted to an electronically excited state, it often finds itself in highly excited \textit{vibrational} states too, for specific modes. The molecule subsequently redistributes this excess of vibrational energy, leading to the cooling of the initially excited mode (IVR).\cite{Karmakar2020IVR} IVR is inherently a quantum process, which might be too rapid when nuclei are described classically.

The IVR is closely related to an artificial phenomenon known as zero-point energy leakage (mentioned above). Using a Wigner sampling to describe the initial state of a molecule (for example, before photoexcitation) accurately maps the initial vibrational state, including the zero-point energy in each mode. In quantum systems, high-frequency modes retain more energy than low-frequency modes, and each mode will retain at least its zero-point energy. In contrast, classical systems assume equal energy distribution across all modes, with no minimal zero-point energy limits. As a result, a classical evolution (based on an initial quantum distribution of the initial vibrational state) tends to cause an unnatural energy flow from high-frequency modes to low-frequency modes -- defining zero-point energy. 

Zero-point energy leakage can affect the results of nonadiabatic molecular dynamics methods based on a swarm of classical trajectories (TSH) when long timescale dynamics are considered,\cite{persico2014overview,PersicoZPE2023} and different corrections for trajectory-based techniques were suggested.\cite{mukherjee2022hessian,PersicoZPE2023}
}

\item{\textit{How about simulating the photoexcitation of a molecule with a continuous-wave (CW) laser?} \\
As stated earlier, short laser pulses are not always used in photochemical experiments. Instead, photochemical reactions are often initiated by radiation at a well-defined wavelength, for example, with nanosecond pulses which can be treated almost as continuous-wave (CW) radiation from a theoretical perspective.\cite{suchan2018importance} Another common scenario of photoexcitation departing from short laser pulses is when the molecule is exposed to thermal radiation, such as sunlight.\cite{barbatti2020simulation} In these cases, the photodynamics can differ significantly from those observed when the reaction is initiated by a short laser pulse generating a nuclear wavepacket in the excited electronic state(s). One may legitimately ask whether such photochemical processes can still be mimicked by using (classical) trajectories. 

While still being an active topic of research, the photochemistry triggered by a CW laser can, in some cases, still be approximated by trajectory-based nonadiabatic molecular dynamics.\cite{suchan2018importance} A CW laser is essentially a specific example of a laser pulse scenario, and one can use the Wigner transform formalism (including the CW laser pulse) to generate a 'filtered' Wigner distribution. This filtered distribution can technically be produced by applying an energy filtering to a ground-state Wigner distribution, which can be sampled using techniques like the QT method combined with ab initio molecular dynamics. However, this approach may become computationally expensive, especially for excitations at the tails of the distribution. An alternative, considered in Ref.~\citenum{suchan2018importance}, is to perform long ground-state dynamical simulations by applying a set of constraints to enforce a given resonance condition (between the desired CW laser frequency and the energy gap between the ground and a given excited electronic state). This strategy is expected to work best for photodissociation (as would the use of a windowing in this particular case).

Thinking of possible future developments in describing state-specific photoexcitation, one could, for example, try to sample a given eigenstate of an excited electronic state by using some of the methodologies described in Section~\ref{sec:gssampling}. Trajectory-based nonadiabatic dynamics could then be launched from these ICs, possibly over long timescales and their associated challenges (see the question above on zero-point energy leakage and Ref.~\citenum{mukherjee2022simulations}). One may also work on an extension of the PDA, invoking the conceptual idea that a CW pulse is equivalent to a pulsed short laser with a high repetition rate.\cite{heller2018semiclassical}
}

\item{\textit{Do I need to account for geometric phase effects in nonadiabatic molecular dynamics?} \\
Geometric (or topological) phase effects~\cite{mead_determination_1979, berry_quantal_1984} are a more definitive signature of a CX. However, when using a trajectory-based approach to nonadiabatic dynamics using the adiabatic representation, geometric phase effects in the vicinity of CXs can usually be neglected, which is due to the locality of classical mechanics and, often, to a `fortuitous' error cancellation in the underlying approximations.\cite{ryabinkin2017geometric,ibele2021diabolical,ryabinkin2014we} In short, the appearance of geometric phases can be understood as follows. When parallel-transported along a path in nuclear configuration space, the (real-valued) adiabatic electronic eigenstate picks up a purely geometric phase factor, $\exp{(i\gamma_n)}$, in addition to the dynamical phase factor, which appears naturally from the solution of the electronic TDSE.\cite{tannor2007approximate} The associated phase, $\gamma_n$, is a geometric quantity given that its value depends on (the geometry of) the path.\cite{daggett2024toward, ibele2023nature} However, if the path encircles a singularity in the nuclear configuration space, highlighted by a pole in the $\nacv{I}{J}$ vector (field), the phase becomes topological~\cite{daggett2024toward} and depends only on the winding number of the path.\cite{ibele2023nature} If the closed path encircles a CX, then the topological phase takes a value of $\pi$; if it does not enclose a CX, then it takes a value of zero.\cite{mead_geometric_1992} In the former case, this results in the adiabatic electronic wavefunction gaining a phase factor of $-$1 (i.e., its sign flips) and, thus, becoming double-valued.\cite{longuet-higgins_studies_1958, herzberg_intersection_1963}  To guarantee that the total molecular wavefunction remains single-valued, either (i) the (real-valued) nuclear wavefunction must also gain a topological phase of $\pi$ upon full rotation around a CX,\cite{tannor2007molecular} or (ii) as suggested by Mead and Truhlar,\cite{mead_determination_1979} the adiabatic electronic Hamiltonian can be transformed to introduce the resolution of identity, $1 = \exp{[i\theta(\R)]} \exp{[-i\theta(\R)]}$, giving now single-valued (but complex) electronic and nuclear wavefunctions.\cite{tannor2007molecular, ryabinkin2017geometric} 
}

\item{\textit{What about using machine learning in nonadiabatic molecular dynamics?} \\
Great progress has been made in the development of machine-learning approaches for excited-state processes, yet with challenges. The interested reader is referred to Refs.~\citenum{westermayr2021MLNAMD,dral2021molecular,li2023MLNAMD} for reviews on the topic. 
}

\item{\textit{What is the practical relevance of CXs in trajectory-based nonadiabatic dynamics simulations?} \\
Despite their theoretical importance, the strict points of degeneracy at CXs are rarely visited directly in TSH or AIMS simulation runs,\cite{levine2007isomerization,curchod2017ab} rather hopping/spawning events occur at very small, but finite, energy gaps in the vicinity of CXs (i.e., at supposed avoided crossings).
This consideration relates to the fact that, as mentioned above, the seam space is two dimensions smaller than the full space of molecular coordinates, and as such, classical trajectories/TBFs simply miss the intersection seam in practice.\cite{curchod2018ab}
Strictly speaking, exact points of degeneracy are unobtainable in computational investigations of real molecules due to the finite numerical accuracy of electronic-structure codes.\cite{zhu_non-adiabaticity_2016,yarkony_conical_2001}
Nonetheless, CXs serve in a practical sense as `PES staples',\cite{todd2024private} bringing adiabatic electronic states in close proximity to one another (and arguably closer than can be achieved by true ACs alone), such that rapid nonadiabatic transitions (described practically through hopping/spawning events) can occur. 
Furthermore, as alluded to above, MECX geometries, in particular, act as mechanistic `PES signposts',\cite{levine2007isomerization,curchod2017ab} which highlight likely photochemical reaction pathways and can, in a number of cases, be indicative of hopping/spawning geometries.
}

\end{itemize}

\section{Checklist}
\label{sec:checklist}

The attached checklist attempts to group all relevant steps and checks for successful and reliable performance of nonadiabatic dynamics. Spanning electronic structure selection, generation of ICs, and running excited-state dynamics, it follows the previous paragraphs and reflects best practices. Ticking all items in the checklist should lead to reliable nonadiabatic dynamics. However, we note that the checklist may miss some important points. 

\begin{Checklists*}[p!]

\begin{checklist}{Electronic-structure selection}
\textbf{Elements to consider when selecting an appropriate electronic-structure method}
\begin{itemize}
\item Review literature for anticipated electronic-state characters and possible photoreactions.
\item Estimate how many excited states are needed.
\item Assess if the system requires a single-reference or multireference method.
\item Consider what orbitals should be included in the active space if multireference treatment is necessary. 
\item Estimate whether dynamic correlation will be important.
\item Check whether double excitations, charge transfer or Rydberg states are to be expected.
\item Check if intersystem crossing is to be expected.
\item Assess the affordability of selected methods for the expected timescale of simulations.
\end{itemize}
\textbf{Benchmarking in the Franck-Condon region}
\begin{itemize}
\item Locate the ground-state minimum and benchmark excited-state characters, excitation energies, and oscillator strengths for several methods. 
\item Benchmark various basis sets on vertical excitations, including effects of polarization and diffuse functions.
\item Benchmark SOC magnitude for various SOC operators, if applicable.
\item Select a proper active space capturing excited states of interest, if applicable.
\item For selected methods, compare absorption spectra from the nuclear ensemble approach or the Franck-Condon Herzberg-Teller method. Contrast them with the experimental spectrum, if available.
\end{itemize}
\textbf{Benchmarking beyond the Franck-Condon region}
\begin{itemize}
\item Locate critical molecular geometries: local minima and transition states for each electronic state, together with minimum-energy conical intersections between the electronic states.
\item Assess the topology and topography of conical intersections.
\item Create interpolation coordinates between the critical points.
\item Benchmark the electronic structure using the interpolated coordinates.
\item Benchmark various active spaces along the interpolated coordinates and select the most stable one, if applicable.
\item Select the final method and recalculate critical geometries and interpolation coordinates.
\end{itemize}
\end{checklist}

\begin{checklist}{Initial conditions}
\textbf{Ground-state density sampling}
\begin{itemize}
\item Locate all relevant local minima with non-negligible Boltzmann weights. 
\item Perform vibrational analysis on the minima and identify possibly problematic modes. 
\item Select an appropriate method for the ground-state density sampling. 
\item \textbf{hW}: Select normal or Cartesian coordinates. 
\item \textbf{hW}: Remove possibly problematic low-frequency modes. 
\item \textbf{MD}: Select the thermostat and set up its parameters. 
\item \textbf{MD}: Check that the simulation is properly thermalized before extracting initial conditions.
\item \textbf{PIMD}: Select a method to generate complementary velocities to the quantum distribution of positions. 
\item Calculate the absorption spectrum to validate the ground-state sampling.
\end{itemize}

\textbf{Laser pulse inclusion} 
\begin{itemize}
    \item Characterize laser field: temporal duration and energy spectrum.
    \item Identify target excited states using the calculated absorption spectrum and pulse energy spectrum. 
    \item Select an approximation for the excitation: vertical excitation, CW, PDA, or explicit laser pulse. 
\end{itemize}
    
\end{checklist}
\end{Checklists*}

\clearpage
\begin{Checklists*}[p!]

\begin{checklist}{Excited-state dynamics}

\textbf{Before simulations}
\begin{itemize}
    \item Prepare the guess for the electronic wavefunction to initiate the trajectories.
    \item Set the time step for nuclear propagation. 
    \item Set the total simulation time based on the expected timescale. 
    \item Select an approach to calculate NACs and set the threshold for their calculations. 
    \item Set the total energy conservation criteria reflecting the electronic-structure methods used. 
    \item \textbf{TSH}: Set the velocity rescaling method and frustrated hops handling.
    \item \textbf{TSH}: Select the decoherence correction (if applicable) and set its parameters.
    \item \textbf{TSH}: Check the integrator for the electronic Schrödinger equation and its time step. 
    \item \textbf{AIMS}: Set the spawning threshold.
    \item \textbf{AIMS}: Set the TBF widths $\alpha$ for elements without default values.
    \item Carefully check and set remaining method-dependent parameters based on literature and tests.
\end{itemize}

\textbf{After simulations}
\begin{itemize}
    \item Check failed simulations and attempt to restart them (carefully checking the continuity of propagated quantities and the sign of NACs). For simulations failing in the ground state, consider carrying on the propagation with adiabatic Born-Oppenheimer MD. 
    \item Check the initial orbitals in each trajectory to confirm the correct initial wavefunction.
    \item Decide whether to include failed trajectories that cannot be restarted in statistics or not.
    \item Analyze individual trajectories: potential, kinetic, and total energies, electronic coefficients/amplitudes, hopping/spawning geometries, etc.
    \item Check the total energy conservation and analyze possible discontinuities.
    \item Check distributions of energy gaps between electronic states at hopping/spawning geometries. Confirm that the threshold for calculating NACs is larger than the energy gap spread.
\end{itemize}

\textbf{Analysis of the results}

\begin{itemize}
    \item Analyze electronic-state populations and estimate their confidence intervals.
    \item Play the molecular movies and analyze the behaviour of the molecule. Contrast the observed mechanism with the mapping of PESs.
    \item Analyze the time evolution of specific molecular parameters, i.e., bond distances, angles, or dihedrals of interest.
    \item Characterize hopping/spawning geometries and contrast them with minimum-energy conical intersections.
    \item Check for hops/spawns over multiple states, e.g., from S\textsubscript{2} to S\textsubscript{0}, as they indicate unusual photochemistry.
    \item Calculate time-dependent experimental observables like diffraction, photoelectron, or absorption spectra.
\end{itemize}
\end{checklist}
\end{Checklists*}
\clearpage

\section{Rationale}
\label{sec:rationale}
In this rationale for the checklist, only specific technical details omitted in the main text are provided. We reference the specific sections touching a particular aspect for further information and to avoid repetition.

\subsection{Electronic-structure selection}

The aspects to consider and guidance on how to select appropriate electronic-structure methods are detailed at the beginning of Section~\ref{sec:banchmarkelstr}. Initial benchmarking in the FC region is rationalized in Section~\ref{sec:fctest}. Subsequent mapping of PESs and benchmarking beyond the FC regions is detailed in Section~\ref{sec:beyondfstest}. A thorough guideline for controlled selection of a proper and stable active space is highlighted in Section~\ref{sec:activespace}. We note that the selection of a proper electronic-structure method and the correct setting of its parameters is a complex and strongly system-dependent task, which may require considerations beyond those presented here. A chemical intuition and thorough literature survey are always essential.

\subsection{Initial conditions}

The rationale for the ground-state density sampling is provided in Section~\ref{sec:gssampling}. Specific options for different ground-state sampling methods are denoted with the following abbreviations: harmonic Wigner (hW), molecular dynamics (MD), and path integral MD (PIMD). 
The list of thermostats available for MD simulations goes beyond that mentioned in this text. While based on different principles, they all require specific parameters for correct functioning, e.g., QT needs GLE matrices while the Nosé-Hoover thermostat requires appropriate coupling constants to balance thermal exchange between the system and the thermostat. Proper thermalization of the simulation before extracting ICs from the MD simulation is also necessary. The ICs should be extracted along the trajectory such that they are uncorrelated, i.e., with sufficient temporal separation between the selected time frames. 

Calculating absorption spectra to validate the ground-state sampling and identifying target excited states is described in Section~\ref{sec:photoabscrosssec}.

For an appropriate description of photoexcitation, a careful characterization of the laser pulse is necessary. The key factors are the pulse temporal duration (pulse intensity profile) and energy spectrum, which dictate the model for excitation to be used and the number of excited states targeted. The choice depends strongly on the pulse and techniques used for nonadiabatic dynamics. The options for laser pulse inclusion are detailed in Section~\ref{sec:initcond}. Note that when the laser pulse is not known before simulations, a sudden vertical excitation constitutes the most straightforward strategy. Moreover, such a simulation can be later post-processed with PDA or PDAW to include a specific laser pulse.

\subsection{Excited-state dynamics}
We note that the focus of the presented checklist is not on selecting the best method for excited-state dynamics. More details on possible strategies can be found in Section~\ref{nonadiabaticdynproblem}. Instead, the checklist highlights best practices when performing simulations with typical trajectory-based methods for nonadiabatic dynamics.

\subsubsection{Before simulations}
We stress that the checks and steps necessary before running nonadiabatic dynamics are highly method-dependent. Hence, the checklist features only the most typical checks and steps that are common for most trajectory-based methods. As discussed in Section~\ref{sec:practicalcons}, it is highly recommended to start by running a small set of testing trajectories and carefully inspect their behaviour before launching a larger swarm. The thresholds for various parameters can be more benevolent for the testing set of trajectories, which allows a more efficient exploration of the performance of each method. Based on the testing set of simulations, tighter thresholds should be explored and set for production simulations.

A detailed discussion about some practical aspects, such as conservation of total energy, calculations of nonadiabatic couplings, and velocity rescaling, is provided in Section~\ref{sec:practicalcons}, and decoherence corrections are mentioned in Section~\ref{sec:classtrajmethod}. We focus here only on some remaining practical aspects of trajectory-based simulations. A crucial step for reliable simulations is a proper initialization of the electronic wavefunction, especially for multiconfigurational and multireference techniques. Either load a precomputed wavefunction guess or ensure the wavefunction guess is correctly configured in the electronic-structure interface. Starting from a HF wavefunction or even from a multireference wavefunction computed at a (different) distorted geometry can lead to orbital rotations, see Section~\ref{sec:activespace}. 
A nuclear time step of 5 atomic units (0.12~fs) typically suffices. Larger time steps like 10–20 atomic units may be computationally cheaper but bear the risk of missing sharp avoided crossings or trivial intersections (if no detection strategy is deployed). These issues might be mitigated by the use of time-derivative couplings in TSH (various implementations were proposed), yet the time-derivative couplings have their own pitfalls and do    not allow for velocity rescaling along the NACVs. The threshold for calculating NACs is typically around 1 or 1.5 eV, but may be strongly system-dependent.
Finally, each method has its specific parameters. For example, AIMS requires more parameters than TSH,\cite{lassmann2023} such as setting up minimum overlap to calculate couplings, minimum population of a TBF to spawn a child TBF, or the TBF widths (not available for all elements but with protocols described to calculate them\cite{THOMPSON201070width,esch2019widthlevine} and a code available.\cite{yorick_lassmann_2022_7382685})

\subsubsection{After simulations}

A thorough inspection of the simulations is mandatory once they are terminated. First, check the failed simulations. If possible, restart them or finish those failing in the ground state with adiabatic Born-Oppenheimer molecular dynamics, see Section~\ref{sec:practicalcons}. If the simulations cannot be restarted, decide whether to include them in the statistics or not, see FAQ in Section~\ref{sec:faq} for discussions. For all simulations, check that the initial electronic wavefunction was loaded correctly. Monitoring of a single TSH trajectory is detailed in Section~\ref{sec:tshoutcome} while analysis of a single AIMS run is discussed in Section~\ref{sec:aimsoutcome}. The total energy conservation and discussion on discontinuities is available in Section~\ref{sec:practicalcons}. Finally, analysis of the energy-gap distribution at hopping or spawning geometries may reveal pathological problems with the dynamics. The distribution should peak around or close to zero and then exponentially decay. If the distribution does not decay before reaching the threshold for computing NACs, increase this threshold and rerun the simulations.

\subsubsection{Analysis of the results}

Once the user has confirmed that the nonadiabatic dynamics were performed successfully and reliably (at least from a numerical perspective), it is then time to analyze the photochemical (or photophysical) outcomes. A detailed discussion on TSH analysis is available in Section~\ref{sec:tshoutcome}, while AIMS analysis is described in Section~\ref{sec:aimsoutcome}.
Analysis of experimental observables is illustrated in Section~\ref{sec:analysisandobservables}. Inspection of hops/spawns over multiple excited states can reveal rare multistate CXs (Section~\ref{ch2_cx}).

\section{Available software packages}
\label{sec:codes}
The following tables list the available software packages for the different ingredients required for nonadiabatic molecular dynamics. The tables focus on the capabilities related to nonadiabatic molecular dynamics, and the reader is referred to the links provided for the full list of functionalities (and interfaces). We expect this list to grow with future iterations of the live article.

\begin{table*}[h!]
    \centering
    \begin{tabular}{|c|c|c|}
    \hline 
    \textbf{Code}  & \textbf{Main capabilities} & \textbf{Link} \\ 
    \hline
      \texttt{ABIN}   & Quantum-thermostat AIMD, Path integral MD & \href{https://github.com/PHOTOX/ABIN}{GitHub} \\
        \texttt{I-PI}   & Quantum-thermostat AIMD, Path integral MD & \href{https://ipi-code.org}{Website} \\
      \texttt{NEWTON-X} & Harmonic Wigner distribution, NEA  & \href{https://newtonx.org}{Website} \\
      \texttt{SHARC}   & Harmonic Wigner distribution, NEA & \href{https://sharc-md.org}{Website}\\
    \texttt{ATMOSPEC}   & Harmonic Wigner distribution, NEA & \href{https://github.com/ispg-group/aiidalab-ispg}{Website}\\
    \texttt{PROMDENS}   & PDA, windowing approaches & \href{https://github.com/PHOTOX/promdens}{GitHub}\\
    \hline
    \end{tabular}
    \caption{Codes of interest for the sampling of initial conditions and the description of the photoexcitation process.}
    \label{tab:IC}
\end{table*}

\begin{table*}[h!]
    \centering
    \begin{tabular}{|c|c|c|}
    \hline 
    \textbf{Code}  & \textbf{Main capabilities} & \textbf{Link} \\ 
    \hline
        \texttt{MOLPRO}   & SA-CASSCF, RASPT2, (X)MS-CASPT2 & \href{https://www.molpro.net}{Website}  \\
        \texttt{OPENMOLCAS}   & SA-CASSCF, RASPT2 (X)MS-CASPT2 &  \href{https://gitlab.com/Molcas/OpenMolcas}{GitLab}  \\
        \texttt{BAGEL}   & SA-CASSCF, (X)MS-CASPT2  & \href{https://nubakery.org}{Website}  \\
        \texttt{TURBOMOLE}   & LR-TDDFT, ADC(2), CC2  &  \href{https://www.turbomole.org}{Website}  \\
        \texttt{TERACHEM}   & LR-TDDFT  & \href{http://www.petachem.com}{Website}  \\
        \texttt{ORCA}   & LR-TDDFT, ADC(2) & \href{https://www.faccts.de/orca/}{Website}  \\
        \texttt{ET}   & CC2, CC3 & \href{https://etprogram.org}{Website}  \\
        \texttt{GAMESS}  & ORMAS, GMCQDPT2 & \href{https://www.msg.chem.iastate.edu/gamess/}{Website} \\
        \texttt{Q-CHEM}  & LR-TDDFT & \href{https://www.q-chem.com}{Website}  \\
        \texttt{COLUMBUS}  & SA-CASSCF, MRCI  & \href{https://columbus-program-system.gitlab.io/columbus/}{Website}  \\
        \texttt{GAUSSIAN}  & LR-TDDFT, SA-CASSCF  & \href{https://gaussian.com}{Website}  \\
        \texttt{MOPAC}  & FOMO-CI & -  \\
        \texttt{ADF}  & LR-TDDFT & \href{https://www.scm.com/amsterdam-modeling-suite/adf/}{Website}  \\
        \texttt{OPEN-QP}  & MR-SF-TDDFT & \href{https://github.com/Open-Quantum-Platform/openqp}{GitHub} \\
      \hline
    \end{tabular}
    \caption{Electronic-structure codes interfaced with nonadiabatic molecular dynamics packages.}
    \label{tab:IC2}
\end{table*}

\begin{table*}[h!]
    \centering
    \begin{tabular}{|c|c|c|}
    \hline 
    \textbf{Code}  & \textbf{Main capabilities} & \textbf{Link} \\ 
    \hline
      \texttt{NEWTON-X} & FSSH  & \href{https://newtonx.org}{Website} \\
      \texttt{SHARC}   & FSSH, global flux TSH, coherent switching with decay of mixing & \href{https://sharc-md.org}{Website}\\
      \texttt{JADE}   & FSSH, Zhu-Nakamura TSH & \href{https://github.com/zglan/JADE-NAMD}{GitHub}\\
      \texttt{ABIN}   & FSSH, Landau-Zener TSH & \href{https://github.com/PHOTOX/ABIN}{GitHub} \\
      \texttt{COBRAMM} & FSSH  & \href{https://site.unibo.it/cobramm/en}{Website} \\
      \texttt{LIBRA} & FSSH,  global flux TSH, Markov state TSH  & \href{https://github.com/Quantum-Dynamics-Hub/Libra-X}{GitHub} \\
      \texttt{QUANTICS} & MCTDH, vMCG, Zagreb TSH & \href{https://www.chem.ucl.ac.uk/quantics/}{Website} \\
    \texttt{PYRAI2MD} & FSSH, Zhu-Nakamura TSH  & \href{https://github.com/mlcclab/PyRAI2MD-hiam}{GitHub} \\   
          \texttt{G-CTMQC} & (G)CTMQC, coupled-trajectory TSH & \href{https://gitlab.com/agostini.work/g-ctmqc}{GitLab} \\
        \texttt{PYUNIXMD} & CTMQC, TSH with exact-factorization-based decoherence corrections, Ehrenfest dynamics, FSSH & \href{https://github.com/skmin-lab/unixmd}{GitHub} \\
       \texttt{MOLCAS} & FSSH & \href{https://www.molcas.org}{Website} \\
        \texttt{TURBOMOLE} & FSSH & \href{https://www.turbomole.org}{Website} \\
        \texttt{Q-CHEM} & FSSH & \href{http://www.q-chem.com}{Website} \\
        \texttt{OCTOPUS} & Ehrenfest dynamics  & \href{https://octopus-code.org/documentation/16/}{Website} \\
        \texttt{CPMD} & FSSH, Ehrenfest dynamics, CTMQC  & \href{https://github.com/CPMD-code}{GitHub} \\
                \texttt{MOPAC}  & FSSH, AIMS & -  \\
        \texttt{PYSURF} & LZSH, FSSH & \href{https://github.com/MFSJMenger/pysurf}{GitHub} \\
        \texttt{FMS90} & AIMS, XFAIMS, GAIMS, SSAIMS, AIMSWISS  & - \\ 
        \texttt{MOLPRO} & AIMS & \href{https://www.molpro.net}{Website} \\
        \texttt{PYSPAWN} & AIMS  & \href{https://github.com/blevine37/pySpawn17}{GitHub} \\ 
        \texttt{LEGION} & AIMS  & \href{https://gitlab.com/light-and-molecules/legion}{GitLab} \\  
    \hline
    \end{tabular}
    \caption{Codes of interest for the nonadiabatic molecular dynamics.}
    \label{tab:NAMD}
\end{table*}

\clearpage

\section*{Author Contributions}

AP and BFEC wrote the main core of the manuscript, with contributions from JTT, JJ, EL, DH, PS, and FA. All authors contributed to the production of the final version of the manuscript.

For a more detailed description of author contributions of subsequent versions of this work, see the GitHub issue tracking and changelog at \githubrepository.

\section*{Potentially Conflicting Interests}

The authors declare no conflict of interest. 

\section*{Funding Information}

BFEC acknowledges funding from the European Research Council (ERC) under the European Union's Horizon 2020 research and innovation programme (Grant agreement No. 803718, project SINDAM) and EPSRC for the grants EP/V026690/1, EP/Y01930X/1, and EP/X026973/1.

\section*{Author Information}
\makeorcid

\end{document}